\documentclass[12pt,a4paper]{article}
\pdfoutput=1
\usepackage[utf8]{inputenc}
\usepackage{ifthen} 
\newboolean{pdflatex}
\setboolean{pdflatex}{true} 

\newboolean{articletitles}
\setboolean{articletitles}{true} 

\newboolean{uprightparticles}
\setboolean{uprightparticles}{false} 

\newboolean{inbibliography}
\setboolean{inbibliography}{false} 

\usepackage{microtype}
\usepackage{lineno}  
\usepackage{xspace} 
\usepackage{caption} 

\usepackage{graphicx}  
\usepackage{color}
\usepackage{colortbl}
\graphicspath{{./figs/}} 

\usepackage{amsmath} 
\usepackage{amssymb}
\usepackage{amsfonts}
\usepackage{upgreek} 
\usepackage{titlesec}
\usepackage{bigstrut}
\usepackage{import}

\newcommand*\patchAmsMathEnvironmentForLineno[1]{%
\expandafter\let\csname old#1\expandafter\endcsname\csname #1\endcsname
\expandafter\let\csname oldend#1\expandafter\endcsname\csname
end#1\endcsname
 \renewenvironment{#1}%
   {\linenomath\csname old#1\endcsname}%
   {\csname oldend#1\endcsname\endlinenomath}%
}
\newcommand*\patchBothAmsMathEnvironmentsForLineno[1]{%
  \patchAmsMathEnvironmentForLineno{#1}%
  \patchAmsMathEnvironmentForLineno{#1*}%
}
\AtBeginDocument{%
\patchBothAmsMathEnvironmentsForLineno{equation}%
\patchBothAmsMathEnvironmentsForLineno{align}%
\patchBothAmsMathEnvironmentsForLineno{flalign}%
\patchBothAmsMathEnvironmentsForLineno{alignat}%
\patchBothAmsMathEnvironmentsForLineno{gather}%
\patchBothAmsMathEnvironmentsForLineno{multline}%
\patchBothAmsMathEnvironmentsForLineno{eqnarray}%
}		
		
\usepackage{hyperref}    
\usepackage[all]{hypcap} 

\def\lhcb {\mbox{LHCb}\xspace}

\def\lhc    {\mbox{LHC}\xspace}

\def\velo   {VELO\xspace}
\def\rich   {RICH\xspace}
\def\richone {RICH1\xspace}
\def\richtwo {RICH2\xspace}
\def\ttracker {TT\xspace}
\def\intr   {IT\xspace}

\def\ot     {OT\xspace}

\def\spd    {SPD\xspace}
\def\presh  {PS\xspace}
\def\ecal   {ECAL\xspace}
\def\hcal   {HCAL\xspace}
\def\muonch {MUON\xspace}
\def\MagUp {\mbox{\em Mag\kern -0.05em Up}\xspace}

\def\hlt    {HLT\xspace}
\def\hltone {HLT1\xspace}
\def\hlttwo {HLT2\xspace}

\ifthenelse{\boolean{uprightparticles}}%
{

 \def\Pmu         {\ensuremath{\upmu}\xspace}

 \def\Ppi         {\ensuremath{\uppi}\xspace}

 \def\Ppsi        {\ensuremath{\uppsi}\xspace}

 \def\PDelta      {\ensuremath{\Delta}\xspace}                 
 \def\PXi      {\ensuremath{\Xi}\xspace}                 
 \def\PLambda      {\ensuremath{\Lambda}\xspace}                 
 \def\PSigma      {\ensuremath{\Sigma}\xspace}                 
 \def\POmega      {\ensuremath{\Omega}\xspace}                 
 \def\PUpsilon      {\ensuremath{\Upsilon}\xspace}                 
 
 \def\PB      {\ensuremath{\mathrm{B}}\xspace}                 
                  
 \def\PD      {\ensuremath{\mathrm{D}}\xspace}

 \def\PJ      {\ensuremath{\mathrm{J}}\xspace}                 
 \def\PK      {\ensuremath{\mathrm{K}}\xspace}

 \def\PW      {\ensuremath{\mathrm{W}}\xspace}

 \def\PZ      {\ensuremath{\mathrm{Z}}\xspace}                 
                  
 \def\Pb      {\ensuremath{\mathrm{b}}\xspace}                 
 \def\Pc      {\ensuremath{\mathrm{c}}\xspace}

 \def\Pi      {\ensuremath{\mathrm{i}}\xspace}

 \def\Ps      {\ensuremath{\mathrm{s}}\xspace}

}
{

 \def\Pmu         {\ensuremath{\mu}\xspace}

 \def\Ppi         {\ensuremath{\pi}\xspace}

 \def\Ppsi        {\ensuremath{\psi}\xspace}                 
                  
 \mathchardef\PDelta="7101
 \mathchardef\PXi="7104
 \mathchardef\PLambda="7103
 \mathchardef\PSigma="7106
 \mathchardef\POmega="710A
 \mathchardef\PUpsilon="7107
                  
 \def\PB      {\ensuremath{B}\xspace}                 
                  
 \def\PD      {\ensuremath{D}\xspace}

 \def\PJ      {\ensuremath{J}\xspace}                 
 \def\PK      {\ensuremath{K}\xspace}

 \def\PW      {\ensuremath{W}\xspace}

 \def\PZ      {\ensuremath{Z}\xspace}                 
                  
 \def\Pb      {\ensuremath{b}\xspace}                 
 \def\Pc      {\ensuremath{c}\xspace}

 \def\Pi      {\ensuremath{i}\xspace}

 \def\Ps      {\ensuremath{s}\xspace}

}

\makeatletter
\ifcase \@ptsize \relax
  \newcommand{\miniscule}{\@setfontsize\miniscule{4}{5}}
\or
  \newcommand{\miniscule}{\@setfontsize\miniscule{5}{6}}
\or
  \newcommand{\miniscule}{\@setfontsize\miniscule{5}{6}}
\fi
\makeatother

\DeclareRobustCommand{\optbar}[1]{\shortstack{{\miniscule (\rule[.5ex]{1.25em}{.18mm})}
  \\ [-.7ex] $#1$}}

\def\mup        {{\ensuremath{\Pmu^+}}\xspace}
\def\mun        {{\ensuremath{\Pmu^-}}\xspace} 
\def\mumu       {{\ensuremath{\Pmu^+\Pmu^-}}\xspace}

\def\W      {{\ensuremath{\PW}}\xspace}

\def\Z      {{\ensuremath{\PZ}}\xspace}

\def\squark    {{\ensuremath{\Ps}}\xspace}

\def\bquark    {{\ensuremath{\Pb}}\xspace}

\def\pion   {{\ensuremath{\Ppi}}\xspace}
\def\piz    {{\ensuremath{\pion^0}}\xspace}

\def\pip    {{\ensuremath{\pion^+}}\xspace}
\def\pim    {{\ensuremath{\pion^-}}\xspace}

\def\kaon    {{\ensuremath{\PK}}\xspace}
  \def\Kbar    {{\kern 0.2em\overline{\kern -0.2em \PK}{}}\xspace}

\def\KorKbar    {\kern 0.18em\optbar{\kern -0.18em K}{}\xspace}

\def\Kp      {{\ensuremath{\kaon^+}}\xspace}
\def\Km      {{\ensuremath{\kaon^-}}\xspace}

\def\KS      {{\ensuremath{\kaon^0_{\rm\scriptscriptstyle S}}}\xspace}

\def\Kstarz  {{\ensuremath{\kaon^{*0}}}\xspace}

  \def\Dbar    {{\kern 0.2em\overline{\kern -0.2em \PD}{}}\xspace}
\def\D       {{\ensuremath{\PD}}\xspace}

\def\DorDbar    {\kern 0.18em\optbar{\kern -0.18em D}{}\xspace}
\def\Dz      {{\ensuremath{\D^0}}\xspace}
\def\Dzb     {{\ensuremath{\Dbar{}^0}}\xspace}

\def\B       {{\ensuremath{\PB}}\xspace}
\def\Bbar    {{\ensuremath{\kern 0.18em\overline{\kern -0.18em \PB}{}}}\xspace}

\def\BorBbar    {\kern 0.18em\optbar{\kern -0.18em B}{}\xspace}
\def\Bz      {{\ensuremath{\B^0}}\xspace}

\def\Bu      {{\ensuremath{\B^+}}\xspace}

\def\Bp      {{\ensuremath{\Bu}}\xspace}

\def\Bs      {{\ensuremath{\B^0_\squark}}\xspace}

\def\jpsi     {{\ensuremath{{\PJ\mskip -3mu/\mskip -2mu\Ppsi\mskip 2mu}}}\xspace}

  \def\Y#1S{\ensuremath{\PUpsilon{(#1S)}}\xspace}

\def\Lz          {{\ensuremath{\PLambda}}\xspace}
\def\Lbar        {{\ensuremath{\kern 0.1em\overline{\kern -0.1em\PLambda}}}\xspace}
\def\LorLbar    {\kern 0.18em\optbar{\kern -0.18em \PLambda}{}\xspace}

\newcommand{\decay}[2]{\ensuremath{#1\!\to #2}\xspace}         

\def\to                 {\ensuremath{\rightarrow}\xspace}

\def\CP                {{\ensuremath{C\!P}}\xspace}

\def\AT#1     {\ensuremath{A_{\mathrm{T}}^{#1}}\xspace}           

\def\C#1      {\ensuremath{\mathcal{C}_{#1}}\xspace}                       
\def\Cp#1     {\ensuremath{\mathcal{C}_{#1}^{'}}\xspace}                    
\def\Ceff#1   {\ensuremath{\mathcal{C}_{#1}^{\mathrm{(eff)}}}\xspace}        
\def\Cpeff#1  {\ensuremath{\mathcal{C}_{#1}^{'\mathrm{(eff)}}}\xspace}       
\def\Ope#1    {\ensuremath{\mathcal{O}_{#1}}\xspace}                       
\def\Opep#1   {\ensuremath{\mathcal{O}_{#1}^{'}}\xspace}                    

\newcommand{\unit}[1]{\ensuremath{\rm\,#1}\xspace}          

\newcommand{\tev}{\ifthenelse{\boolean{inbibliography}}{\ensuremath{~T\kern -0.05em eV}\xspace}{\ensuremath{\mathrm{\,Te\kern -0.1em V}}}\xspace}
\newcommand{\TByte}{\ensuremath{\mathrm{\,T\kern -0.1em B}}\xspace}
\newcommand{\PByte}{\ensuremath{\mathrm{\,P\kern -0.1em B}}\xspace}
\newcommand{\gev}{\ensuremath{\mathrm{\,Ge\kern -0.1em V}}\xspace}
\newcommand{\mev}{\ensuremath{\mathrm{\,Me\kern -0.1em V}}\xspace}
\newcommand{\kev}{\ensuremath{\mathrm{\,ke\kern -0.1em V}}\xspace}
\newcommand{\ev}{\ensuremath{\mathrm{\,e\kern -0.1em V}}\xspace}
\newcommand{\gevc}{\ensuremath{{\mathrm{\,Ge\kern -0.1em V\!/}c}}\xspace}
\newcommand{\mevc}{\ensuremath{{\mathrm{\,Me\kern -0.1em V\!/}c}}\xspace}
\newcommand{\gevcc}{\ensuremath{{\mathrm{\,Ge\kern -0.1em V\!/}c^2}}\xspace}
\newcommand{\gevgevcccc}{\ensuremath{{\mathrm{\,Ge\kern -0.1em V^2\!/}c^4}}\xspace}
\newcommand{\mevcc}{\ensuremath{{\mathrm{\,Me\kern -0.1em V\!/}c^2}}\xspace}

\def\mum  {\ensuremath{{\,\upmu\rm m}}\xspace}

\def\mus  {\ensuremath{{\,\upmu{\rm s}}}\xspace}
\def\ns   {\ensuremath{{\rm \,ns}}\xspace}

\def\fs   {\ensuremath{\rm \,fs}\xspace}

\def\gsim{{~\raise.15em\hbox{$>$}\kern-.85em
          \lower.35em\hbox{$\sim$}~}\xspace}
\def\lsim{{~\raise.15em\hbox{$<$}\kern-.85em
          \lower.35em\hbox{$\sim$}~}\xspace}

\def\ptot       {\mbox{$p$}\xspace}
\def\pt         {\mbox{$p_{\rm T}$}\xspace}
\def\et         {\mbox{$E_{\rm T}$}\xspace}

\def\evtgen     {\mbox{\textsc{EvtGen}}\xspace}

\def\geant      {\mbox{\textsc{Geant4}}\xspace}

\def\photos     {\mbox{\textsc{Photos}}\xspace}

\def\pythia     {\mbox{\textsc{Pythia}}\xspace}

\def\tell1  {TELL1\xspace}
\def\ukl1   {UKL1\xspace}

\def\lz {\texttt{L0}\xspace}

\def\lzmuon {\texttt{L0Muon}\xspace}
\def\lzdimuon {\texttt{L0DiMuon}\xspace}

\def\hltTrack1MVA {\texttt{Hlt1Track1MVA}\xspace}
\def\hltTrack1MVA {\texttt{Hlt1Track2MVA}\xspace}

\newcommand{\eg}{\mbox{\itshape e.g.}\xspace}

\usepackage{cite} 
\usepackage{mciteplus}

\definecolor{bleudefrance}{rgb}{0.19, 0.55, 0.91}

\usepackage{longtable} 
\usepackage{overpic}

\usepackage{tikz}
\usepackage{units}
\usetikzlibrary{arrows, calc}

\begin{document}

\definecolor{greenp1}{rgb}{0, 0.8, 0}
\newcommand{\mw}[1]{\textbf{\textcolor{greenp1}{[MW: #1]}}\xspace}

\renewcommand{\thefootnote}{\fnsymbol{footnote}}
\setcounter{footnote}{1}


\begin{titlepage}
\pagenumbering{roman}

\vspace*{-1.5cm}
\centerline{\large EUROPEAN ORGANIZATION FOR NUCLEAR RESEARCH (CERN)}
\vspace*{1.5cm}
\noindent
\begin{tabular*}{\linewidth}{lc@{\extracolsep{\fill}}r@{\extracolsep{0pt}}}
\ifthenelse{\boolean{pdflatex}}
{\vspace*{-2.7cm}\mbox{\!\!\!\includegraphics[width=.14\textwidth]{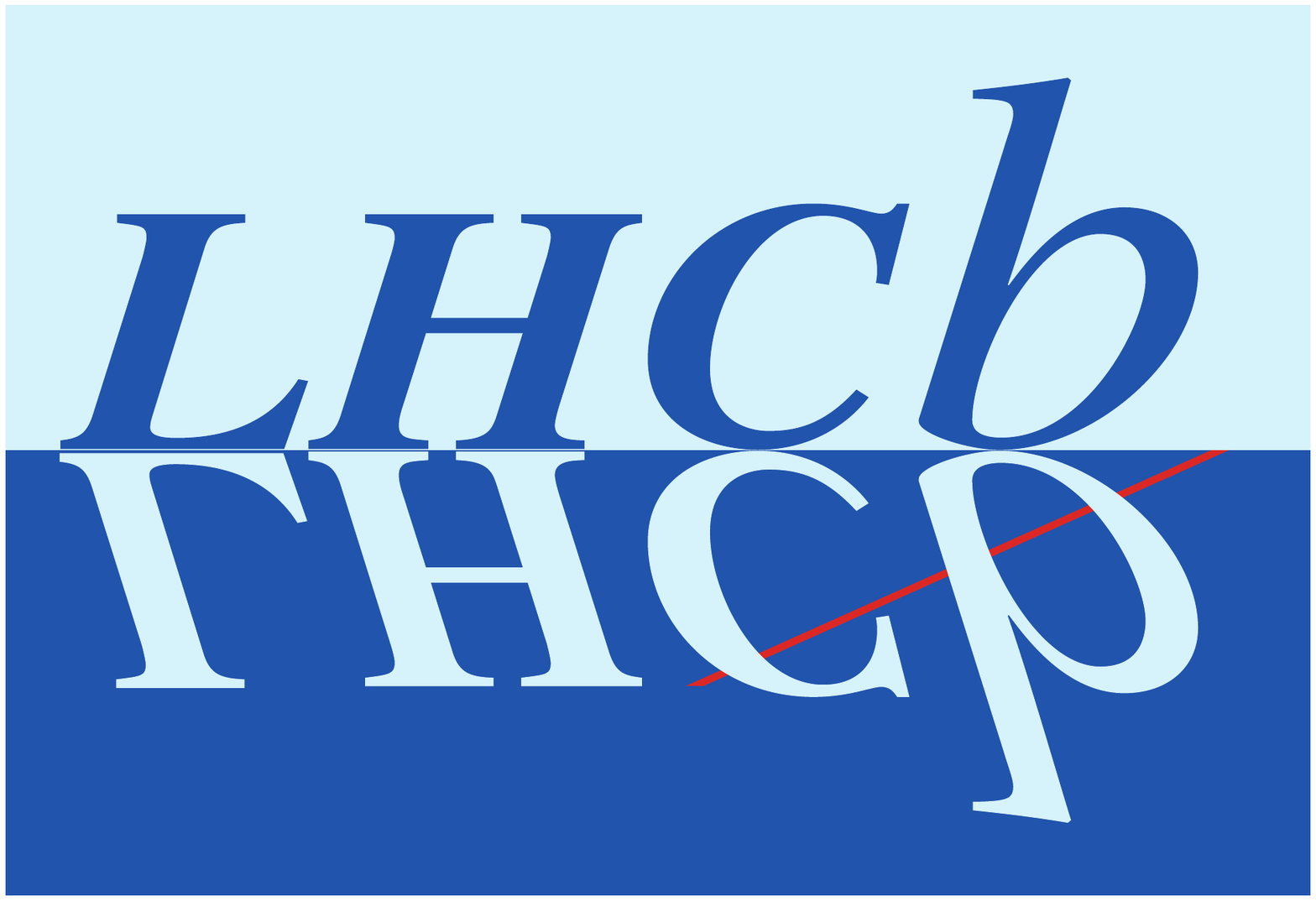}} & &}%
{\vspace*{-1.2cm}\mbox{\!\!\!\includegraphics[width=.12\textwidth]{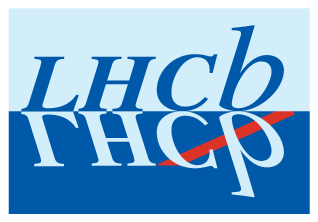}} & &}%
\\
 & & CERN-LHCb-DP-2019-001 \\  
 & & \today \\ 
 & & \\
\end{tabular*}

\vspace*{3.0cm}

{\bf\boldmath\huge
\begin{center}
 Design and performance of the LHCb trigger and full real-time reconstruction in Run 2 of the LHC
\end{center}
}

\vspace*{2.0cm}

\begin{center}
Author list on the following page
\end{center}

\vspace*{1.0cm}

\begin{abstract}
  \noindent
  The LHCb collaboration has redesigned its trigger to enable the full offline detector reconstruction
  to be performed in real time. Together with the real-time alignment and calibration of the detector, and a
  software infrastructure to make persistent the high-level physics objects produced during real-time processing,
  this redesign enabled the widespread deployment of real-time analysis during Run~2.
  We describe the design of the Run~2 trigger and real-time reconstruction, and present
  data-driven performance measurements for a representative sample of LHCb's physics programme.

\end{abstract}

\vspace*{2.0cm}

\begin{center}
 Submitted to JINST 
\end{center}

\vspace{\fill}

{\footnotesize
\centerline{\copyright~CERN on behalf of the \lhcb collaboration, licence \href{http://creativecommons.org/licenses/by/4.0/}{CC-BY-4.0}.}}
\vspace*{2mm}

\end{titlepage}


\newpage
\begin{flushleft}
\small
R.~Aaij$^{19}$,
S.~Akar$^{40,7\dagger}$,
J.~Albrecht$^{11}$,
M.~Alexander$^{34}$,
A.~Alfonso~Albero$^{24}$,
S.~Amerio$^{17}$,
L.~Anderlini$^{15}$,
P.~d'Argent$^{12}$,
A.~Baranov$^{22}$,
W.~Barter$^{26\dagger,36}$
% ...

{\footnotesize \it

$ ^{1}$Universit\'{e} catholique de Louvain, Louvain, Belgium\\
$ ^{2}$Center for High Energy Physics, Tsinghua University, Beijing, China\\
$ ^{3}$University of Chinese Academy of Sciences, Beijing, China\\
$ ^{4}$Institute of Particle Physics, Central China Normal University, Wuhan, Hubei, China\\
$ ^{5}$Univ. Grenoble Alpes, Univ. Savoie Mont Blanc, CNRS, IN2P3-LAPP, Annecy, France\\
$ ^{6}$Clermont Universit\'{e}, Universit\'{e} Blaise Pascal, CNRS/IN2P3, LPC, Clermont-Ferrand, France\\
$ ^{7}$Aix Marseille Univ, CNRS/IN2P3, CPPM, Marseille, France\\
$ ^{8}$LAL, Univ. Paris-Sud, CNRS/IN2P3, Universit{\'e} Paris-Saclay, Orsay, France\\
$ ^{9}$LPNHE, Sorbonne Universit{\'e}, Paris Diderot Sorbonne Paris Cit{\'e}, CNRS/IN2P3, Paris, France\\
$ ^{10}$I. Physikalisches Institut, RWTH Aachen University, Aachen, Germany\\
$ ^{11}$Fakult{\"a}t Physik, Technische Universit{\"a}t Dortmund, Dortmund, Germany\\
$ ^{12}$Physikalisches Institut, Ruprecht-Karls-Universit{\"a}t Heidelberg, Heidelberg, Germany\\
$ ^{13}$INFN Sezione di Bologna, Bologna, Italy\\
$ ^{14}$INFN Sezione di Cagliari, Monserrato, Italy\\
$ ^{15}$INFN Sezione di Firenze, Firenze, Italy\\
$ ^{16}$INFN Laboratori Nazionali di Frascati, Frascati, Italy\\
$ ^{17}$INFN Sezione di Padova, Padova, Italy\\
$ ^{18}$INFN Sezione di Pisa, Pisa, Italy\\
$ ^{19}$Nikhef National Institute for Subatomic Physics, Amsterdam, Netherlands\\
$ ^{20}$Nikhef National Institute for Subatomic Physics and VU University Amsterdam, Amsterdam, Netherlands\\
$ ^{21}$Henryk Niewodniczanski Institute of Nuclear Physics  Polish Academy of Sciences, Krak{\'o}w, Poland\\
$ ^{22}$Yandex School of Data Analysis, Moscow, Russia\\
$ ^{23}$Institute for High Energy Physics (IHEP), Protvino, Russia\\
$ ^{24}$ICCUB, Universitat de Barcelona, Barcelona, Spain\\
$ ^{25}$Instituto Galego de F{\'\i}sica de Altas Enerx{\'\i}as (IGFAE), Universidade de Santiago de Compostela, Santiago de Compostela, Spain\\
$ ^{26}$European Organization for Nuclear Research (CERN), Geneva, Switzerland\\
$ ^{27}$Institute of Physics, Ecole Polytechnique  F{\'e}d{\'e}rale de Lausanne (EPFL), Lausanne, Switzerland\\
$ ^{28}$Physik-Institut, Universit{\"a}t Z{\"u}rich, Z{\"u}rich, Switzerland\\
$ ^{29}$University of Birmingham, Birmingham, United Kingdom\\
$ ^{30}$H.H. Wills Physics Laboratory, University of Bristol, Bristol, United Kingdom\\
$ ^{31}$Cavendish Laboratory, University of Cambridge, Cambridge, United Kingdom\\
$ ^{32}$Department of Physics, University of Warwick, Coventry, United Kingdom\\
$ ^{33}$School of Physics and Astronomy, University of Edinburgh, Edinburgh, United Kingdom\\
$ ^{34}$School of Physics and Astronomy, University of Glasgow, Glasgow, United Kingdom\\
$ ^{35}$Oliver Lodge Laboratory, University of Liverpool, Liverpool, United Kingdom\\
$ ^{36}$Imperial College London, London, United Kingdom\\
$ ^{37}$School of Physics and Astronomy, University of Manchester, Manchester, United Kingdom\\
$ ^{38}$Department of Physics, University of Oxford, Oxford, United Kingdom\\
$ ^{39}$Massachusetts Institute of Technology, Cambridge, MA, United States\\
$ ^{40}$University of Cincinnati, Cincinnati, OH, United States\\
\bigskip
$^{a}$Scuola Normale Superiore, Pisa, Italy\\
$^{b}$Dunnhumby Ltd., Hammersmith, United Kingdom\\
$^{c}$La Salle, Universitat Ramon Llull, Barcelona, Spain\\
$^{d}$Institute of Nuclear Physics, Moscow State University (SINP MSU), Moscow, Russia\\
\medskip
$^{\dagger}$ Author was at institute at time work was performed.
}
\end{flushleft}

\setcounter{page}{2}
\mbox{~}
%
%
%
%

\cleardoublepage


\renewcommand{\thefootnote}{\arabic{footnote}}
\setcounter{footnote}{0}



\pagestyle{plain} 
\setcounter{page}{1}
\pagenumbering{arabic}


%

\section{Introduction}
The \lhcb experiment is a dedicated heavy-flavour physics experiment at the LHC, focused on the
reconstruction of particles containing \Pc~and~\Pb~quarks. During Run~1, the LHCb physics programme
was extended to electroweak, soft QCD and even heavy-ion physics.
This was made possible in large part due to a versatile real-time reconstruction and trigger system,
which is responsible for reducing the rate of collisions saved for
off\-line analysis by three orders of magnitude. The trigger used by LHCb in Run~1~\cite{LHCb-DP-2012-004} executed a simplified two-stage
version of the full off\-line reconstruction. In the first stage, only charged particles with at least $\sim\! 1$~\gevc of transverse
momentum (\pt) and displaced from the primary vertex (PV) were available; the \pt threshold was somewhat lower for muons,
which in addition were not required to be displaced. This first stage reconstruction enabled the bunch crossing rate to be reduced efficiently
by roughly one order of magnitude. In the following second stage, most charged particles with $\pt \gtrsim 300\mevc$
were available to classify the bunch crossings (hereafter ``events''). Particle-identification information
and neutral particles such as photons or \piz mesons were available on-demand to specific
classification algorithms. Although this trigger enabled the majority of the LHCb physics
programme, the lack of low-momentum charged particles at the first stage and full particle identification
at the second stage limited the performance for \Pc-hadron physics in particular. In addition,
resolution differences between the online and off\-line reconstructions made it difficult to precisely understand
absolute trigger efficiencies.

For these reasons, the LHCb trigger system was redesigned during 2013--2015 to
perform the full off\-line event reconstruction. The entire data processing framework was
redesigned to enable a single coherent real-time detector alignment and calibration, as well
as real-time analyses using information directly from the trigger system. The key objectives
of this redesign were twofold: firstly, to enable the full off\-line reconstruction to run in the trigger, greatly
increasing the efficiency with which charm- and strange-hadron decays could be selected; and
secondly, to achieve the same quality of alignment and calibration within the trigger as was achieved
off\-line in Run~1, enabling the final signal selection to be performed at the trigger level. 

\usetikzlibrary{arrows.meta,automata, decorations.pathreplacing}

\tikzset{
	lhcbnode/.style={draw, thick, fill=white,
		align=center, font=\small,
		rectangle, rounded corners,
		inner sep=3pt,
		minimum width=4.5cm, minimum height=.6cm},
	calibnode/.style={draw, thick, fill=white,
		align=center, font=\small,
		rectangle, rounded corners,
		inner sep=3pt,
		minimum width=1.8cm, minimum height=.6cm},
	tracknode/.style={draw, thick, fill=white, text=white, 
		align=center, font=\small,
		rectangle, rounded corners,
		minimum width=6.0cm, minimum height=1.0cm},
	track2node/.style={draw, thick, fill=white, text=white, 
		align=center, font=\small,
		rectangle, rounded corners,
		minimum width=6.0cm, minimum height=1.5cm},
	startnode/.style={fill=red!10},
	endnode/.style={fill=blue!10},
	greennode/.style={fill=green!10},
	bluenode/.style={fill={rgb,255:red,68; green,114; blue,196}, text=white, font=\bfseries},
	pvnode/.style={fill={rgb,255:red,132; green,180; blue,223}, text=white, font=\bfseries},
	fitnode/.style={fill={rgb,255:red,112; green,173; blue,71}, text=white, font=\bfseries},
	hlt1node/.style={fill={rgb,255:red,237; green,125; blue,49}, text=white, font=\bfseries},
	outputnode/.style={fill={rgb,255:red,255; green,223; blue,128}, 
	                   text={rgb,255:red,1; green,25; blue,147}, 
			   double,
			   font=\bfseries},
	outputpvnode/.style={fill={rgb,255:red,255; green,255; blue,215}, 
	                     text={rgb,255:red,1; green,25; blue,147}, 
			     double,
			     font=\bfseries},
	orange/.style={fill=orange!60!yellow!70!white, text=brown!60!black},
	rate/.style={right=.1cm, midway},
	mbrate/.style={left=.1cm, midway},
	widearr/.style={ultra thick, -latex},
	bluearr/.style={fill={rgb,255:red,68; green,114; blue,196},
	                draw={rgb,255:red,68; green,114; blue,196},
			line width=1.5mm},
	pvarr/.style={fill={rgb,255:red,132; green,180; blue,223},
	              draw={rgb,255:red,132; green,180; blue,223},
		      line width=1.5mm},
	greenarr/.style={fill={rgb,255:red,112; green,173; blue,71},
	                 draw={rgb,255:red,112; green,173; blue,71},
			 line width=1.5mm},
	hlt1arr/.style={fill={rgb,255:red,237; green,125; blue,49},
	                 draw={rgb,255:red,237; green,125; blue,49},
			 line width=1.5mm},
	grarr/.style={fill={rgb,255:red,192; green,192; blue,192},
	                 draw={rgb,255:red,192; green,192; blue,192},
			 line width=1.5mm},
	darkgreen/.style={fill=green!40!black, text=white, font=\footnotesize}, 
	myviolet/.style={fill=violet!60!black, text=white, font=\footnotesize},
	lightblue/.style={fill=blue!30!white}
}

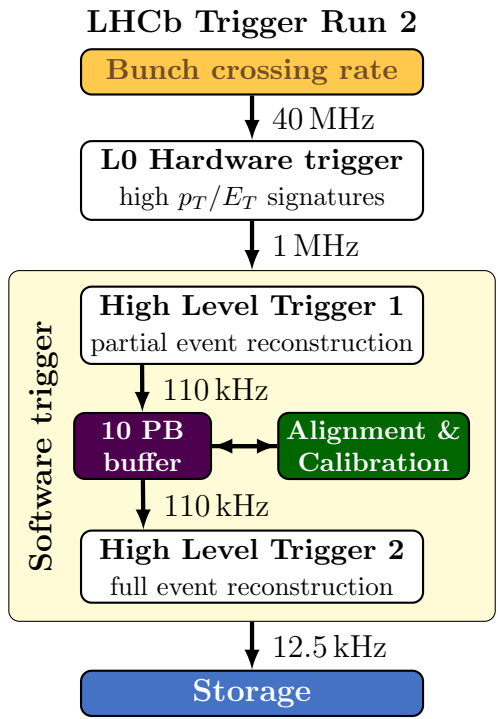
\begin{figure}[t]
	\centering

  {%
    \begin{tikzpicture}[scale=1.6]

    \node (softtrig) at (-.0,-3.4) [fill=yellow!20, rounded corners, draw,
      minimum width=6.4cm, minimum height=4.65cm] {};
	    \node at (-1.65, -3.4) [right, align=center, anchor=base, rotate=90] {\textbf{Software trigger}};

    \node at (0,.1) {\textbf{LHCb Trigger Run~2}};
	    \node (coll) at (0,-.3) [lhcbnode, orange] {\textbf{Bunch crossing rate}};
	    \node (L0) at (0,-1.2) [lhcbnode] {\textbf{L0 Hardware trigger} \\ \footnotesize{high $p_T/E_T$ signatures}};
	    \node (HLT1) at (0,-2.4) [lhcbnode] { \textbf{High Level Trigger 1} \\ \footnotesize{partial event reconstruction} };
	    \node (buff) at (-0.9,-3.4) [calibnode, myviolet]{\textbf{10 PB} \\ \textbf{buffer}};

     \node (calib) at (1.0,-3.4) [calibnode, darkgreen]
	     {\textbf{Alignment \&} \\ \textbf{Calibration} };
	    \node (HLT2) at (0,-4.4) [lhcbnode] { \textbf{High Level Trigger 2} \\ \footnotesize{full event reconstruction}};
	    \node (storage) at (0,-5.45) [lhcbnode, endnode, bluenode] {\textbf{Storage}};

    \draw [widearr] (coll) -- (L0)
      node [rate] {\unit[40]{MHz}};
    \draw [widearr] (L0) -- (L0 |- softtrig.north)
      node [rate] {\unit[1]{MHz}};
    
    \draw [widearr] (HLT1.200) -- (buff)
      node [rate] {\unit[110]{kHz}};

    \draw [widearr] (buff) -- (HLT2.161)
        node [rate] {\unit[110]{kHz}};

    \draw [widearr] (storage |- softtrig.south) -- (storage)
      node [rate] {\unit[12.5]{kHz}};

    \draw [widearr] (buff) -- (calib);
    \draw [widearr] (calib) -- (buff);

    \end{tikzpicture}%
  }%
  \caption{Overview of the \lhcb trigger system.}
\label{fig:trigger2015}
\end{figure}

A schematic diagram showing the trigger data flow in Run~2 is depicted in Fig.~\ref{fig:trigger2015}.
The LHCb trigger is designed to allow datataking with minimal deadtime at the full LHC bunch crossing rate of 40~MHz.
The maximum rate at which all LHCb subdetectors can be read out is imposed by the bandwidth and frequency of the front-end
electronics, and corresponds to around 1.1\,MHz when running at the designed rate of visible interactions per bunch crossing in LHCb of $\mu=0.4$.
During Run~2 LHCb operated at $\mu=1.1$ in order to collect a greater integrated luminosity, which limited
the actual readout rate to about 1~MHz.
A system of field-programmable gate arrays with a
fixed latency of $4\,\mus$ (the \lz trigger) determines which events are kept. Information from the electromagnetic calorimeter,
hadronic calorimeter, and muon stations
is used in separate \lz trigger lines.


The High Level trigger (\hlt) is divided into two stages, \hltone and \hlttwo.
The first level of the software trigger performs an inclusive selection of
events based on one- or two-track signatures, on the presence of muon tracks
displaced from the PVs, or on dimuon combinations in the event.
Events selected by the \hltone trigger are buffered to disk storage in the online system.
This is done for two purposes: events can be processed further during inter-fill periods, and
the detector can be calibrated and aligned run-by-run before the \hlttwo stage.
Once the detector is aligned and calibrated, events are passed to \hlttwo, where
a full event reconstruction is performed. This allows for a wide range of inclusive and exclusive
final states to trigger the event and obviates the need for further offline processing. 

This paper describes the design and performance of the Run~2 LHCb trigger system, including the
real-time reconstruction which runs in the \hlt. 
The software framework enabling real-time analysis (``TURBO'') has been described in detail elsewhere. The initial 
proof-of-concept deployed in 2015~\cite{LHCb-DP-2016-001} allowed off\-line-quality signal candidates selected in the 
trigger to be written to permanent storage. It also allowed physics analysts to use the off\-line analysis tools when 
working with these candidates, which was crucial in enabling LHCb to rapidly produce a number of publications 
proving that real-time analysis was possible without losing precision or introducing additional systematics. 
Subsequent developments ~\cite{Aaij:2019uij} generalized this approach to allow not only the 
signal candidate but also information about other, related, particles in the event to be saved. 
These developments also transformed the proof-of-concept implementation into a scalable solution which will 
now form the basis of LHCb's upgrade computing model~\cite{LHCbCollaboration:2319756}.

\section{The LHCb detector}
\label{sec:Detector}
\begin{figure}[t]
  \centering
  \includegraphics[width=0.75\textwidth]{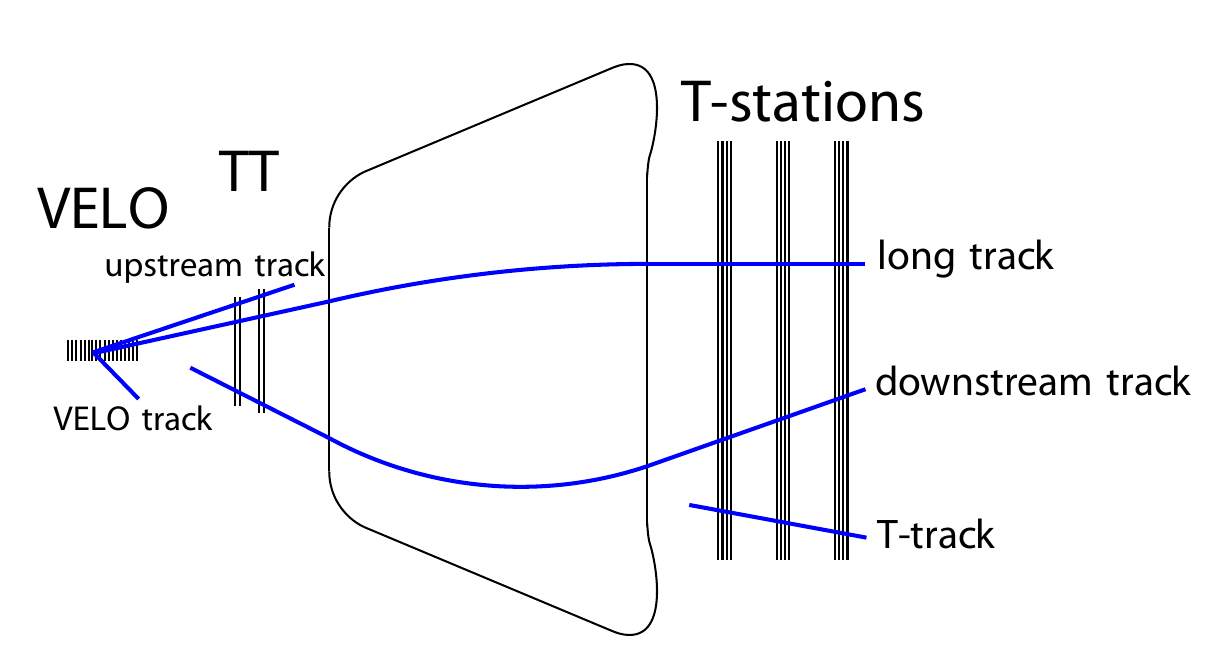}
  \caption{Sketch of the different types of tracks within \lhcb.}
  \label{fig:trackTypes}
\end{figure}

The \lhcb detector~\cite{LHCb,LHCb-DP-2014-002} is a single-arm forward
spectrometer covering the \mbox{pseudorapidity} range $2<\eta <5$.
The detector coordinate system is such that $z$ is along the beam line and
$x$ is the direction in which charged particle trajectories are deflected by the magnetic field.
The detector includes a high-precision tracking system
consisting of a silicon-strip vertex detector (\velo) surrounding the $pp$
interaction region~\cite{LHCb-DP-2014-001}, a large-area silicon-strip detector (TT) located
upstream of a dipole magnet with a bending power of about
$4{\rm\,Tm}$, and three stations of silicon-strip detectors (IT) 
 and straw
drift tubes~\cite{LHCb-DP-2013-003} (\ot) placed downstream of the magnet. These are collectively referred to as the T-stations.
The tracking system provides a measurement of momentum, \ptot, of charged particles with
a relative uncertainty that varies from 0.5\% at low momentum to 1.0\% at 200\gevc.
A sketch of the various track types relevant in LHCb is shown in Fig.~\ref{fig:trackTypes}.

The minimum distance of a track to a PV,
the impact parameter, is measured with a resolution of $(15+[29\gevc]/\pt)\mum$.
Different types of charged hadrons are distinguished using information
from two ring-imaging Cherenkov detectors~\cite{LHCb-DP-2012-003}.
Photons, electrons and hadrons are identified by a calorimeter system consisting of
scintillating-pad (\spd) and preshower detectors (\presh), an electromagnetic
calorimeter (\ecal) and a hadronic calorimeter (\hcal). Muons are identified by a
system composed of alternating layers of iron and multiwire
proportional chambers (\muonch)~\cite{LHCb-DP-2012-002}.

The \lhcb detector data taking is divided into fills and runs. A fill is a single period
of collisions delimited by the announcement of stable beam conditions and
the dumping of the beam by the \lhc, and typically lasts around twelve hours.
A fill is subdivided into runs, each of which lasts a maximum of one hour. The downtime
associated with run changes is negligible compared to other sources of downtime.

Detector simulation has been used in the tuning of most reconstruction
and selection algorithms discussed in this paper. In simulated LHCb events,
$pp$ collisions are generated using \pythia~\cite{Sjostrand:2006za,*Sjostrand:2007gs}
with a specific \lhcb configuration~\cite{LHCb-PROC-2010-056}.  Decays of hadronic particles
are described by \evtgen~\cite{Lange:2001uf}, in which final-state
radiation is generated using \photos~\cite{Golonka:2005pn}. The
interaction of the generated particles with the detector, and its response,
are implemented using the \geant
toolkit~\cite{Allison:2006ve, *Agostinelli:2002hh} as described in
Ref.~\cite{LHCb-PROC-2011-006}.


\section{Data acquisition and the \lhcb trigger}
\label{sec:daqandtrigger}
All trigger systems consist of a set of algorithms that classify events (or parts thereof) as either interesting or uninteresting for further
analysis, so that the data rate can be reduced to a manageable level by keeping only interesting events or interesting parts of them.
It is conventional to refer to a single trigger classification algorithm as a ``line'', so that a trigger consists of a set of
trigger lines.

\subsection{Hardware trigger}
The energies deposited in the  \spd, \presh, \ecal and \hcal
are used in the \lz-calorimeter system to trigger the selection of events.
All detector components are segmented transverse to the beam axis into cells of different size.
The decision to trigger an event is based on the transverse energy
deposited in clusters of $2\times2$ cells in the \ecal and \hcal.
The transverse energy of a cluster is defined as
\begin{equation}
      \et=\sum_{i=1}^4 E_i \sin{\theta_{i}} \;,
\label{trig-eq1}
\end{equation}
where $E_i$ is the energy deposited in cell $i$ and $\theta_{i}$ is the angle between the $z$-axis and 
a line from the cell centre to the average $pp$ interaction point (for more details, see Ref.~\cite{LHCb-DP-2012-004}).
Additionally, information from the \spd and \presh systems is used to distinguish between hadron, photon and electron candidates.

The \lz-muon trigger searches for straight-line tracks in the five muon stations.
Each muon station is sub-divided into logical pads in the $x$-$y$~plane.
The pad size scales with the distance to the beam line.
The track direction is used to estimate the \pt
of a muon candidate, assuming that the particle originated from the interaction point and received a single kick from the magnetic field.
The \pt resolution of the \lz-muon trigger is about $25\,\%$ averaged over the relevant \pt range.
The trigger decision is based on the two muon candidates with the largest \pt:
either the largest \pt must be above the \lzmuon threshold, or
the product of the largest and second largest \pt values must be above the \lzdimuon threshold.
In addition there are special trigger lines that select events with low particle multiplicity 
to study central exclusive production and inclusive jet trigger lines for QCD measurements.

To reduce the complexity of events and, hence, to enable a faster reconstruction in the subsequent software stage,
a requirement is placed on the maximum number of \spd hits in most \lz trigger lines.
The \lzdimuon trigger accepts a low rate of events, and therefore, only a loose SPD requirement is applied,
while no SPD requirement is applied in the high \pt \lzmuon trigger
used for electroweak production analyses in order to avoid systematic uncertainties associated with the determination
of the corresponding efficiency.
The thresholds used to take the majority of the data are listed in Table~\ref{tab:LzThresholds} as a function of the year of data taking.
Note that while the use of \spd requirements selects
simpler and faster-to-reconstruct events, it does not result in a significant loss of absolute signal efficiency compared to
a strategy using only \et and \pt requirements.
This is because the \lz signal-background discrimination deteriorates rapidly with increasing event complexity
for all but the dimuon and electroweak trigger lines. Note that the 2017 thresholds are looser than the 2016 thresholds because the
maximum number of colliding bunches, and hence, the collision rate of the LHC was significantly lower in 2017, due to difficulties with
part of the injection chain. The optimization of the \lz criteria is described in more detail in Sec.~\ref{sec:hlt}.

\begin{table}
  \caption{The \lz thresholds for the different trigger lines used to take the majority of the data for each indicated year. Technical trigger lines
  and those used for special areas of the physics programme are excluded for brevity. The Hadron, Photon, and Electron trigger lines select events
based on the \et of reconstructed \ecal and \hcal clusters. 
The Muon, Muon High, and Dimon trigger lines select events based on the \pt reconstruced \muonch stubs,
where the Dimuon selection is based on the product of the largest and second largest \pt stubs found in the event. 
As some of the subdetectors also read out hits associated to other bunch crossings, 
the use of bandwidth is further optimised in most of the L0 lines by rejecting events with a large $\et$ ($>24\gev$) 
for the previous bunch crossing~\cite{LHCb-DP-2017-001}.}
  \label{tab:LzThresholds}
  \begin{center}
    \begin{tabular}{l|ccc|c}
      \lz trigger      & \multicolumn{3}{c|}{\et/\pt threshold \bigstrut[b]} & \spd threshold\\
            & 2015 & 2016 & 2017 & \\
      \hline
      Hadron & $>3.6$~GeV & $>3.7$~GeV & $>3.46$~GeV & $<450$\\
      Photon & $>2.7$~GeV & $>2.78$~GeV & $>2.47$~GeV & $<450$\\
      Electron & $>2.7$~GeV & $>2.4$~GeV & $>2.11$~GeV & $<450$ \\
      Muon & $>2.8$~GeV & $>1.8$~GeV & $>1.35$~GeV & $<450$\\
      Muon high \pt& $>6.0$~GeV & $>6.0$~GeV & $>6.0$~GeV & none\\
      Dimuon & $>1.69$~GeV$^{2}$ & $>2.25$~GeV$^{2}$ & $>1.69$~GeV$^{2}$ & $<900$\\
    \end{tabular}
  \end{center}
\end{table}

\subsection{High level trigger}
Events selected by \lz are transferred to the Event Filter Farm (EFF) for further selection.
The EFF consists of approximately 1700 nodes, 800 of which were added for Run 2, with 27000 physical cores.
The EFF can accommodate $\approx50000$ single-threaded processes using hyper-threading technology.

The \hlt is written in the same framework as the software
used in the off\-line reconstruction of events for physics analyses.
This allows for off\-line software to be easily incorporated into the trigger.
As detailed later, the increased EFF capacity and
improvements in the software allowed the off\-line reconstruction to be performed in the \hlt in Run~2.

The total disk buffer of the EFF is 10~PB, distributed such that farm nodes with
faster processors get a larger portion of the disk buffer. At an average event size of 55~kB
passing \hltone, this buffer allows for up to two weeks of  consecutive \hltone data taking before
\hlttwo has to be executed. Therefore, it is large enough to accommodate both regular running (where,
as we will see, the alignment and calibration is completed in a matter of minutes) and to serve
as a safety mechanism to delay \hlttwo processing in case of problems with the detector or calibration.

Around 40\% of the trigger output rate is dedicated to inclusive topological trigger lines, another 40\% is
dedicated to exclusive \Pc-hadron trigger lines, with the rest divided among dimuon lines, trigger lines
for electroweak physics, searches for exotic new particles, and other exclusive trigger lines
for specific analyses. There are in total around 20 \hltone and 500 \hlttwo trigger lines.

\subsection{Real-time alignment and calibration}
\label{sec:align}

The computing power available in the Run~2 EFF allows for automated
alignment and calibration tasks, providing off\-line quality information to the trigger
reconstruction and selections, as described in Ref.~\cite{Dujany,Borghi:2017hfp}. A more detailed description
of this real-time alignment and calibration procedure will be the topic of a separate publication.

Dedicated samples selected by \hltone are used to align and calibrate the detector in real time.
The alignment and calibrations are performed at regular intervals, and
the resulting alignment and calibration constants are updated only if they differ significantly
from the current  values.

\begin{figure}[t]
	\centering
	\includegraphics[width=.9\textwidth]{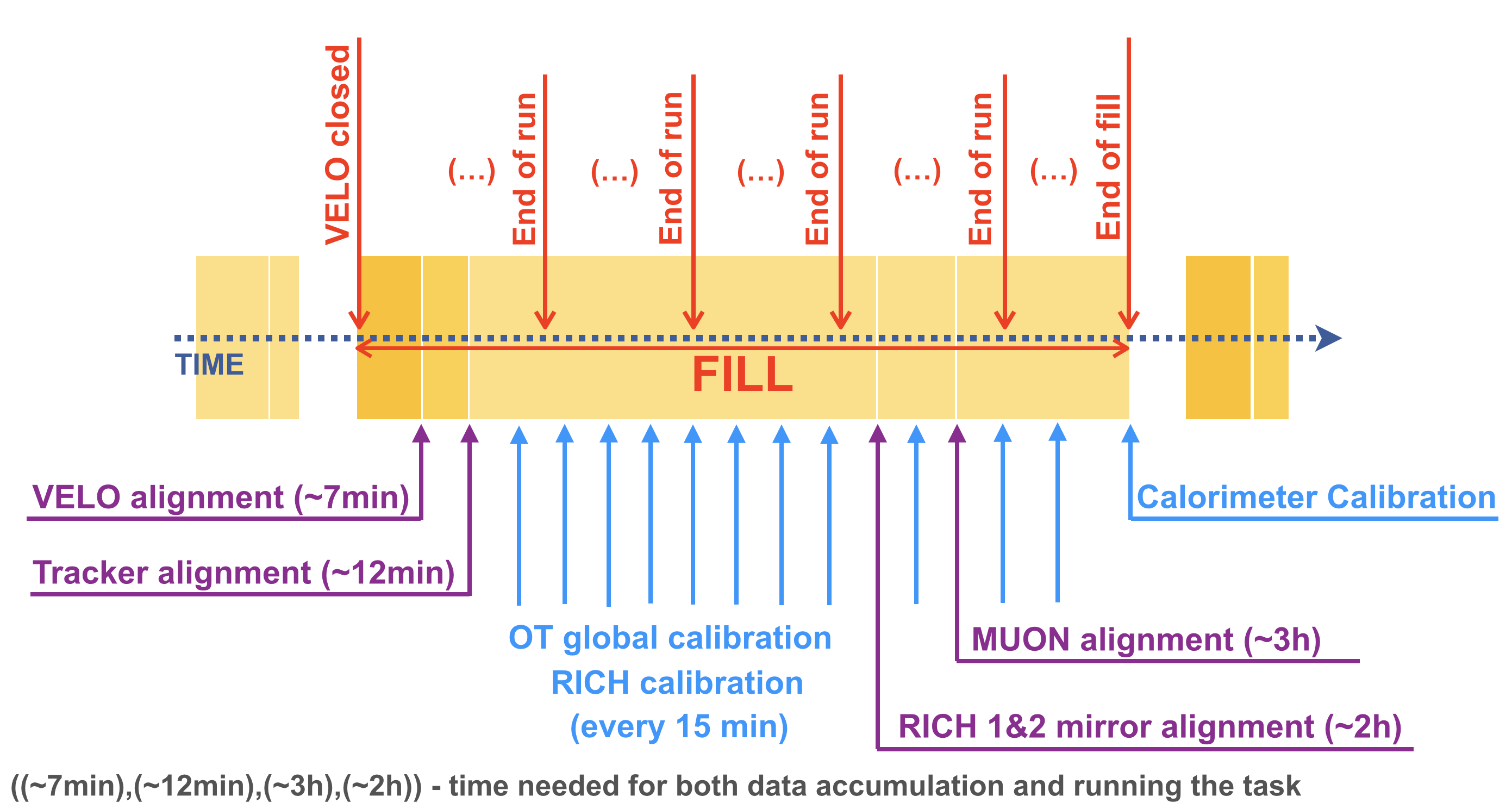}
	\caption{Schematic view of the real-time alignment and calibration procedure starting at the beginning
	of each fill, as used for 2018 data taking.}
	\label{fig:align}
\end{figure}

The major detector alignment and calibration tasks consist of:
\begin{itemize}
\item the \velo alignment, followed by the alignment of the tracking stations;
\item the \muonch alignment;
\item alignment of the rotations around various local axes in both \rich detectors of the primary and secondary mirrors;
\item global time calibration of the \ot;
\item \rich gas refractive-index calibration;
\item \rich Hybrid Photon Detectors calibration;
\item \ecal LED (relative) and $\pi^0$ (absolute) calibrations.
\end{itemize}
Each of these tasks has a dedicated \hltone trigger line which supplies it with the types of events required.
When the required sample sizes have been collected,
the selected events are saved to the disk buffer of the EFF, and calibration and alignment tasks are
performed in parallel within the EFF. A schematic view of the alignment and calibration procedure
is shown in Fig.~\ref{fig:align}, together with the time when the tasks are launched and the typical time taken to complete them.

\section{\hltone partial event reconstruction}
\label{sec:hlt1}

\hltone reconstructs the trajectories of charged particles traversing the full \lhcb tracking system, called long tracks,
with a \pt larger than 500\mevc.
In addition, a precise reconstruction of the PV is performed.
 The details of both steps are presented in Sec.~\ref{sec:hlt1_tracking}.

Tight timing constraints in \hltone mean that most 
particle-identification algorithms cannot be executed.
The exception is muon identification, which due to
a clean signature produced by muons in the \lhcb detector can be performed
already in \hltone,
as described in Sec.~\ref{sec:muonid}.
As discussed in Sec.~\ref{sec:hlt1_calib}, a subset of specially selected \hltone events 
serves as input to the alignment and calibrations tasks.

\subsection{Track and vertex reconstruction in \hltone}
\label{sec:hlt1_tracking}

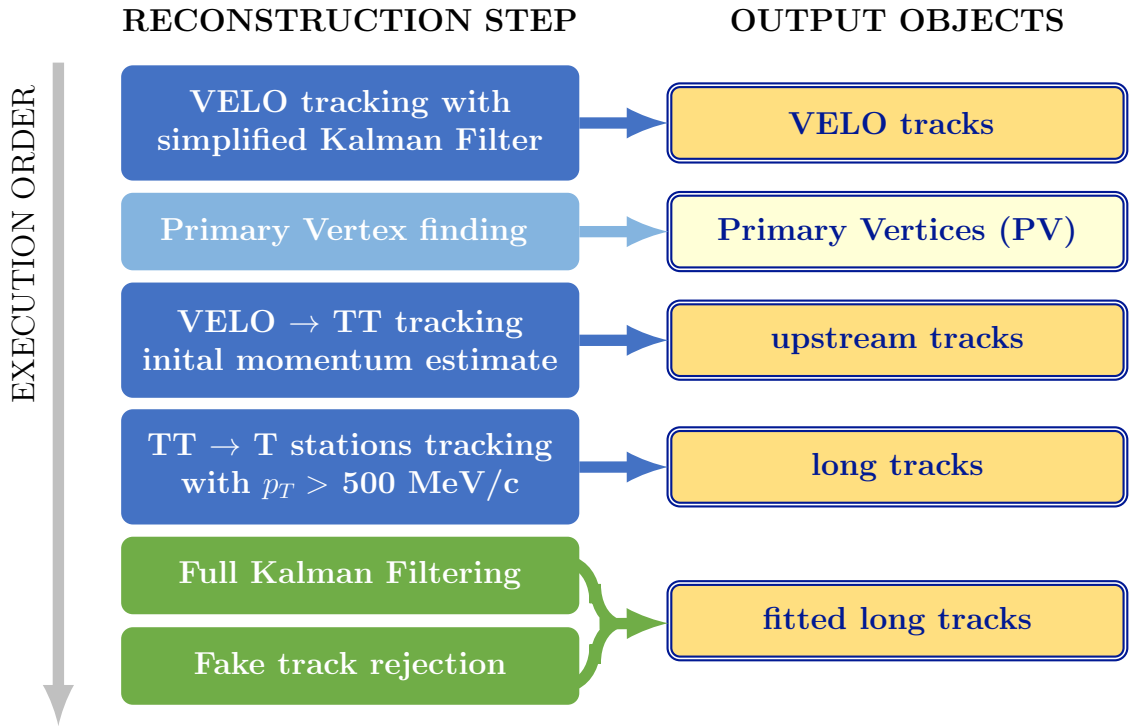
\begin{figure}[t]
	\centering
    \begin{tikzpicture}[scale=1.6]  

	    \node at (-1.,.1) {\textbf{RECONSTRUCTION STEP}};
	    \node at (3.5,.1) {\textbf{OUTPUT OBJECTS}};
	    \node at (-3.6, -1.75) [right, align=center, anchor=base, rotate=90] {EXECUTION ORDER};

	    \node (velo) [color={rgb,255:red,68; green,114; blue,196}] at (-1.,-.75)   [track2node, bluenode]
	          {\textbf{VELO tracking with} \\ \textbf{simplified Kalman Filter}};
	    \node (pv)   [color={rgb,255:red,132; green,180; blue,223}]  at (-1.,-1.65)   [tracknode, pvnode]
	          {\textbf{Primary Vertex finding} };
	    \node (ut)   [color={rgb,255:red,68; green,114; blue,196}]  at (-1.,-2.55)    [track2node, bluenode] 
	          {\textbf{VELO $\to$ TT tracking} \\ \textbf{inital momentum estimate}};
	    \node (fwd)   [color={rgb,255:red,68; green,114; blue,196}] at (-1.,-3.6)  [track2node, bluenode]
	          {\textbf{TT $\to$ T stations tracking} \\ \textbf{with $p_T >$ 500 MeV/c}};
	    \node (kal)   [color={rgb,255:red,112; green,173; blue,71}] at (-1.,-4.5)   [tracknode, fitnode] 
	          {\textbf{Full Kalman Filtering}};
	    \node (ghost) [color={rgb,255:red,112; green,173; blue,71}] at (-1.,-5.25)[tracknode, fitnode] 
	          {\textbf{Fake track rejection}};

	    \node (velotracks) [color = {rgb,255:red,1; green,25; blue,147}] at (3.5,-.75)    [tracknode, outputnode] { VELO tracks };
	    \node (pvs)        [color = {rgb,255:red,1; green,25; blue,147}] at (3.5,-1.65)   [tracknode, outputpvnode] 
	          { Primary Vertices (PV)};
            \node (upstream)   [color = {rgb,255:red,1; green,25; blue,147}] at (3.5,-2.55)   [tracknode, outputnode] { upstream tracks};
            \node (long)       [color = {rgb,255:red,1; green,25; blue,147}] at (3.5,-3.6)    [tracknode, outputnode] { long tracks};
            \node (fitted)     [color = {rgb,255:red,1; green,25; blue,147}] at (3.5,-4.875)  [tracknode, outputnode] { fitted long tracks};

	    \draw [widearr,bluearr] (velo) -- (velotracks);
	    \draw [widearr,pvarr] (pv) -- (pvs);
	    \draw [widearr,bluearr] (ut) -- (upstream);
	    \draw [widearr,bluearr] (fwd) -- (long);

	    \draw [widearr,grarr] (-3.4,-0.25) -- (-3.4,-5.75);
	    \draw [decorate,decoration={brace,amplitude=15pt}, 
	           color={rgb,255:red,112; green,173; blue,71},
		   line width =1.5mm]  (0.85,-4.4) -- (0.85,-5.4);
	    \draw [widearr,greenarr] (1.05,-4.9) -- (fitted);

    \end{tikzpicture}%
 
  \caption{Sketch of the \hltone track and vertex reconstruction.}
\label{fig:HLT1Reco}
\end{figure}

The sequence of \hltone algorithms which reconstruct vertices and long tracks is shown in Fig.~\ref{fig:HLT1Reco}.
The pattern recognition deployed in \hltone consists of three main steps: reconstructing the \velo tracks,
extrapolating them to the TT~stations to form upstream tracks, and finally extending them further to the T~stations
to produce long tracks.
Next, the long tracks are fitted using a Kalman Filtering and the fake trajectories are rejected.
The set of fitted \velo tracks is re-used to determine the positions of the PVs.

\subsubsection{Pattern recognition of high-momentum tracks}

The hits in the \velo are combined to form straight lines loosely pointing towards the beam line~\cite{FastVelo}.
Next, at least three hits in the \ttracker are required in a small region around a straight-line extrapolation from the \velo
~\cite{VeloTT} to form so-called upstream tracks.
The \ttracker is located in the fringe field of the \lhcb dipole magnet,
which allows the momentum to be determined with a relative resolution of about 20\%.
This momentum estimate is used to reject low \pt tracks. Matching
long tracks with \ttracker hits additionally reduces the number of fake \velo tracks.
Due to the limited acceptance of the \ttracker,
\velo tracks pointing to the region around the beampipe do not deposit charge in the \ttracker; therefore, they are passed on without requiring \ttracker hits.

The search window in the \intr and \ot is defined by the maximum possible deflection of charged particles with
\pt  larger than $500\mevc$. The search is also restricted to one side of the straight-line extrapolation
by the charge estimate of the upstream track. The use of the charge estimate is new in Run 2
and enabled the \pt threshold to be lowered from 1.2\gevc to 500\mevc with respect to Run 1.
For a given slope and position upstream of the magnet and a single hit in the tracking detectors downstream of the magnet, \intr and \ot,
the momentum is fixed and hits are projected along this trajectory into a common plane.
A search is made for clusters of hits in this plane which are then used to define the final long track~\cite{forward}.
In 2016 two artificial neural nets were implemented to increase the purity and the efficiency of the track candidates in the pattern recognition~\cite{Stahl:2260684}. 

\subsubsection{Track fitting and fake-track rejection}

Subsequently, all tracks are fitted with a Kalman filter to obtain the optimal parameter estimate. 
The settings of the online and off\-line reconstruction are harmonised in Run~2 to obtain
identical results for a given track. Previously, the online Kalman filter performend only a single iteration which
did not allow the ambiguities from the drift-time measurement of \ot hits on a track to be resolved.
In Run~2 the online fit runs until convergence or maximally 10 iterations.
Furthermore the possibility to remove up to two outliers has been added in the online reconstruction.
The off\-line reconstruction is changed to use clusters which are faster to decode
but have less information, and to employ a Kalman filter that utilizes
a simplified geometry description of the \lhcb detector.  
This significantly speeds up the calculation of material corrections in the filtering procedure.
For Run 2 the calculation of material corrections due to multiple scattering has been improved.
The new description is additive for many small scatterers resulting in more standard normal pull distributions.
The changes made to the off\-line Kalman filter enable running the same algorithm in both the \hlt and off\-line.
These changes neither affect the impact parameter resolution nor the momentum resolution as shown in Sec.~\ref{sec:resolutions}.

Since 2016 the fake track rejection, described in details in Sec.~\ref{sec:hlt2_tracking}, has been
used in \hltone reducing the rate of events passing this stage by 4\%.

\subsubsection{Primary vertex reconstruction}

Many \lhcb analyses require a precise knowledge of the PV position
and this information is used early in the selection of displaced particles.
The full set of \velo tracks is available in \hltone. Therefore, the PVs in Run~2 are
reconstructed using \velo tracks only, neglecting the additional momentum information on long tracks
which is only available later. This does not result in a degradation in resolution compared to
using a mixture of \velo and long tracks. Furthermore, this approach produces a consistent PV position
from the beginning to the end of the analysis chain which reduces systematic effects.

As there is no magnetic field in the \velo, the Kalman filter for \velo tracks uses a linear
propagation, allowing for a single scattering at each detector plane, tuned using simulation.
This simplification results in no loss of precision compared to a more detailed material description, but significantly reduces
the amount of time spent in the filtering phase, as no expensive computations are necessary.
A byproduct of this simpler track fit is that the PV covariance matrix is more
accurate than that used off\-line in Run~1, with pull distributions more compatible with unit widths in all three dimensions.

\begin{figure}
  \centering
  \includegraphics[width=0.45\textwidth]{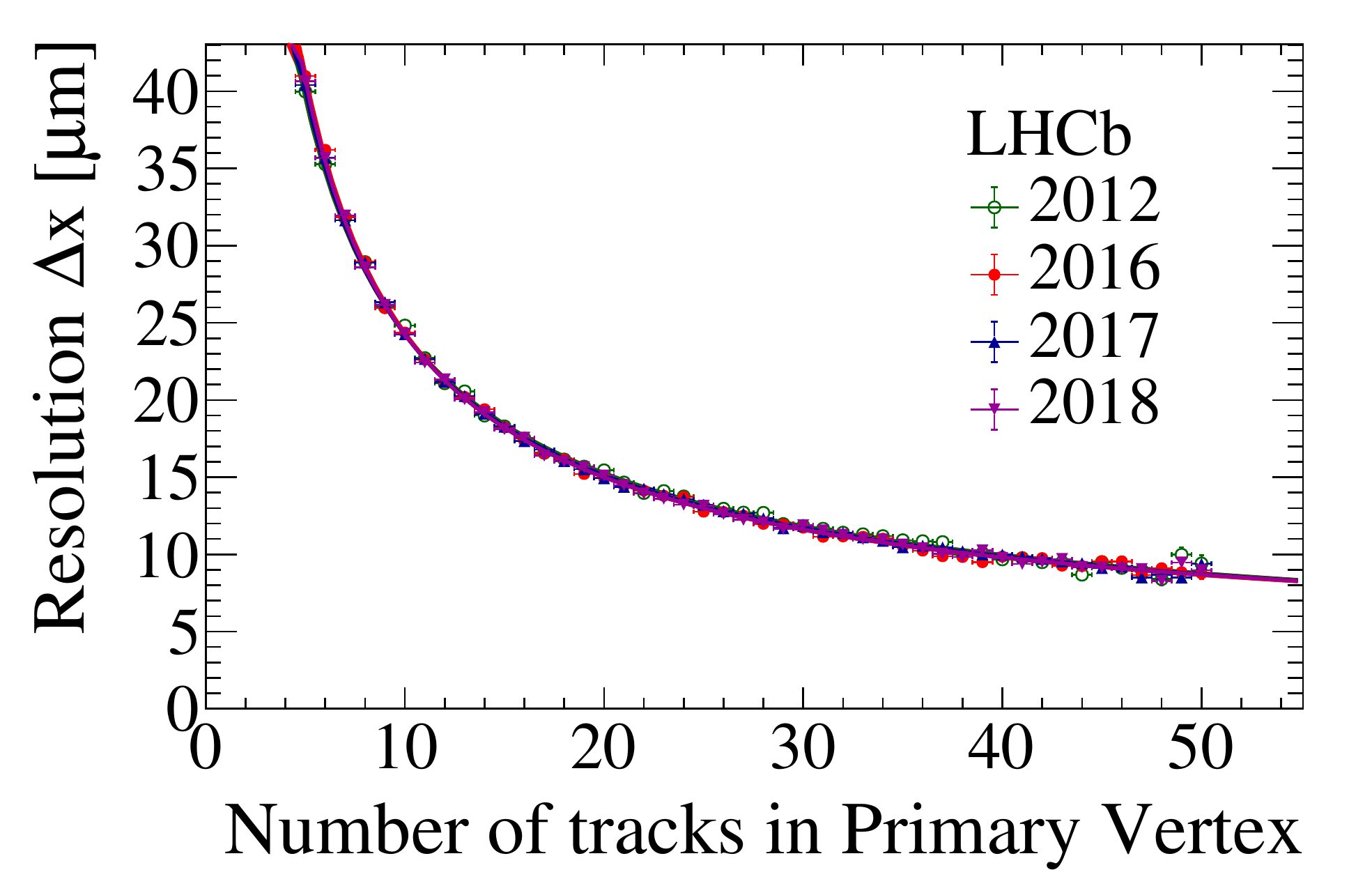}
  \includegraphics[width=0.45\textwidth]{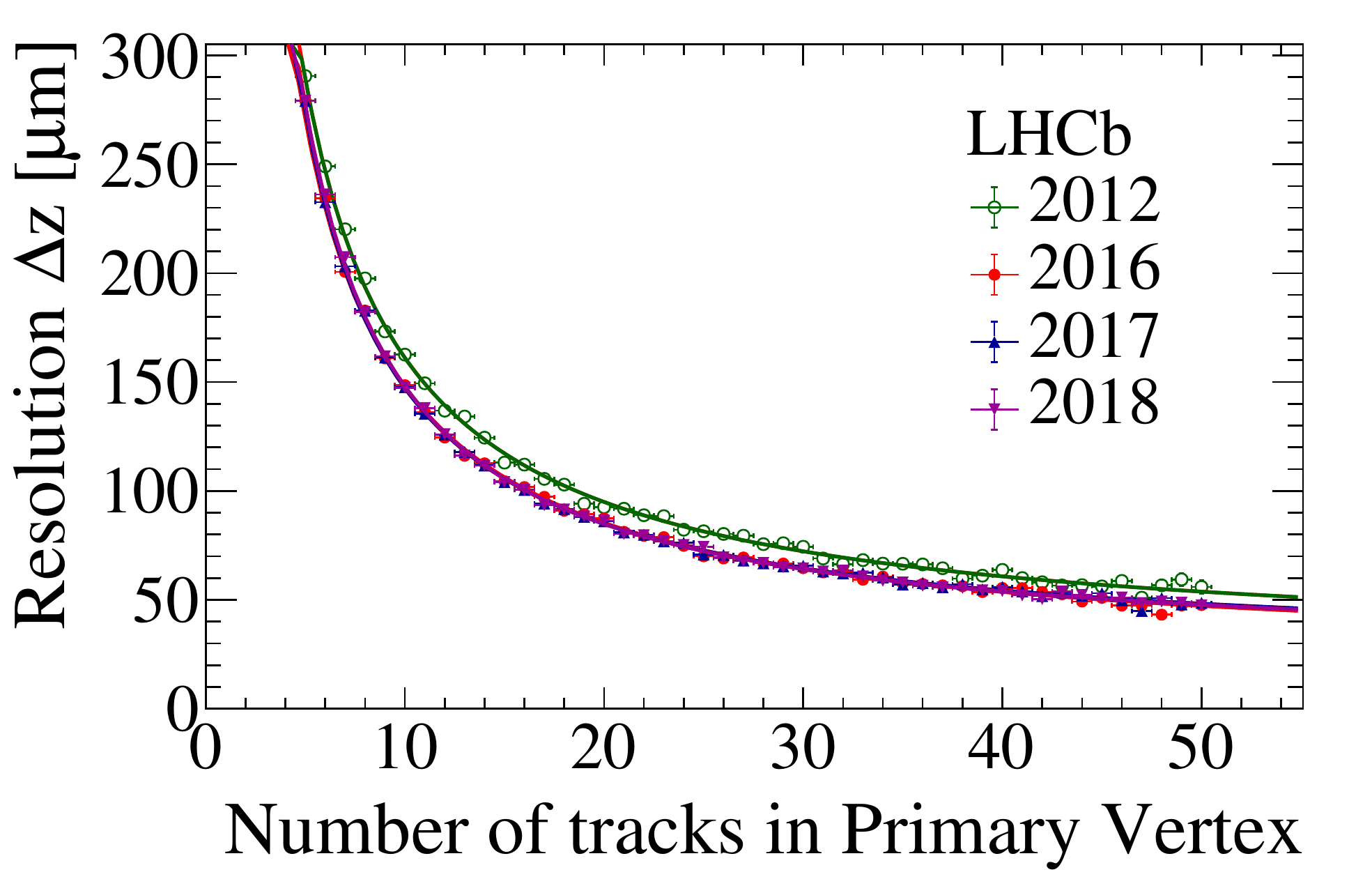}
  \caption{The PV $x$ (left) and $z$ (right) resolution as a function of the number of tracks in the PV for the
Run~1 off\-line and Run~2 (used both off\-line and online) PV reconstruction algorithms.}
  \label{fig:pvreco_resolutions}
\end{figure}

The PV resolution is obtained
by randomly splitting the input VELO tracks into two subsets.
The PV algorithm is executed independently on each subset, and 
the PVs found in each subset are matched based on the distance between them.
The width of the distribution of the difference of matched PV positions in a given dimension,
corrected by a factor of $\sqrt{2}$, gives the corresponding PV position resolution.
The resolution of the PV reconstruction  for  Run~2
is shown in Fig.~\ref{fig:pvreco_resolutions}
compared to the Run~1 (2012) off\-line reconstruction algorithm.  The new algorithm performs equally well
for the $x$ ($y$) coordinate, while with respect to Run 1 the resolution on the $z$ coordinate is improved by about 10\%.

Additionally, the parameters of the PV reconstruction have been retuned to
give a higher efficiency and smaller fake rate~\cite{Dziurda:2115353}.
The resulting improvement in efficiency of reconstructing PVs is 0.5\% for PVs associated with the production of a b quark pair,  1.3\% for those associated with the production of a c quark pair, and 6.6\% for light quarks production.
Simultaneously, the fraction of fake PVs, for example due to material interactions or the decay vertices of
long-lived particles, is reduced from $3.5\%$ to $1\%$.

\subsection{Muon identification}
\label{sec:muonid}
The muon identification starts with fully fitted tracks.
Hits in the \muonch stations are searched for in momentum-dependent regions of interest
around the track extrapolation.
Tracks with $p<3\gevc$ cannot be identified as muons, as they would not be able to reach the \muonch stations. 
Below a momentum of 6\gevc the muon identification algorithm requires hits in the first two stations after the calorimeters.
Between 6 and 10\gevc an additional hit is required in one of the last two stations.
Above 10\gev, hits are required in all the four \muonch stations. 
This same algorithm is used in \hltone, \hlttwo and off\-line.

In \hltone, the track reconstruction is only performed for tracks with \pt above 500~\mevc.
For particles with lower \pt, a complementary muon-identification algorithm has been devised, 
which is more similar to the \hltone muon identification performed in Run 1.
Upstream track segments are extrapolated directly to the \muonch stations, where hits are searched for around the track extrapolation.
The regions of interest used in this search are larger than those used for otherwise-reconstructed long tracks.
If hits are found in the muon system, the VELO-TT segment is extrapolated through the magnetic field using the momentum estimate
and matched to hits not already used in the \hltone long-track reconstruction.
This procedure extends the muon-identification capabilities down to a \pt of 80\mevc for a small additional resource cost,
significantly improving the performance for lower-momentum muons which are important in several measurements~\cite{Aaij:2253050}.

\begin{figure}[t]
  \centering
  \begin{overpic}[width=0.475\textwidth]{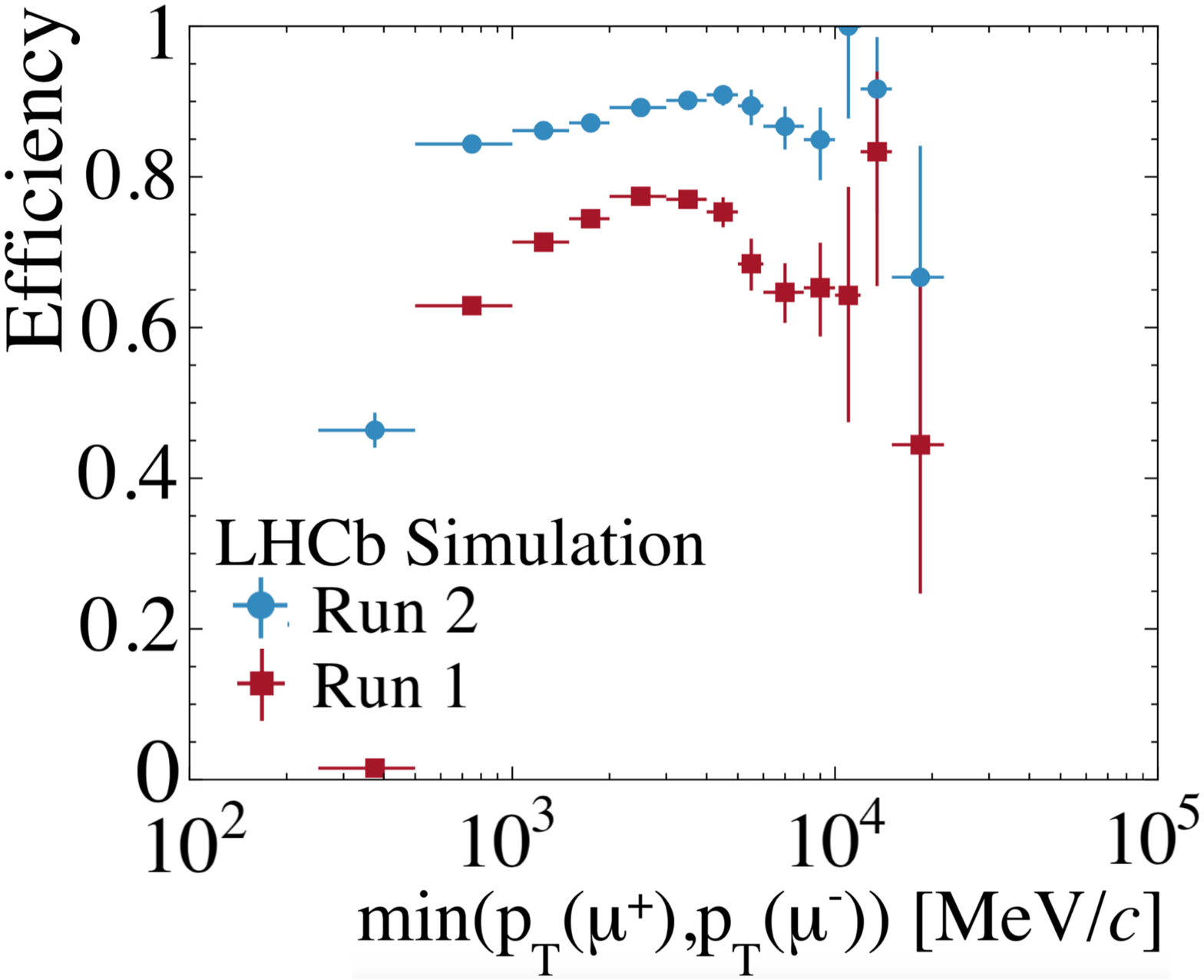}
  \end{overpic}
  \caption{\hltone dimuon efficiency as a function of the minimum \pt of the two muons.
  A large gain, especially at low \pt, can be seen from the comparison of the Run~1 and Run~2 algorithms.}
  \label{fig:muonideff}
\end{figure}

\begin{figure}[t]
\centering
\includegraphics[width = 0.49\textwidth]{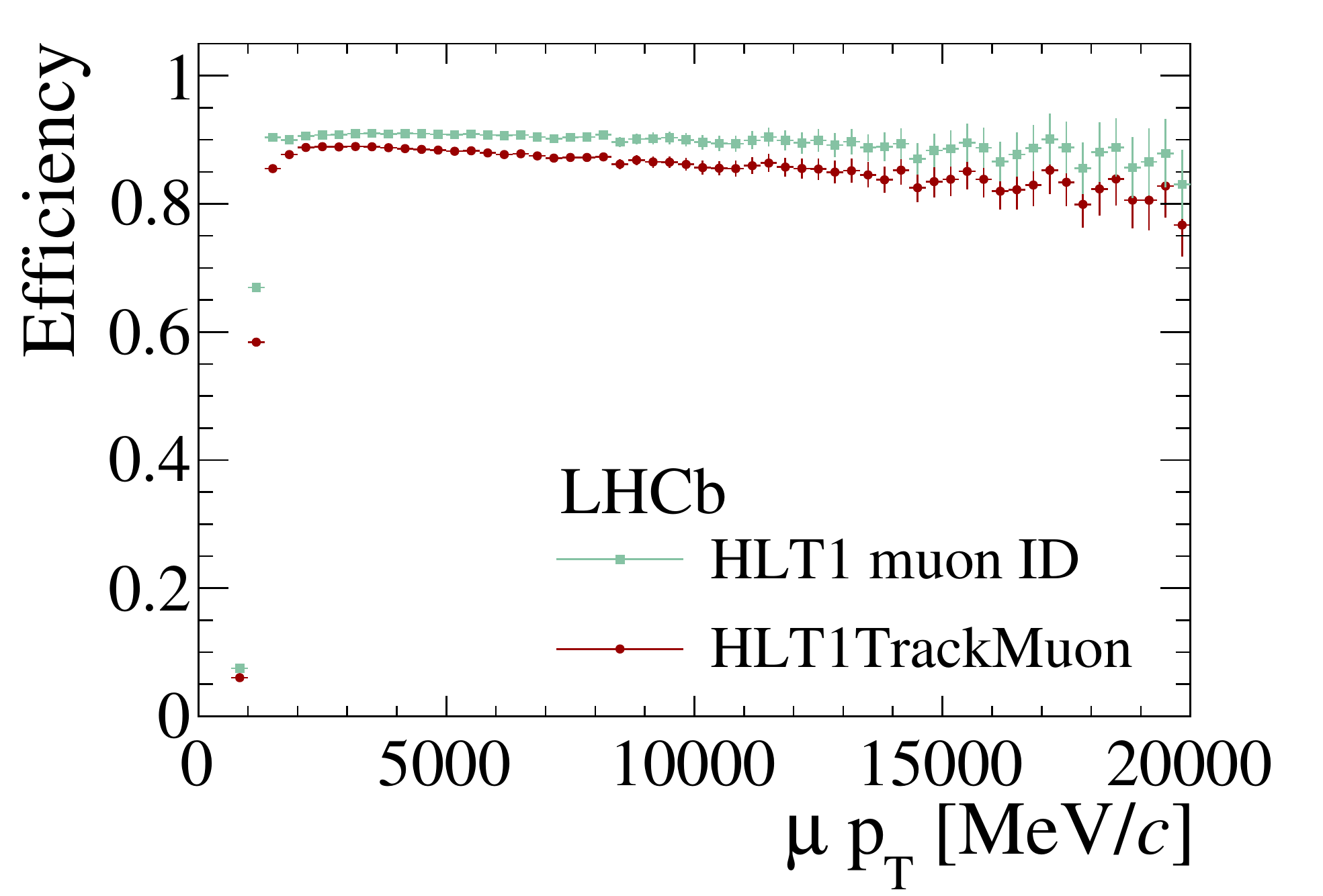}
\includegraphics[width = 0.49\textwidth]{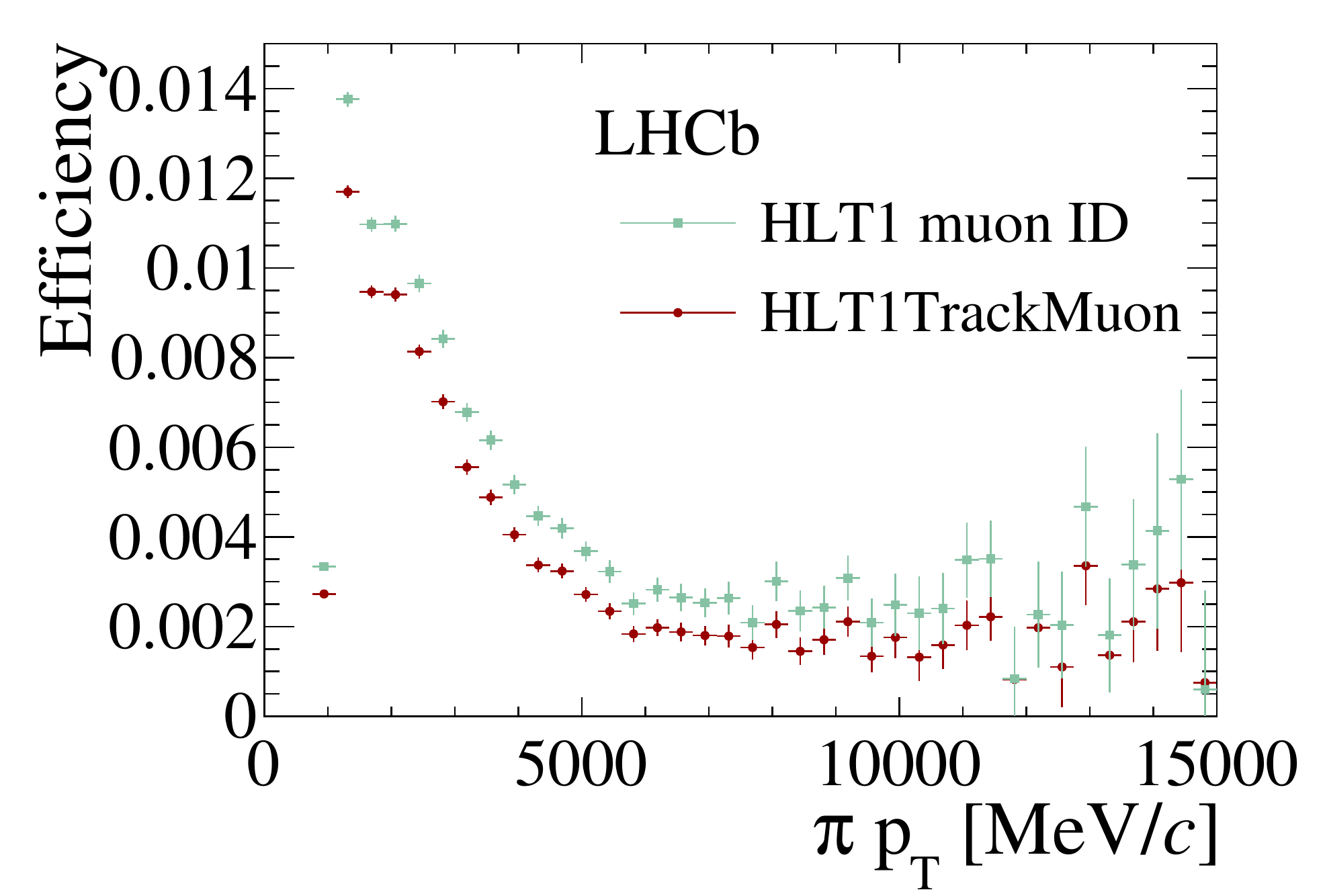}
	\caption{\hltone muon identification efficiency for (left) muons from \decay{\jpsi}{\mumu} decays
and (right) pions from  \decay{D^0}{\Km\pip} decays.
Green circles show only the identification efficiency (HLT1 Muon ID) while red squares show the efficiency of the additional trigger line (named HLT1TrackMuon) requirements (see text). 
}\label{fig:hlt1trackmuoneff}
\end{figure}

The muon identification code has been reoptimized for Run~2, gaining significant efficiency,
in particular at small \pt. 
This is demonstrated using LHCb simulation in Fig.~\ref{fig:muonideff}. 
The performance of the muon identification in \hltone is shown in Fig.~\ref{fig:hlt1trackmuoneff} (left)
as determined from unbiased \decay{\jpsi}{\mumu} decays using the tag-and-probe method.
This performance is obtained by studying the efficiency of the single-muon \hltone trigger,
which includes requirements on the displacement of the muon and on the minimum momentum (6\gevc) and
\pt (1.1\gevc).
Analogously in Fig.~\ref{fig:hlt1trackmuoneff} (right) the misidentification efficiency
with the same criteria is shown for pions from \decay{D^0}{\Km\pip} decays.
The muon identification efficiency of the single-muon \hltone trigger is slightly reduced by
the displacement and (transverse) momentum requirements, at the benefit of a lower misidentification
probability.

\section{\hlttwo full event reconstruction}
\label{sec:hlt2}

The \hlttwo full event reconstruction consists of three major steps: the track reconstruction of charged particles, 
the reconstruction of neutral particles, and particle identification (PID).
The \hlttwo track reconstruction exploits the full information from the tracking sub-detectors,
performing additional steps of the pattern recognition which are not possible in \hltone due to strict time constraints.
As a result high-quality long and downstream tracks are found with the most precise momentum estimate achievable.
Similarly, the most precise neutral cluster reconstruction algorithms are executed in the \hlttwo reconstruction.
Finally, in addition to the muon identification available in \hltone, \hlttwo exploits the full particle identification
from the RICH detectors and calorimeter system.
The code of all reconstruction algorithms has been optimized for Run 2 to better exploit the
capabilities of modern CPUs. Together with the algorithmic changes described in the following sections, this results
in a two times faster execution time while delivering the same or in several cases better physics performance than that
achieved off\-line in Run~1.

\subsection{The track reconstruction of charged particles}
\label{sec:hlt2_tracking}

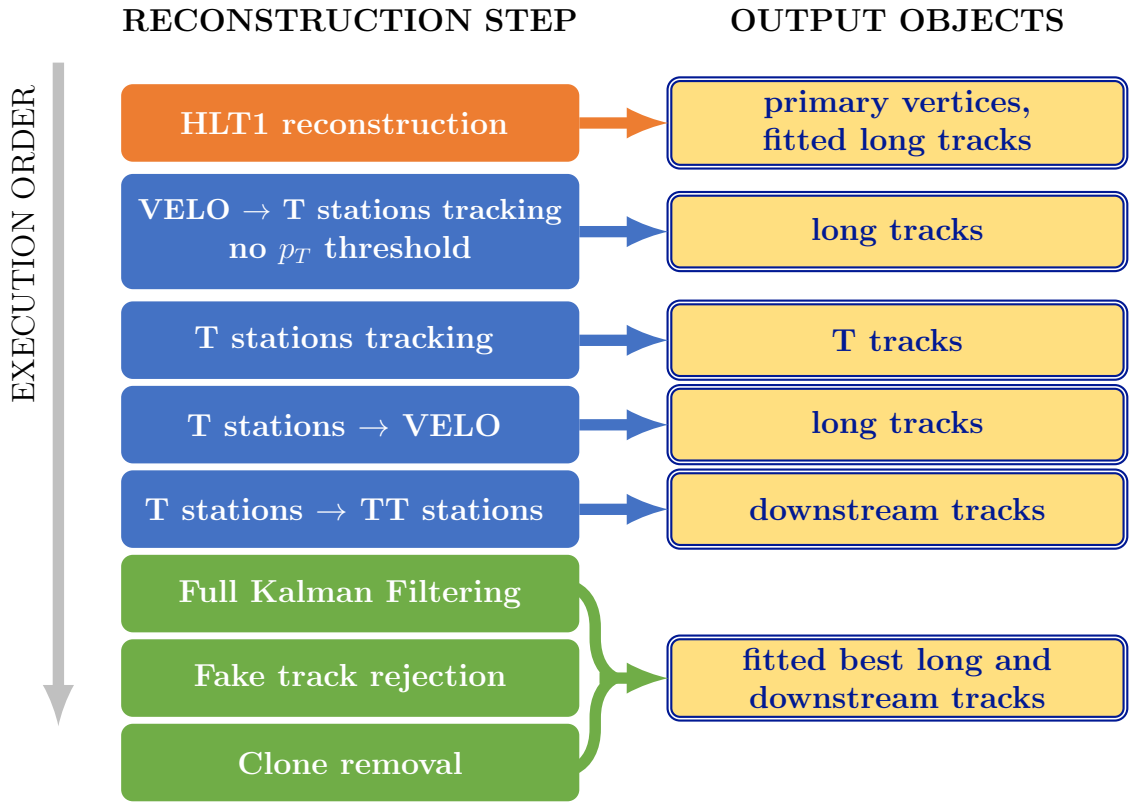
\begin{figure}[t]
	\centering
    \begin{tikzpicture}[scale=1.6]  

	    \node at (-1.,.1) {\textbf{RECONSTRUCTION STEP}};
	    \node at (3.5,.1) {\textbf{OUTPUT OBJECTS}};
	    \node at (-3.6, -1.75) [right, align=center, anchor=base, rotate=90] {EXECUTION ORDER};

	    \node (hlt1)   [color={rgb,255:red,237; green,125; blue,49}]  at (-1.,-0.75)   [tracknode, hlt1node]
		  {\textbf{HLT1 reconstruction} };
            \node (velo)   [color={rgb,255:red,68; green,114; blue,196}]  at (-1.,-1.65)   [track2node, bluenode] 
	          {\small \textbf{VELO $\to$ T stations tracking} \\ \textbf{no $p_T$ threshold}};
	    \node (seed)   [color={rgb,255:red,68; green,114; blue,196}]  at (-1.,-2.55)   [tracknode, bluenode]
		  {\textbf{T stations tracking} };
	    \node (match)   [color={rgb,255:red,68; green,114; blue,196}]  at (-1.,-3.25)   [tracknode, bluenode]
		  {\textbf{T stations $\to$ VELO} };
	    \node (down)   [color={rgb,255:red,68; green,114; blue,196}]  at (-1.,-3.95)   [tracknode, bluenode]
		  {\textbf{T stations $\to$ TT stations} };
	    \node (kal)   [color={rgb,255:red,112; green,173; blue,71}] at (-1.,-4.65)   [tracknode, fitnode] 
	          {\textbf{Full Kalman Filtering}};
            \node (ghost) [color={rgb,255:red,112; green,173; blue,71}] at (-1.,-5.35) [tracknode, fitnode] 
	          {\textbf{Fake track rejection}};
	    \node (clone) [color={rgb,255:red,112; green,173; blue,71}] at (-1.,-6.05) [tracknode, fitnode] 
	          {\textbf{Clone removal}};

	    \node (hlt1output) [color = {rgb,255:red,1; green,25; blue,147}] at (3.5,-.75)    [tracknode, outputnode] 
	         { primary vertices, \\ fitted long tracks };
            \node (long)       [color = {rgb,255:red,1; green,25; blue,147}] at (3.5,-1.65)  [tracknode, outputnode] { long tracks};
            \node (ttracks)     [color = {rgb,255:red,1; green,25; blue,147}] at (3.5,-2.55)  [tracknode, outputnode] { T tracks};
	    \node (long2)      [color = {rgb,255:red,1; green,25; blue,147}] at (3.5,-3.25)  [tracknode, outputnode] { long tracks};
            \node (downtr)      [color = {rgb,255:red,1; green,25; blue,147}] at (3.5,-3.95)  [tracknode, outputnode] { downstream tracks};
            
	    \node (best)      [color = {rgb,255:red,1; green,25; blue,147}] at (3.5,-5.35)  [tracknode, outputnode] 
	          { fitted best long and \\ downstream tracks};

	    \draw [widearr,hlt1arr] (hlt1) -- (hlt1output);
	    \draw [widearr,bluearr] (velo) -- (long);
	    \draw [widearr,bluearr] (seed) -- (ttracks);
	    \draw [widearr,bluearr] (match) -- (long2);
            \draw [widearr,bluearr] (down) -- (downtr);


	    \draw [decorate,decoration={brace,amplitude=15pt}, 
	           color={rgb,255:red,112; green,173; blue,71},
		   line width =1.5mm]  (0.85,-4.6) -- (0.85,-6.1);
	    \draw [widearr,greenarr] (1.05,-5.35) -- (best);

	    \draw [widearr,grarr] (-3.4,-0.25) -- (-3.4,-5.75);

    \end{tikzpicture}%
 
  \caption{Sketch of the \hlttwo track and vertex reconstruction sequence.}
\label{fig:HLT2Reco}
\end{figure}


A sketch of the track and vertex reconstruction sequence in \hlttwo is shown in Fig.~\ref{fig:HLT2Reco}.
The goal is to reconstruct  all tracks without a minimal \pt
requirement.
This is particularly important for the study of the decays of lighter particles, such as charmed or strange hadrons, whose abundance
means that only some of the fully reconstructed and exclusively selected final states fit into the available trigger bandwidth.
Often, not all of the decay products of a charm- or strange-hadron decay pass the 500\mevc \pt
requirement of \hltone, particularly for decays into three or more final-state particles. Therefore, to efficiently
trigger these decays, it is necessary to also reconstruct the lower-momentum tracks within the \lhcb acceptance.

In a first step, the track reconstruction of \hltone is repeated. A second step is then used to reconstruct
the lower-momentum tracks which had not been found in \hltone due to the kinematic thresholds in the reconstruction. Those \velo tracks
and T-station clusters used to reconstruct long tracks with a good fit quality in the first step are disregarded for this second step.
A similar procedure as in \hltone is employed: the remaining \velo tracks are extrapolated through the magnet
to the T-stations using the same algorithm, where the search window is
now defined by the maximal possible deflection of a particle with \pt larger than 80~\mevc.
No \ttracker hits are required for the second step to avoid the loss of track efficiency induced by acceptance gaps in the \ttracker.
The new Run~2 track finding optimization results in 27\% fewer fake tracks and a reconstruction
efficiency gain of 0.5\% for long tracks. 
In addition, a standalone search for tracks in the T~stations is performed~\cite{seeding},
and these standalone tracks are then combined with \velo tracks to form long tracks~\cite{matching1,matching2}.
The redundancy of the two long-track algorithms increases the efficiency by a few percent.

Tracks produced in the decays of long-lived particles like \Lz or \KS that
decay outside the \velo are reconstructed using T-station segments that are extrapolated backwards
through the magnetic field and combined with hits in the \ttracker. For Run 2, a new algorithm was used to reconstruct these
tracks\cite{patllt}. It uses two multivariate classifiers, one to reject fakes, and another
to select the final set of hits in the \ttracker in case several sets are compatible with the same T-station segment. In combination with
other improvements, this results in a higher efficiency and a lower fake rate compared to the corresponding Run~1 algorithm.


The next step in the reconstruction chain is the rejection of fake tracks.
These fakes result from random combinations of hits or a mismatch of track segments upstream and
downstream of the magnet. They are reduced using two techniques. First, all tracks are fitted using the same Kalman
filter. In Run 1, the only selection to reject fake tracks was based on a reduced $\chi^{2}$.
For Run 2, the upper limit on this $\chi^{2}$ selection was increased to allow
for a better overall track reconstruction efficiency. To offset the corresponding increase in the number of fake tracks, a neural network
was trained using the TMVA~\cite{TMVA4,Hocker:2007ht} package to efficiently remove these tracks\cite{ghostprob}. Its input variables are the
overall $\chi^{2}$ of the Kalman filter, the $\chi^{2}$ values of the fits for the different track
segments, the numbers of hits in the different tracking detectors, and the \pt of the track.

A previous version of the neural network was widely and successfully used in Run 1 to discard fake tracks at the analysis level.
The use of a different set of variables, whose computations are less time consuming, together with optimization of the code made it possible
to deploy this classifier in the Run~2 trigger without any significant impact on the execution time.
The evaluation uses only about 0.2\% of the total CPU budget. Furthermore, the better performance  of this fake track rejection in both stages of
the HLT leads to 16\% less CPU consumption in the entire software trigger.
The neural network was trained on simulated tracks. The working point was chosen
such that it rejects 60\% of all fake tracks, while maintaining an efficiency of about 99\%.

\begin{figure}
  \centering
    \includegraphics[width=0.45\textwidth]{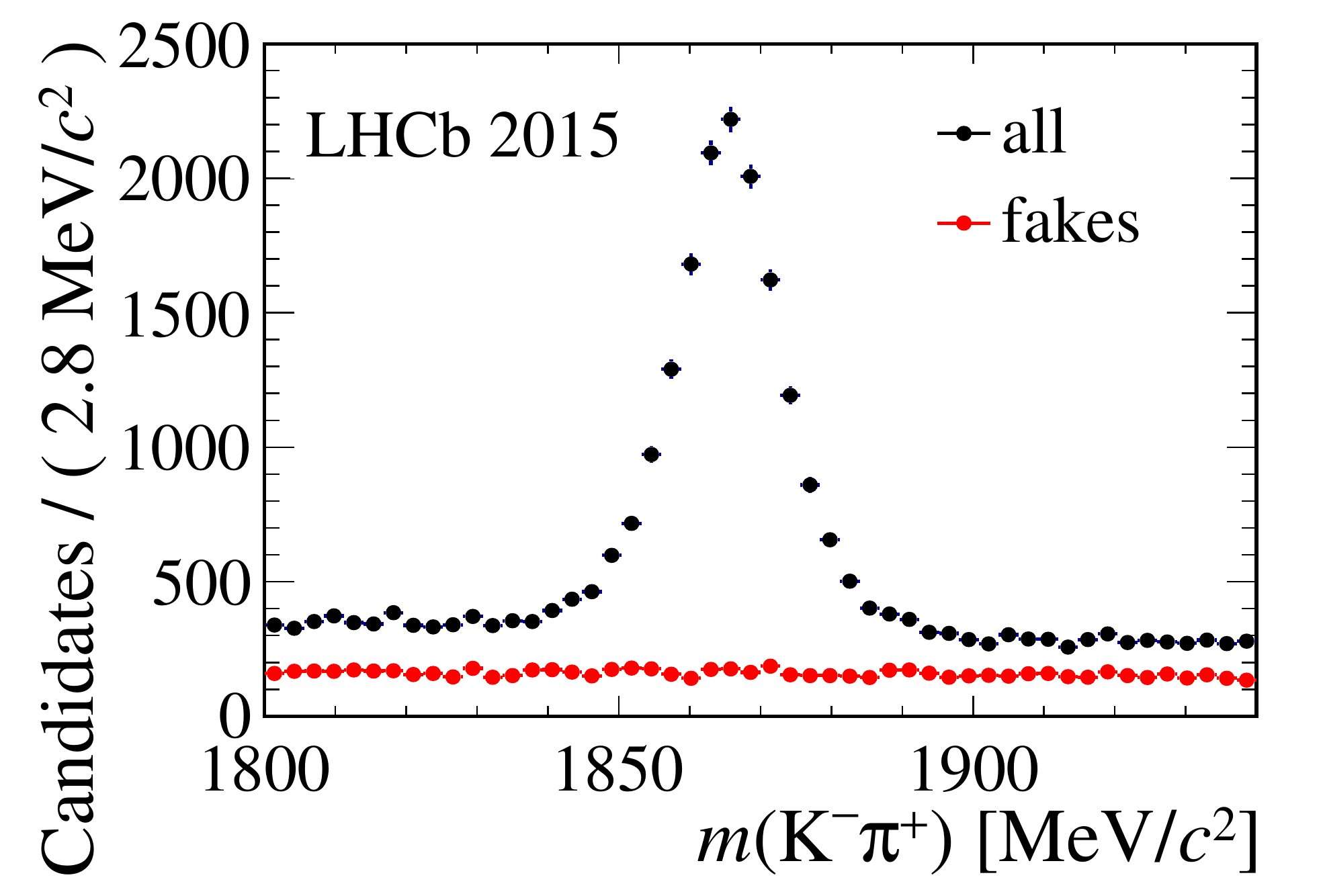}
  \includegraphics[width=0.45\textwidth]{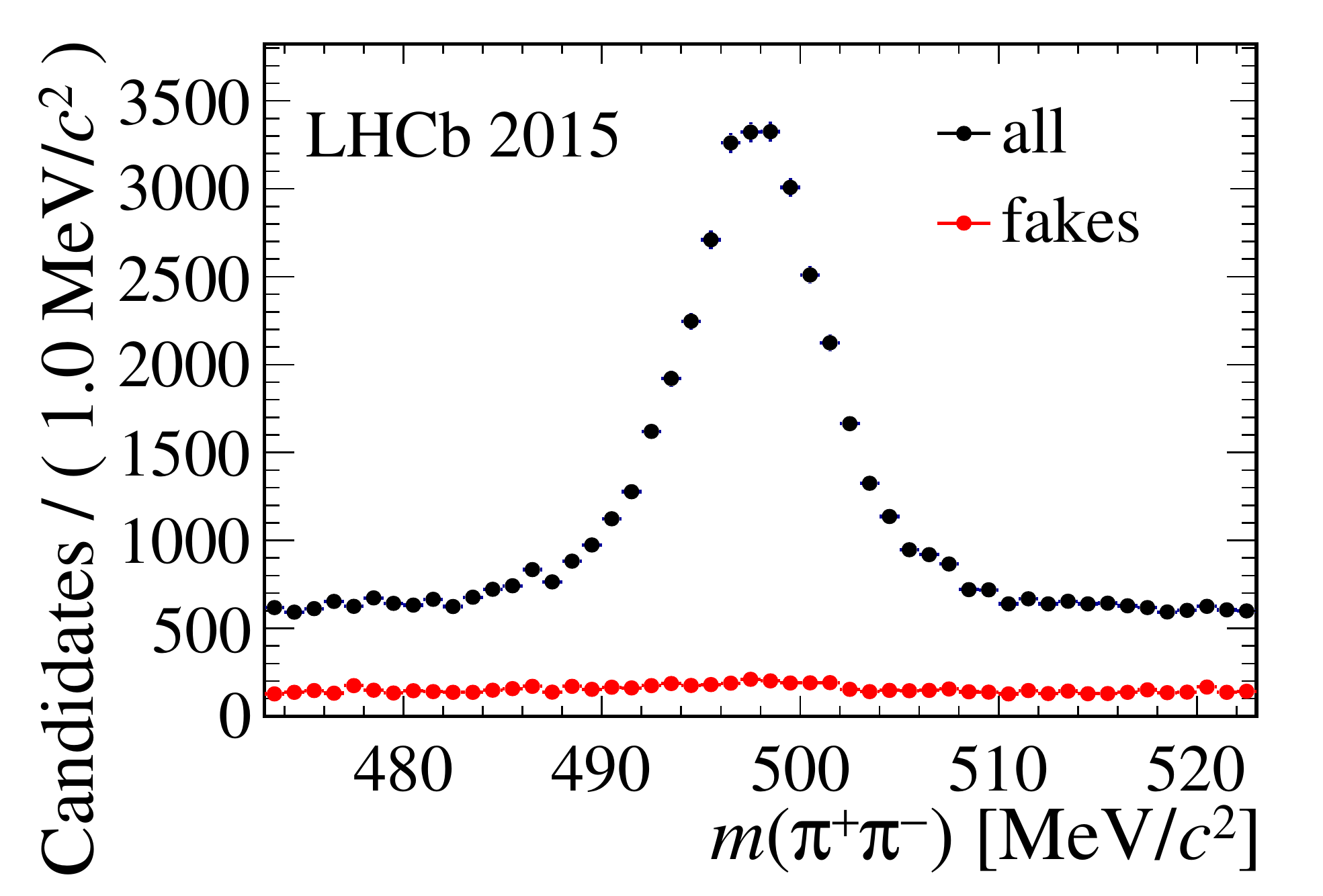}
  \caption{Performance of the fake-track classifier on (left) \decay{D}{\Km\pip} and (right) \decay{\KS}{\pim\pip} decays. For these plots, the clones have been removed.}
  \label{fig:GP}
\end{figure}

The performance of the fake track removal was validated on first collision data in 2015 to ensure a uniform response
over a large area of the phase space. As an example, the
performance for \decay{D^0}{\Km\pip} and \decay{\KS}{\pim\pi} decays is shown in Fig.~\ref{fig:GP}.

After the removal of fake tracks, the remaining tracks are filtered to remove so-called clones.
Clones can be created inside a single pattern-recognition algorithm or, more commonly, originate
from the redundancy in the pattern-recognition algorithms.
Two tracks are defined as clones of each other if they share enough
hits in each subdetector. Only the subdetectors where both tracks have hits
are considered. The track with more hits in total is kept and the other is discarded.
This final list of tracks is subsequently used to select events as discussed in
Sec.~\ref{sec:hlt}.

\subsubsection{Tracking efficiency}
\label{sec:hlt2_tracking_eff}

\begin{figure}
  \centering
  \includegraphics[width=0.45\textwidth]{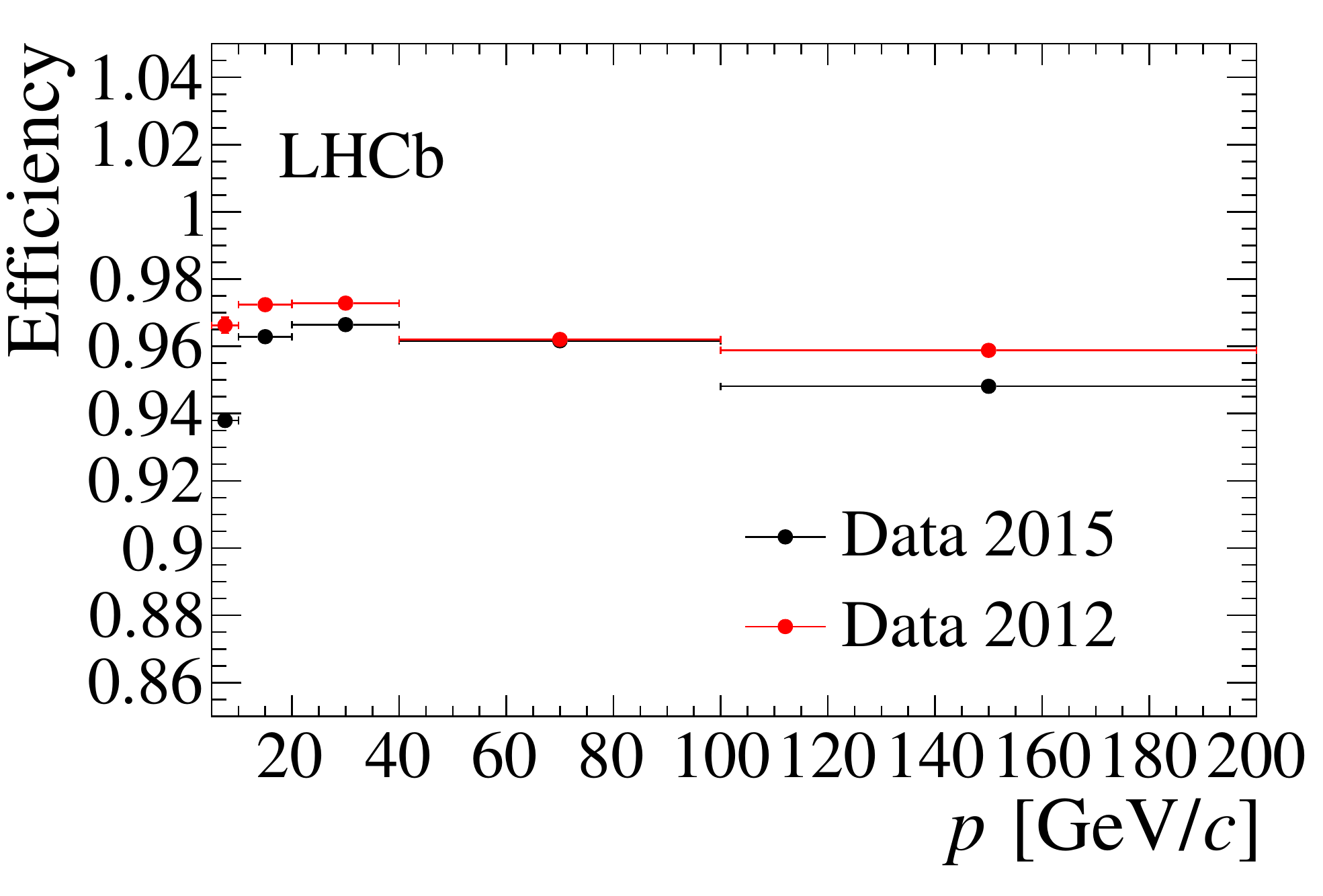}
  \includegraphics[width=0.45\textwidth]{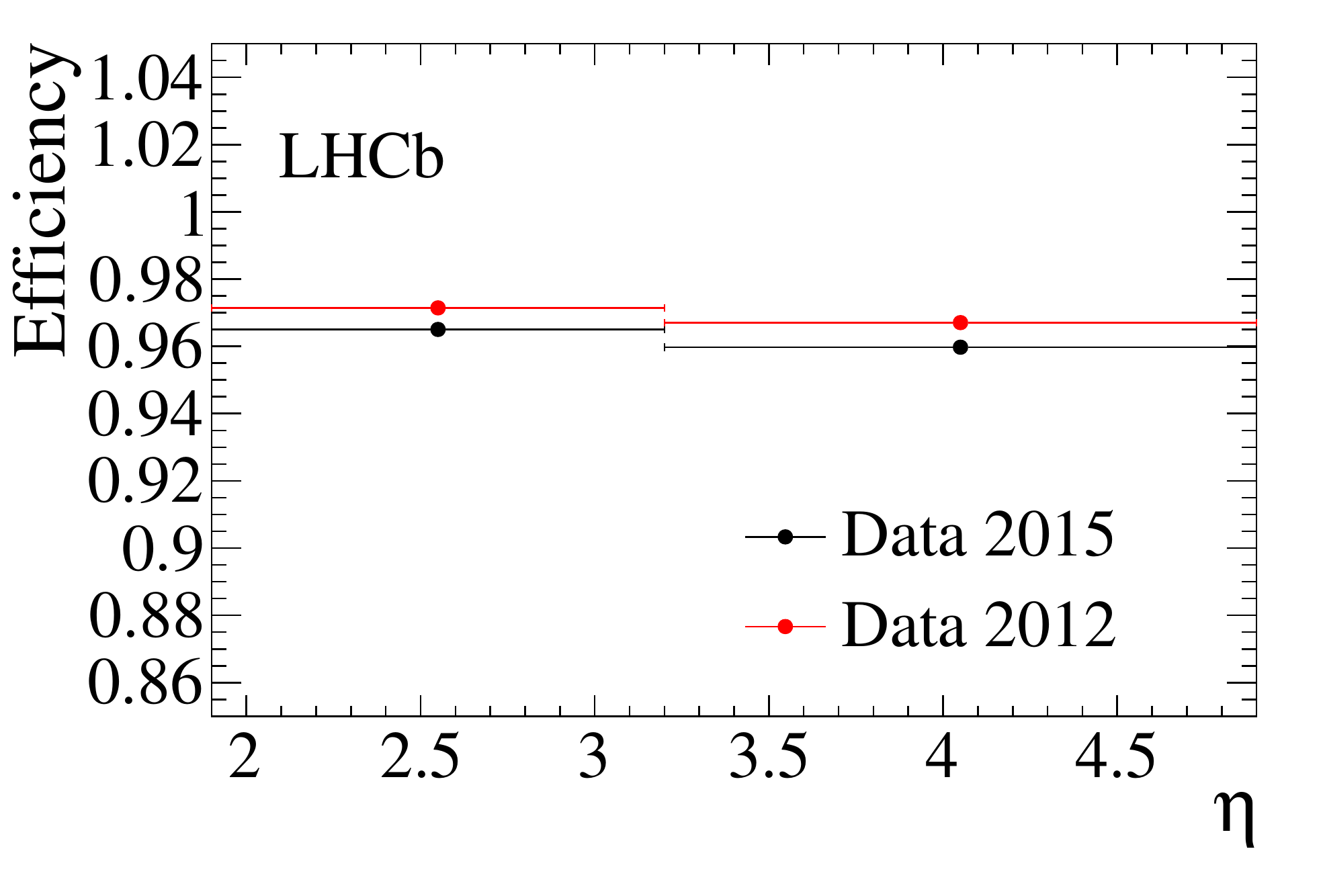}
  \caption{Comparison of the track reconstruction efficiency in 2015  and 2012 data as a function of the momentum (left) and pseudorapidity (right).}
  \label{fig:trackingefficiency}
\end{figure}

The track reconstruction efficiency is determined using a tag-and-probe method on \decay{\jpsi}{\mumu} decays that originate from the decays of \Pb-hadrons~\cite{LHCb-DP-2013-002}. One muon is reconstructed using the full reconstruction, while the other muon is reconstructed using only specific subdetectors, making it possible to probe the others. For Run 2, the track reconstruction efficiency is determined in \hlttwo using  the data collected by specific trigger lines, see Sec.~\ref{sec:hlt2_calib}.
The performance compared to Run 1 is shown in Fig.~\ref{fig:trackingefficiency}.
Given that the OT has a readout window which is larger than 25\ns and therefore is prone to spillover effects when reducing the bunch spacing from 50\ns to 25\ns, a small reduction in the track reconstruction efficiency is observed in 2015.

\subsubsection{Invariant mass resolution}
\label{sec:invariant_mass}

\begin{figure}
  \centering
   \includegraphics[width=0.45\textwidth]{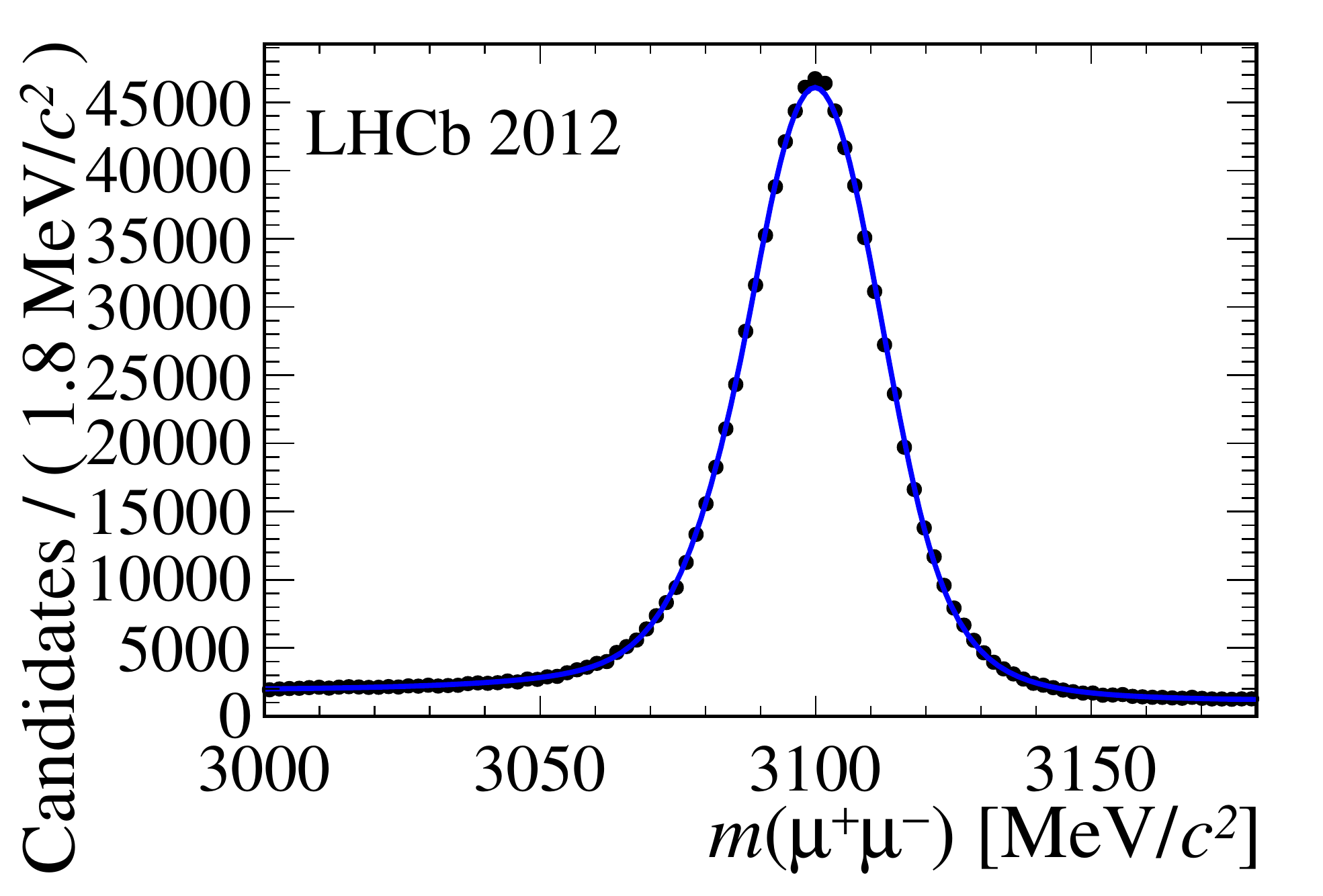}
   \includegraphics[width=0.45\textwidth]{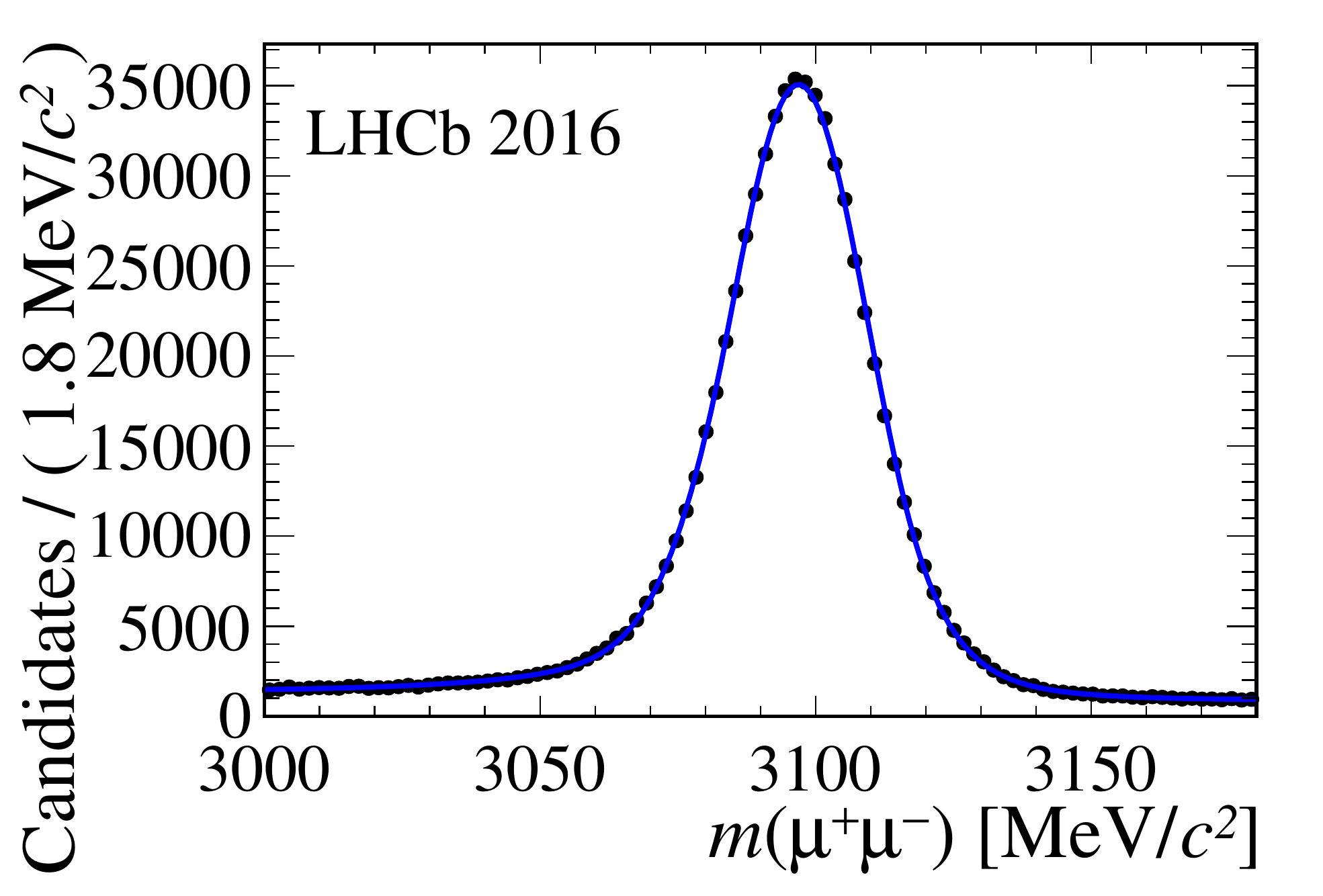}
  \caption{Comparison of the invariant mass distributions for a subset of the 2012 (left) and 2016 (right) data set, using \decay{\jpsi}{\mumu} decays, with the \jpsi originating from a \bquark-hadron.}
  \label{fig:massresolution}
\end{figure}

The invariant mass resolution is determined on a sample of \decay{\jpsi}{\mumu} decays, where the \jpsi originates from a \Pb-hadron decay.
The dimuon invariant mass distribution is modelled using a double Crystal Ball function~\cite{Skwarnicki:1986xj},
where the weighted mean of the standard deviations of the two Gaussian components is used to estimate the resolution.
The distributions for subsamples of the 2012 and 2016 data can be seen in Fig.~\ref{fig:massresolution},
the resolutions are $12.4$~\mevcc for the 2012 data sample and $12.7$~\mevcc for the 2016 data sample. The difference comes from a slightly higher-momentum spectrum in 2016, due to the larger beam energy in Run~2, and a small degradation in the performance due to the use of a simplified description of the
detector geometry throughout Run~2.

\label{sec:resolutions}
\subsubsection{Impact parameter and decay time resolutions}
\label{sec:ip-timeres}

\begin{figure}
  \centering
	\includegraphics[width=0.475\textwidth]{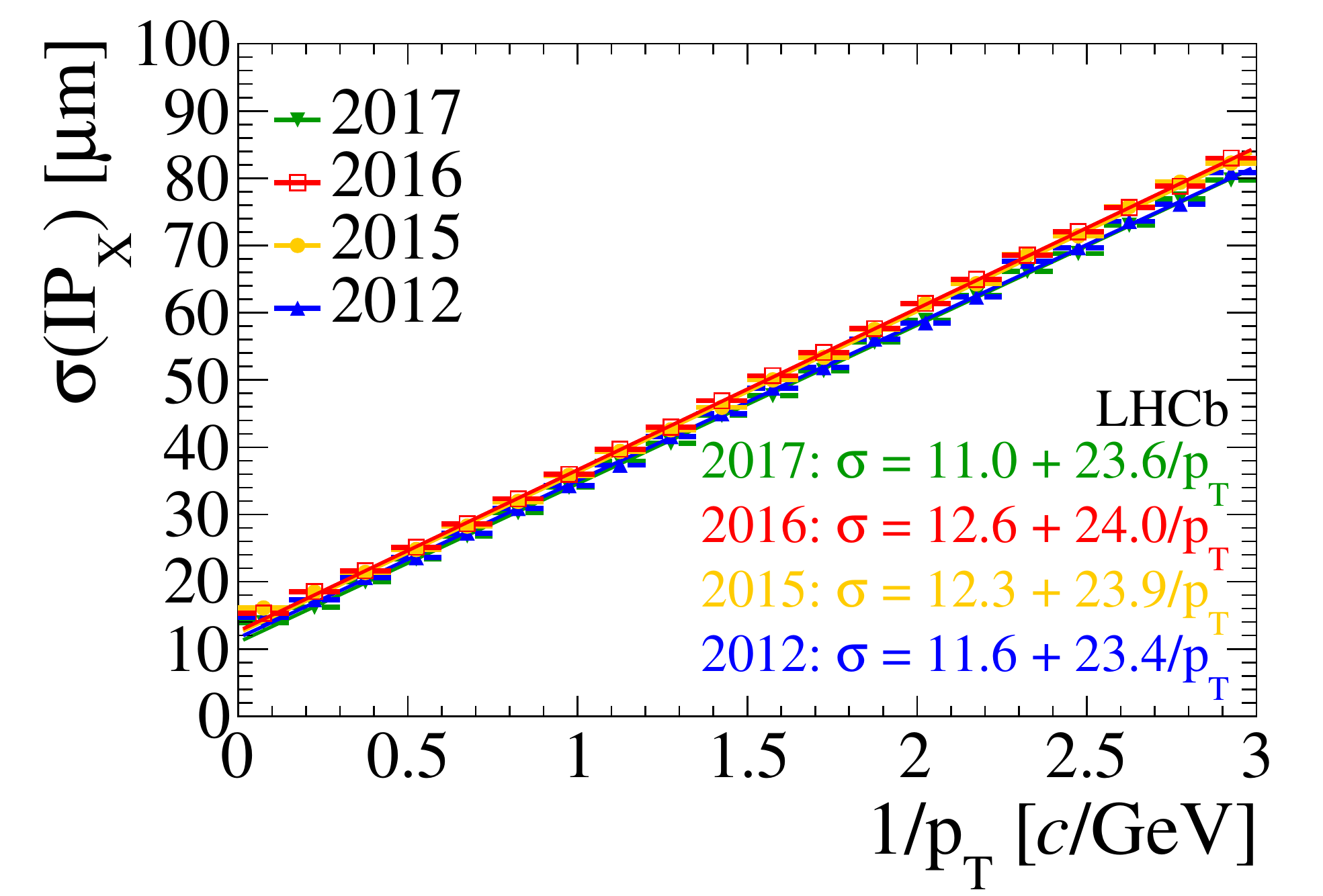}
	\includegraphics[width=0.475\textwidth]{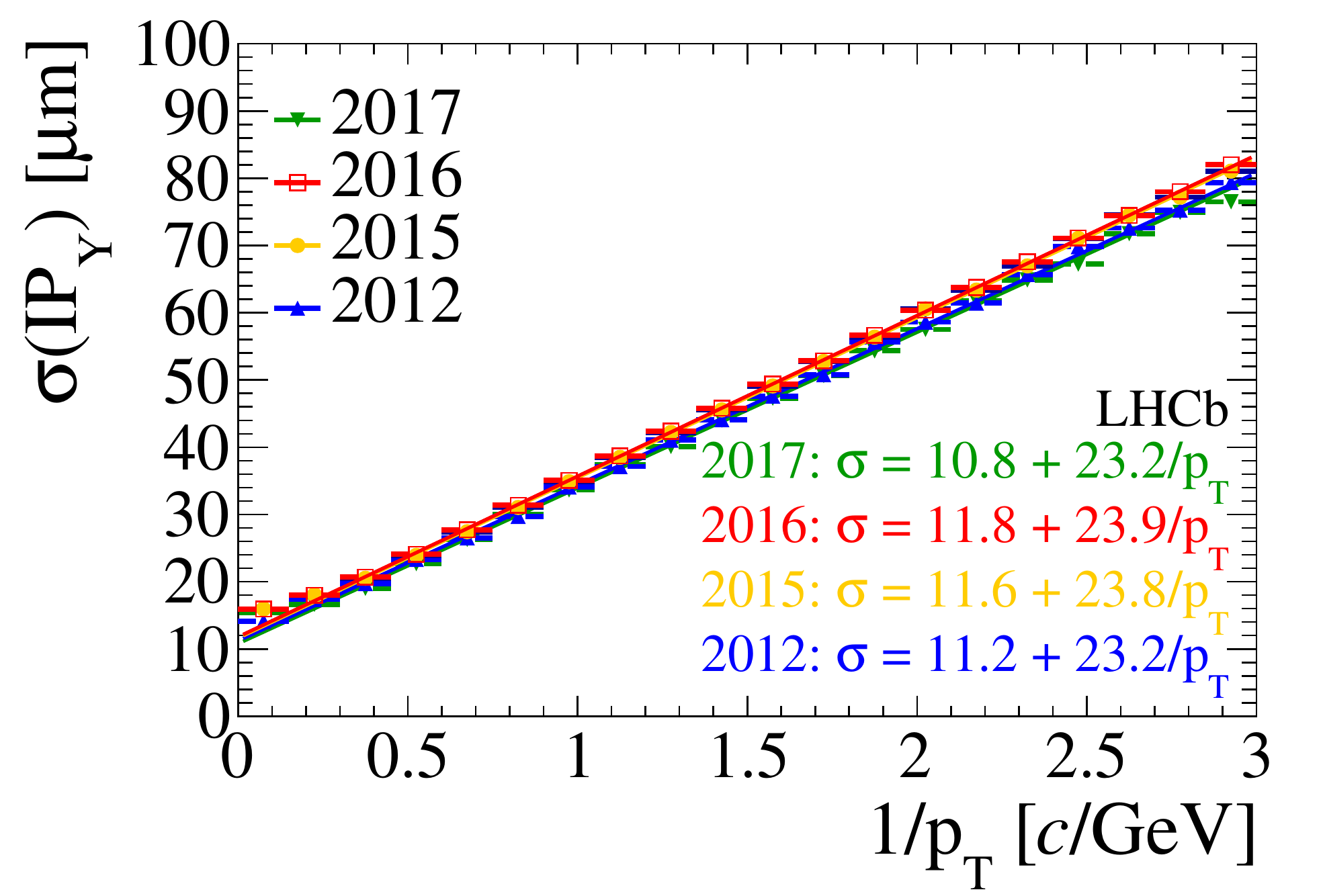}
	\caption{Resolution of the $x$ (left) and $y$ (right) components of the impact parameter comparing the 2012 (blue), 2015 (orange), 2016 (red)
	and 2017 (green) data-taking periods. The resolution as a function of \pt is given in the bottom right
	corner.}
  \label{fig:ipresolution}
\end{figure}

The impact parameter and decay time resolutions are extracted with data-driven methods
which are described in more detail elsewhere\cite{LHCb-DP-2014-002}.
The impact parameter is defined as the distance between a particle trajectory
and a given PV. It is one of the main discriminants between particles produced directly in the primary
interaction and particles originating from the decays of long-lived hadrons.
The impact parameter resolution as a function of $1/\pt$ is shown
in Fig~\ref{fig:ipresolution}.
Only events with one reconstructed PV are used, and the PV fit is rerun excluding each track in turn.
The resulting PV is required to have at least 25 tracks to minimise the contribution from the PV resolution.
Multiple scattering induces a linear dependence on $1/\pt$.
For high \pt particles, the impact parameter resolution is roughly 12\mum in both the $x$ and $y$ directions.
The observed improvement of about 1\mum in 2017 data taking is due to
the use of an updated \velo error parametrisation.

\begin{figure}
  \centering
  \includegraphics[width=0.5\textwidth]{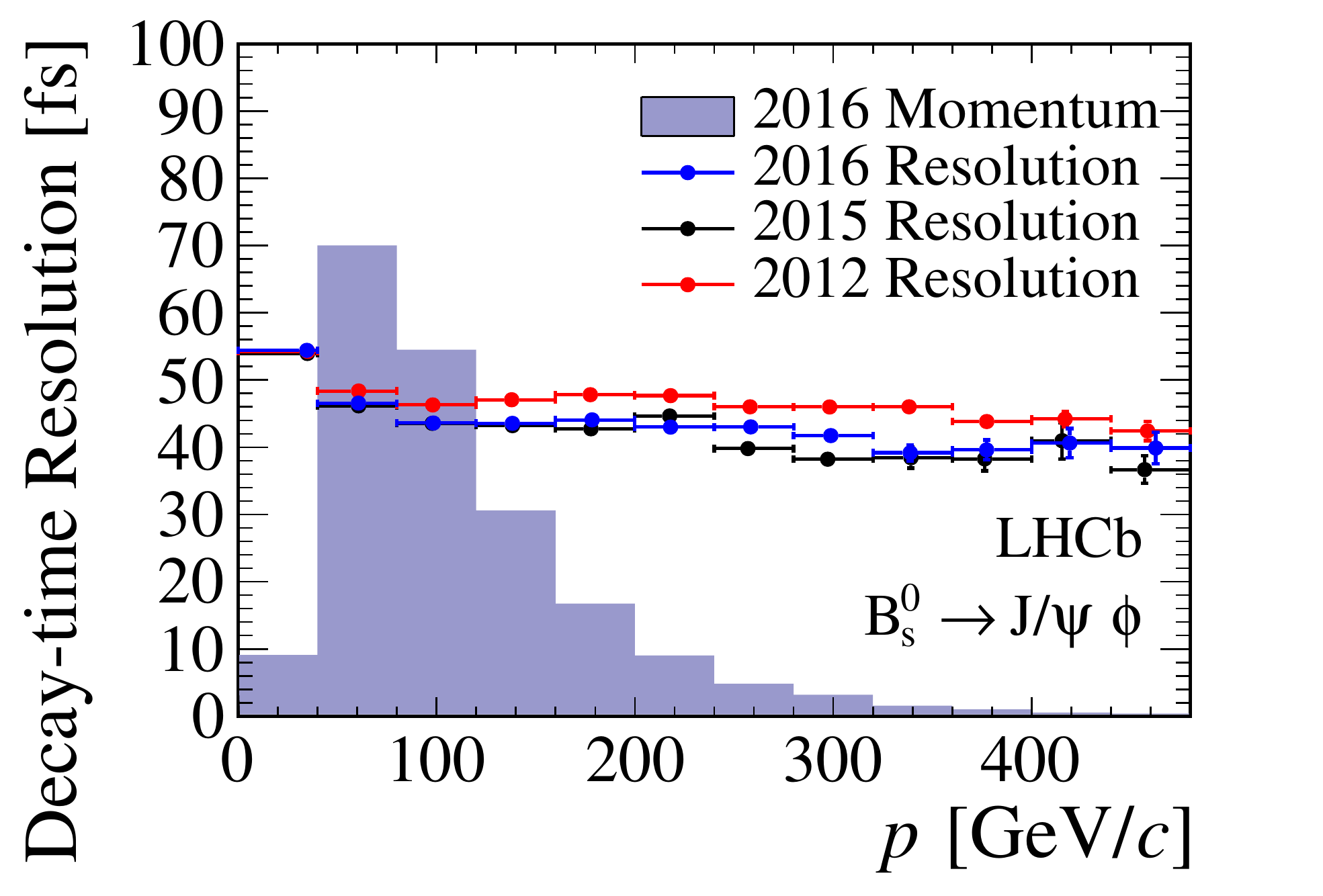}
  \caption{Decay time resolution for \decay{\Bs}{\jpsi\phi} decays (in their rest frame) as a function of momentum. The filled histogram shows the distribution of \Bs meson momenta, to give an idea of the relative importance of the different resolution bins for the analysis sensitivity.}
  \label{fig:decaytimeresolution}
\end{figure}

The decay time of a particle is determined from the distance between the PV and the secondary decay vertex.
An excellent decay time resolution is a key ingredient of time-dependent mixing and \CP violation measurements.
The resolution is determined from \jpsi decays combined with two random tracks which mimic \decay{\Bs}{\jpsi\phi} decays.
In the absence of any impact parameter requirements these combinations come mainly from prompt particles
and, therefore, the expected decay time is zero. The width of the distribution is thus
a measure of the decay time resolution. A comparison of the decay time resolution as a function of momentum for Run 1, 2015, and 2016 data taking is shown in Fig.~\ref{fig:decaytimeresolution}. For Run 2  the average resolution is about 45\fs for a 4-track vertex.

%
%




\subsection{Muon reconstruction}

As mentioned in Sec.~\ref{sec:muonid}, the same muon identification algorithm is used
in \hlttwo and \hltone, apart from the fact that the \hlttwo algorithm takes as its input the full set of fitted tracks
available after the \hlttwo reconstruction.

\subsection{RICH reconstruction}
\begin{figure}
  \centering
  \includegraphics[width=0.49\textwidth]{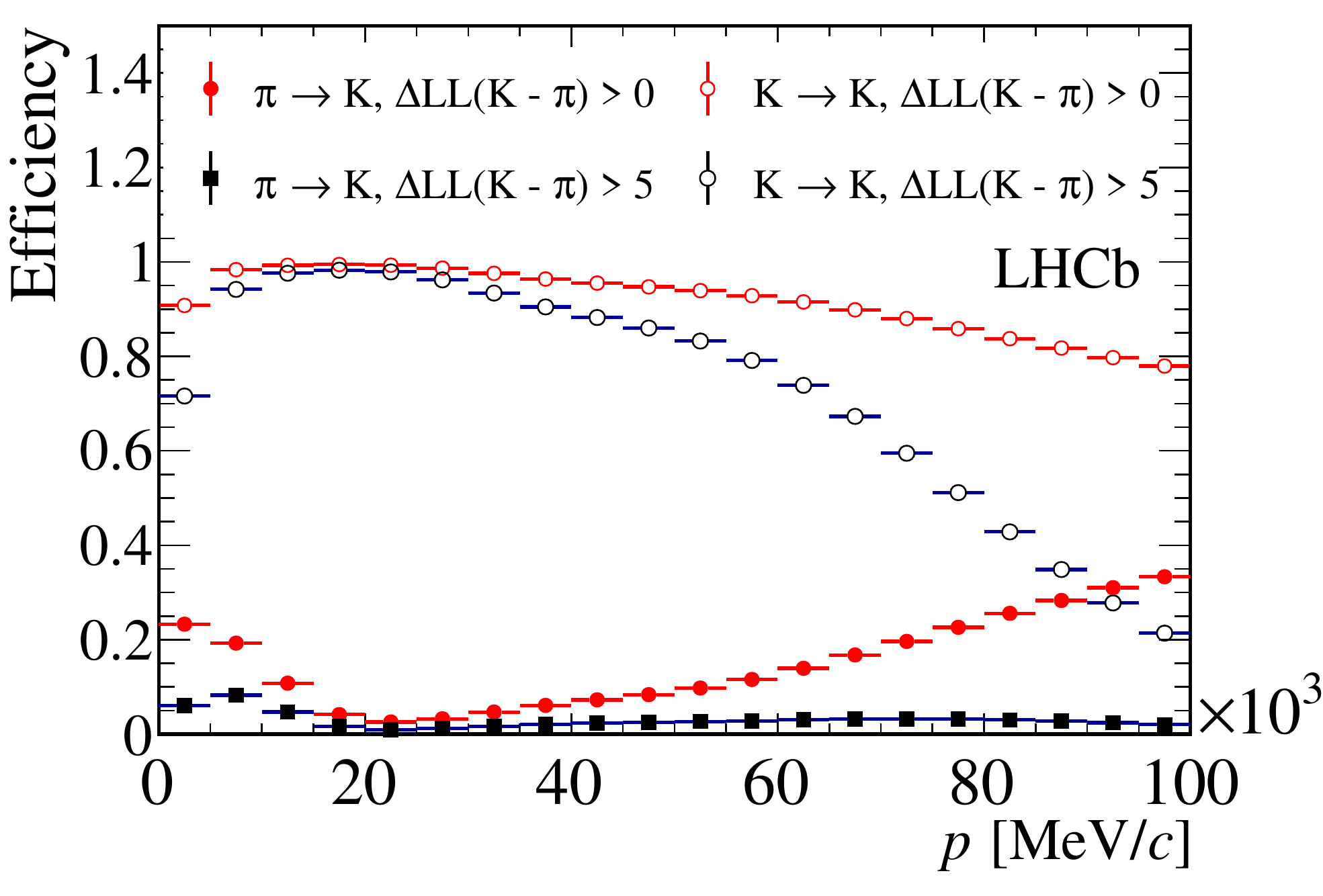}
   \includegraphics[width=0.49\textwidth]{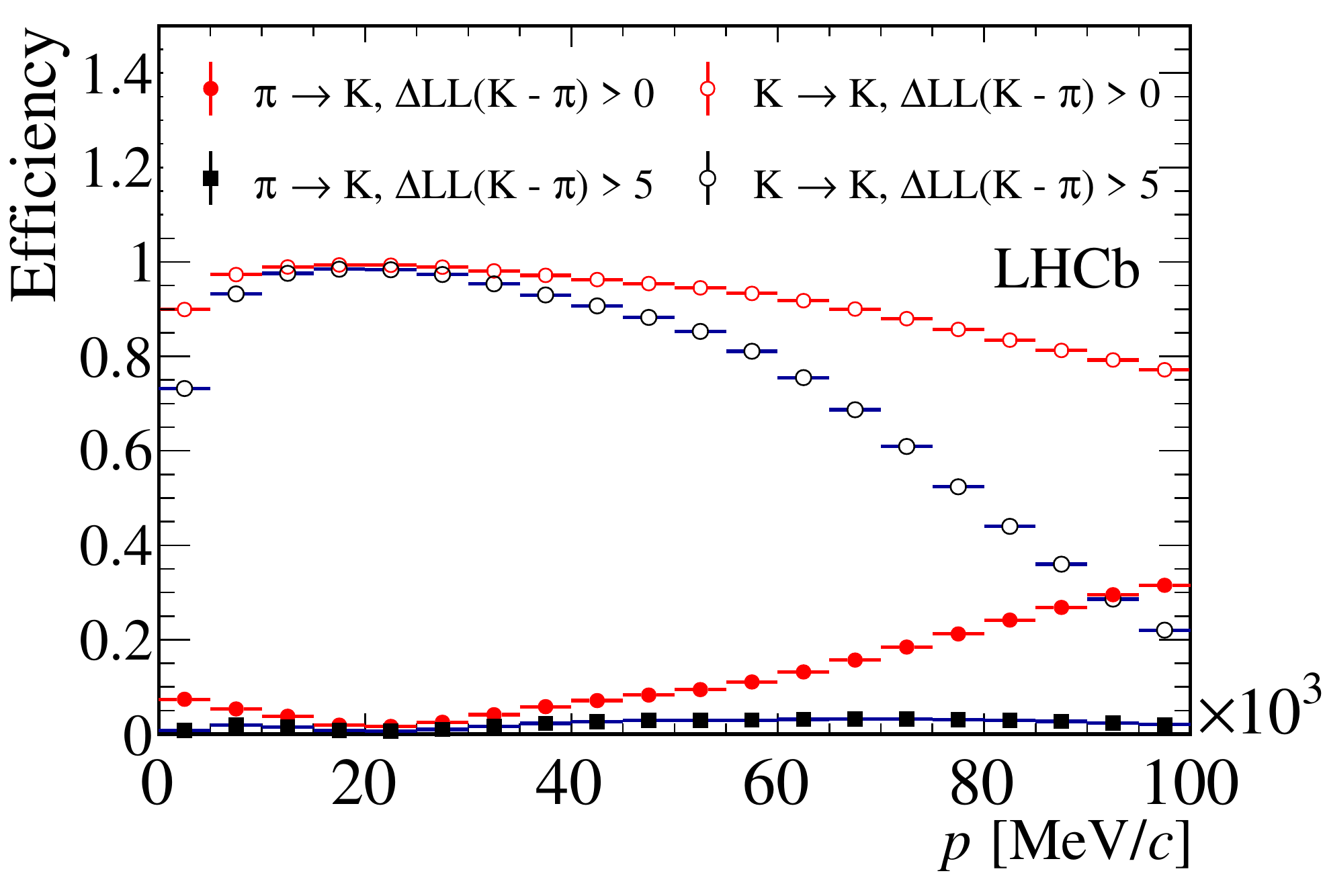}
  \caption{Efficiency and fake rate of the \rich identification for the 2012 (left) and the 2016 (right) data.}
  \label{fig:richperformance}
\end{figure}

The identification of different particle species
is crucial across LHCb's physics programme.
The \rich detectors provide the main
discrimination between deuterons, kaons, pions, and protons.
Cherenkov light is emitted in a cone around the flight direction
of a charged particle, where the cone width depends on the velocity of the particle.
The photon yields, expected Cherenkov angles, and estimates of the per-track Cherenkov angle resolution
are computed under each of the deuteron, proton, kaon, pion, muon and electron mass hypotheses.
The \rich reconstruction considers simultaneously all reconstructed tracks and
all Cherenkov photons in \richone and \richtwo in each event. The reconstruction algorithm
provides a likelihood for each mass hypothesis.
As the \rich reconstruction consumes significant computing power, it could not be run for every track
in the Run~1 real-time reconstruction. Improvements in the \hlt and in the \rich reconstruction itself made
it possible, however, to run the full algorithm in the Run~2 \hlttwo.
The performance of the \rich particle identification is shown in Fig.~\ref{fig:richperformance} for the 2012 and 2016 data.
A small improvement is obtained in Run~2 for particles below 15\gevc of momentum.

\subsection{Calorimeter reconstruction}
The reconstruction of electromagnetic particles (photons, electrons, and \piz mesons) is performed by the calorimeters.
A cellular automaton algorithm 
is used to build clusters from the energy deposits in the different calorimeter subsystems,
which are combined to determine the total energy of each particle~\cite{Breton:681262}.
Neutral particles are then identified according to their isolation with respect to the reconstructed tracks.
Electron identification is also provided by combining information from the isolation of clusters in the \ecal, the presence of clusters in the \presh, the energy deposited in the \hcal, and the position of possible Bremsstrahlung photons.

High-$E_{\text{T}}$ \piz mesons and photons are indistinguishable at the trigger level, as they both appear as a single cluster,
while low-$E_{\text{T}}$ \piz mesons are built by combining resolved pairs of well-separated photons.
The neutral-cluster reconstruction algorithm run in \hlttwo is the same as that run off\-line.

The identification of these clusters as either neutral objects or electrons uses information from both the PS/SPD detectors,
and a matching between reconstructed tracks and calorimeter clusters. Early in Run~2 this online identification was not identical to the off\-line
version because the HLT did not reconstruct T-tracks (see Fig.~\ref{fig:trackTypes}), since these are not
directly used by physics analyses. They are, however, relevant for neutral-particle identification. This misalignment
was gradually reduced as Run~2 progressed, first by adding the reconstruction of T-tracks and then by subsequently applying a Kalman filter to them
to align the algorithm to the off\-line reconstruction sequence.

\begin{figure}
  \centering
   \includegraphics[width=0.475\textwidth]{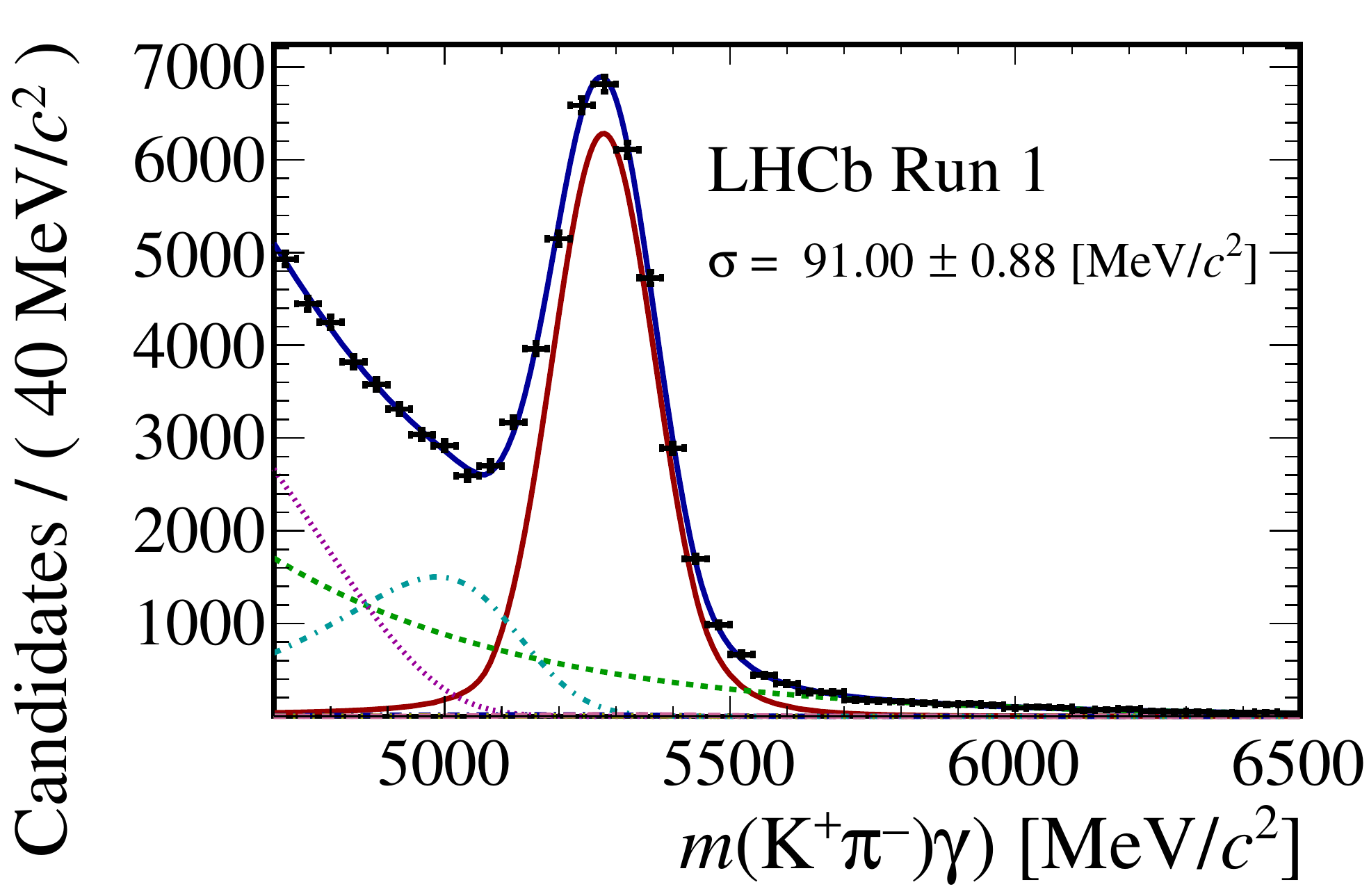}
   \includegraphics[width=0.475\textwidth]{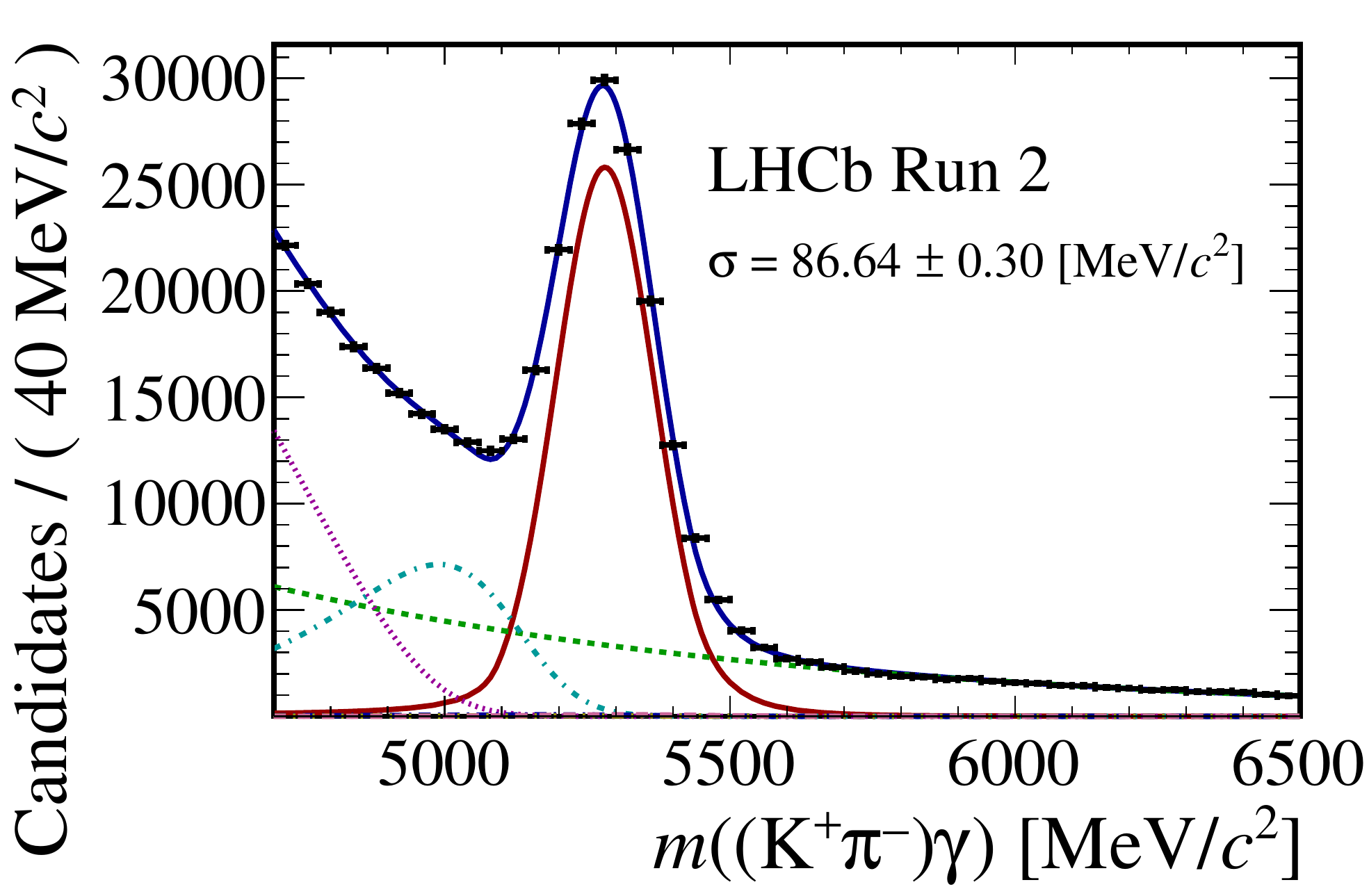}

	\caption{Invariant mass of $B^0 \to (K^+ \pi^-) \gamma$ candidates in Run 1 (left) and Run 2 (right). The fit model includes the (red) signal component, 
(dashed green) combinatorial background, (dot-dashed turqoise) misidentified physics backgrounds (e.g. \decay{\Bs}{\phi\gamma} where a kaon is 
misidentified as a pion) and (dotted magenta) partially reconstructed physics backgrounds.}
  \label{fig:caloperformance}
\end{figure}

A fully automated \ecal calibration was introduced in 2018.
The automatic LED calibration is performed for fills longer than 3.5 hours
as indicated in Fig.~\ref{fig:align}, while the absolute $\pi^0$ calibration
is processed once per month when sufficient data (amounting to 300M events) is collected. 
The performance of the calorimeter reconstruction is shown in Fig.~\ref{fig:caloperformance}
using $B^0 \to (K^+ \pi^-) \gamma$  decays.
The invariant mass resolution has been improved with respect to Run 1 from about $91$\mevcc to
$87$\mevcc.

\section{Trigger performance}
\label{sec:hlt}
The LHCb trigger performance is optimized around two key metrics: the \lz kinematic and occupancy thresholds for
each of the main trigger lines (muon, dimuon, electron, photon, and hadron); and the optimization of the \hlt timing
budget, which defines the maximum allowed \hltone output rate. An automated procedure is used to divide the \lz bandwidth among a set
of representative signal channels. It has been significantly improved with respect to Run~1 and is described here.
The procedure for determining the \hlt processing budget is also described, and the \lz and \hlt performance
is evaluated using a tag-and-probe approach on Run~2 data.

\subsection{\lz bandwidth division}
The relative simplicity of the information available for the \lz trigger decision means that the trigger lines listed in Table~\ref{tab:LzThresholds} cover the majority of the LHCb physics programme. Out of these, the high-\pt muon trigger consumes a relatively negligible rate and is insensitive to the running conditions. The remaining five trigger lines: hadron, muon, electron, photon and dimuon, must have their \pt and \et thresholds tuned to maximize signal efficiency under different LHC conditions. In particular, during the luminosity ramp-up of the LHC, signal efficiencies can be improved by reducing the thresholds to maintain an \lz output rate close to the maximal readout rate of 1~MHz.

Once the LHC reaches its nominal number of colliding bunches in a given year, determining the optimal division of rate between the \lz channels is important for achieving the physics goals of the experiment. This
so-called ``bandwidth division'' is performed using a genetic algorithm to minimise the following pseudo-$\chi^{2}$ for a broad range of simulated signal samples that are representative of the LHCb physics programme:
	\begin{equation}
        \chi^{2}(r) = \sum^{N}_{i}w_{i}\times \left ( 1 - \frac{\varepsilon(r)_{i}}{\varepsilon(r)^{\textrm{max}}_{i}}\right )^{2}\, .
\end{equation}
\noindent The sum is over the $N$ signal samples, $\varepsilon(r)_{i}$ is the efficiency including detector dead time of the $i^{\textrm{th}}$ data set, and $\varepsilon^{\textrm{max}}$ is the efficiency including detector dead time when all of the bandwidth is allocated to this data set.
The ratio of dead-time-corrected efficiencies is designed to ensure that inefficient signal samples contribute more to the $\chi^{2}$, {\em i.e.}\ the algorithm prioritizes improving the efficiency of signal samples which start with a low absolute efficiency over making identical absolute efficiency improvements for signals with high efficiencies to begin with. The weight assigned to each data set, $w_{i}$, is predetermined by the LHCb collaboration and is designed to grant more bandwidth to higher-priority physics channels.

The dead-time correction to the signal efficiency acts as a rate limiter and is dependent upon the filling scheme: 
\begin{equation}
        \varepsilon(r) = \epsilon \times \left [1-\delta^{\text{phys}}(r)\right ] \times\left [1-\delta^{\text{tech}}(r)\right ].
\end{equation}
\noindent  Here $\epsilon$ is the overall \lz signal efficiency and $r$ is the retention of collisions collected using random trigger lines
(henceforth ``nobias''). The physics dead time $\delta^{\text{phys}}(r)$ is zero below $r_{\text{limit}}=1.1$~MHz, which is the maximum HLT1 throughput, and $r/r_{\text{limit}}$ above this. The technical dead time, $\delta^{\text{tech}}(r)$ is determined from a model trained on a filling-scheme dependent emulation of the detector readout dead time.

\begin{figure}
  \centering
  \includegraphics[width=0.80\textwidth]{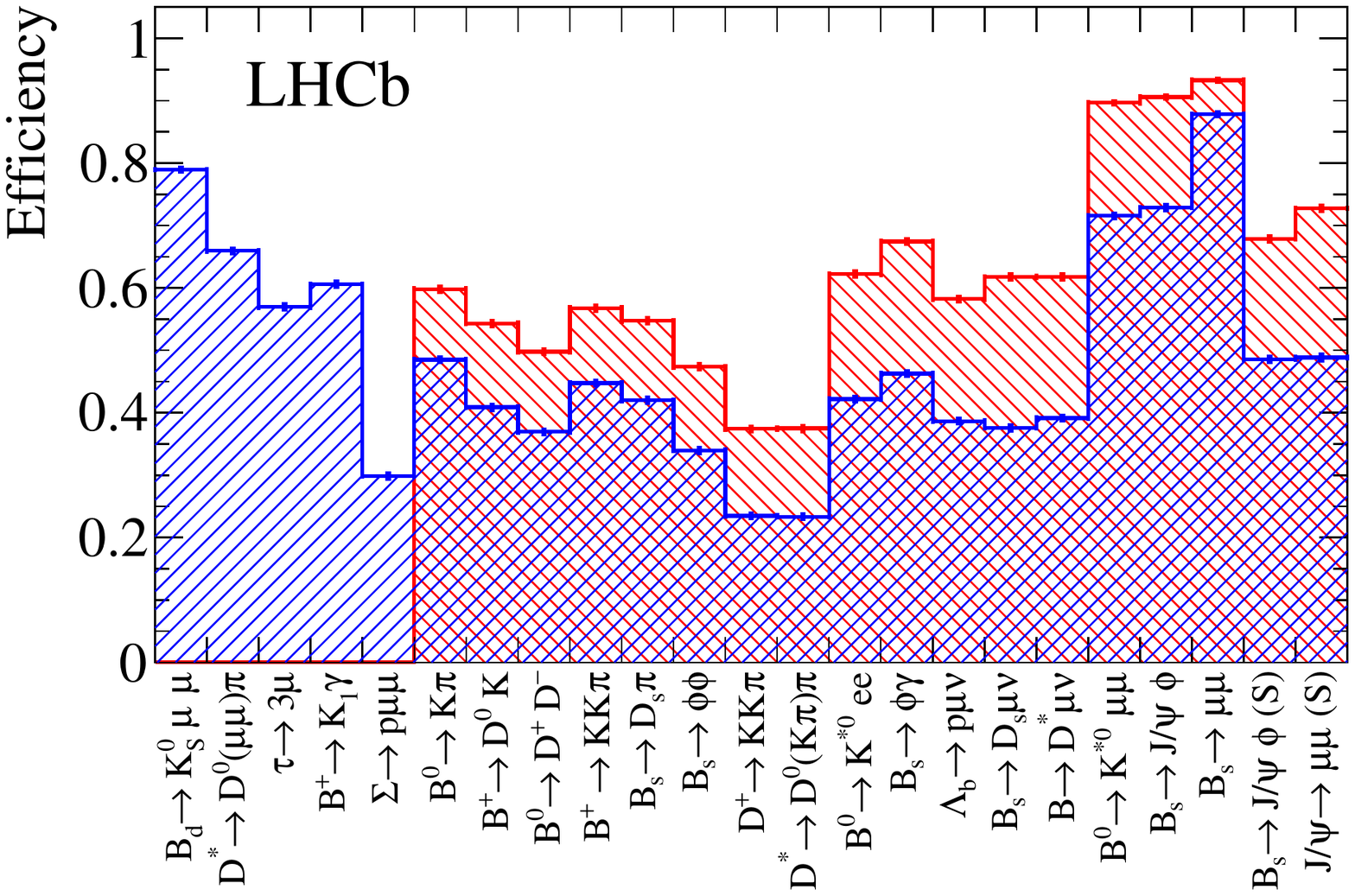}
  \includegraphics[width=0.80\textwidth]{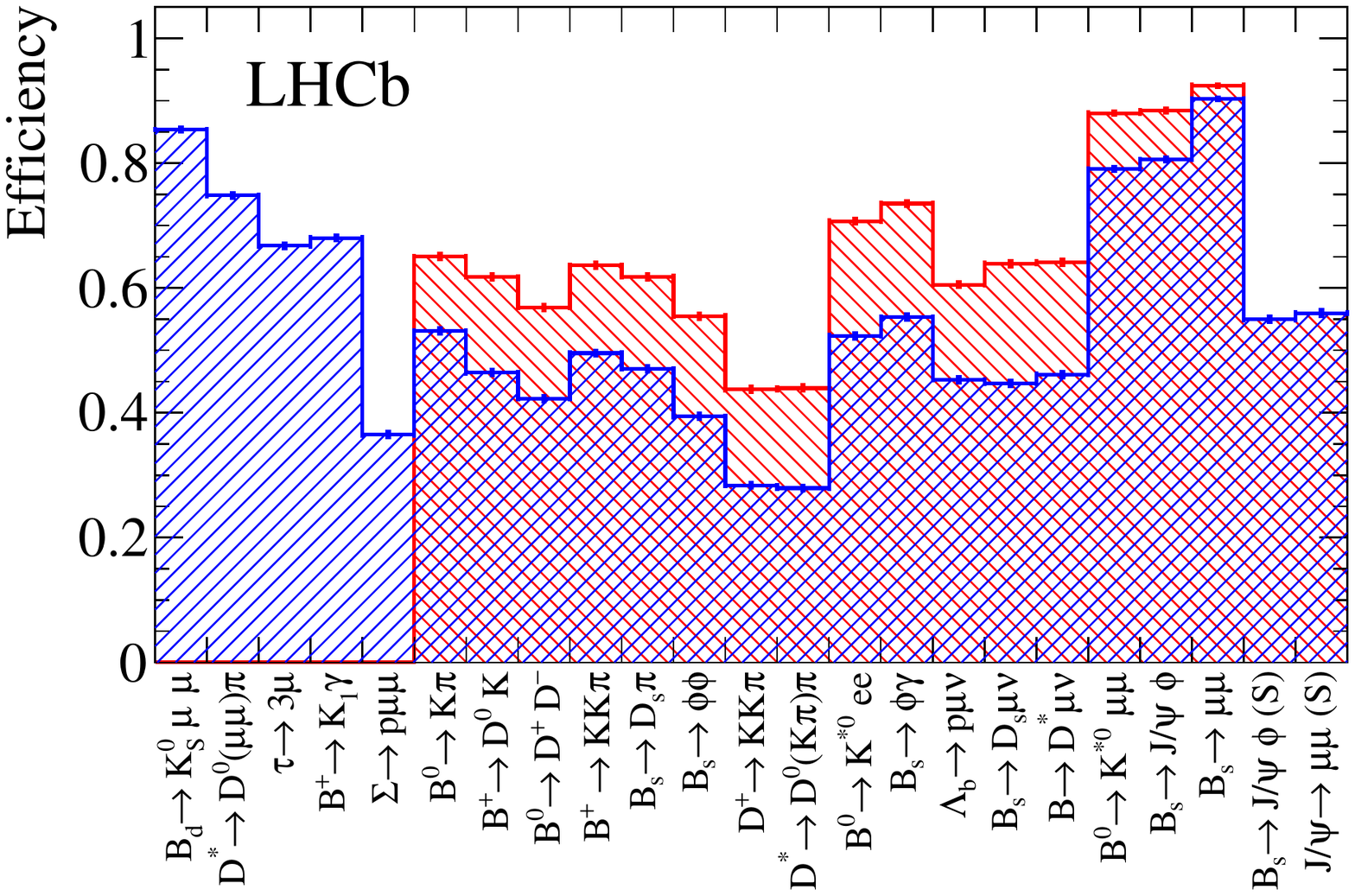}
  \caption{Efficiencies per signal mode for (top) 2016 and (bottom) 2017 data-taking periods measured in simulation.
Red (left-slanted) hatched plots are when the entire \lz bandwidth is granted to this signal mode, whereas blue (right-slanted) hatched plots are following the
bandwidth division. Signals which appear only
in blue are used for performance validation and are not part of the optimization itself. Channels followed by ``(S)'' are selected
in a kinematic and geometric volume which is particularly important for spectroscopy studies.} 
  \label{fig:bwdiv}
\end{figure}

The results of the \lz bandwidth optimization are shown in Fig.~\ref{fig:bwdiv} for the 2016 and 2017 data-taking conditions.
The different optimal points are mostly connected to the different LHC running conditions in the two years,
in particular to problems in 2017 
which limited the maximum possible number of bunches in the
LHC and hence led to lower trigger thresholds.

\subsection{Measuring the \hlt processing speed}
Trigger configurations are tested for processing speed and memory usage on 13\tev nobias
collected at the same average number of visible interactions per bunch crossing as in regular data taking.
As the \lz trigger conditions affect the complexity of events
processed by the \hlt, these tests are repeated for each \lz configuration. Nobias events passing the \lz configuration
are processed by a dedicated ``average'' EFF node loaded with the same number of total tasks as in the online data-taking configuration.
The timing is measured separately for \hltone running on events passing \lz, and \hlttwo running on events passing both \lz and \hltone.
Each \hltone and \hlttwo task processes an independent sample of around 10,000 events during this test, with the number of
events chosen to balance robustness and turnaround speed.
The processing speed of each of the individual \hltone and \hlttwo tasks is then averaged in order to remove fluctuations due to the limited number of test events, and these values are compared to the available budget.
In addition, the memory usage is plotted as a function of the event number to verify that there are no memory leaks.

These tests give confidence that the \hlt is running within its budget, and they are particularly important
for spotting problems after any major changes are made to a stable configuration.
They do not, however, give a perfect reflection of the performance expected on the full farm. In
particular, the effect of calibration and alignment tasks which run in parallel with the \hltone and \hlttwo tasks is neglected,
as are the I/O issues associated with sending events from \lz to the \hltone tasks, the buffering of \hltone events
and the action of reading them back into \hlttwo tasks,
and the overhead from sending the events accepted by \hlttwo to the off\-line storage. For this reason it makes little sense to quote detailed
performance numbers for the \hlt from these tests.

\clearpage

\subsection{Optimization of the \hlt timing and disk buffers}
The timing budget of the \hlt is defined as the average time available for each \hlt task to process an event, when
the processing farm is fully loaded.\footnote{The processing farm consists of processors with a certain number of physical cores,
but a single task will not generally fully load a single physical core. For this reason the number of logical \hlt tasks which
are launched for each physical CPU core is optimized by measuring the overall throughput of events in the farm. For the 2015
\hlt farm, this number is typically around $1.6$, depending on the node in question.}
With a traditional single-stage \hlt, as was the case in Run~1, the timing budget is easily determined because the events must always
be processed as they arrive during the collider runtime.  Therefore, it is simply the number of \hlt tasks divided by the input
event rate, which amounted to about 50~ms for a farm with around 50000 logical cores as was available in 2015.
The calculation is more complicated for the two-stage \hlt used in Run~2. Since the
second stage is deferred and can occur during LHC downtime there are two timing budgets, one for
the \hltone stage and one for the \hlttwo stage. These budgets depend on the assumptions made about the length
and distribution of the LHC runtime and downtime periods.

\begin{figure}[t]
  \centering
  \includegraphics[width=0.49\textwidth]{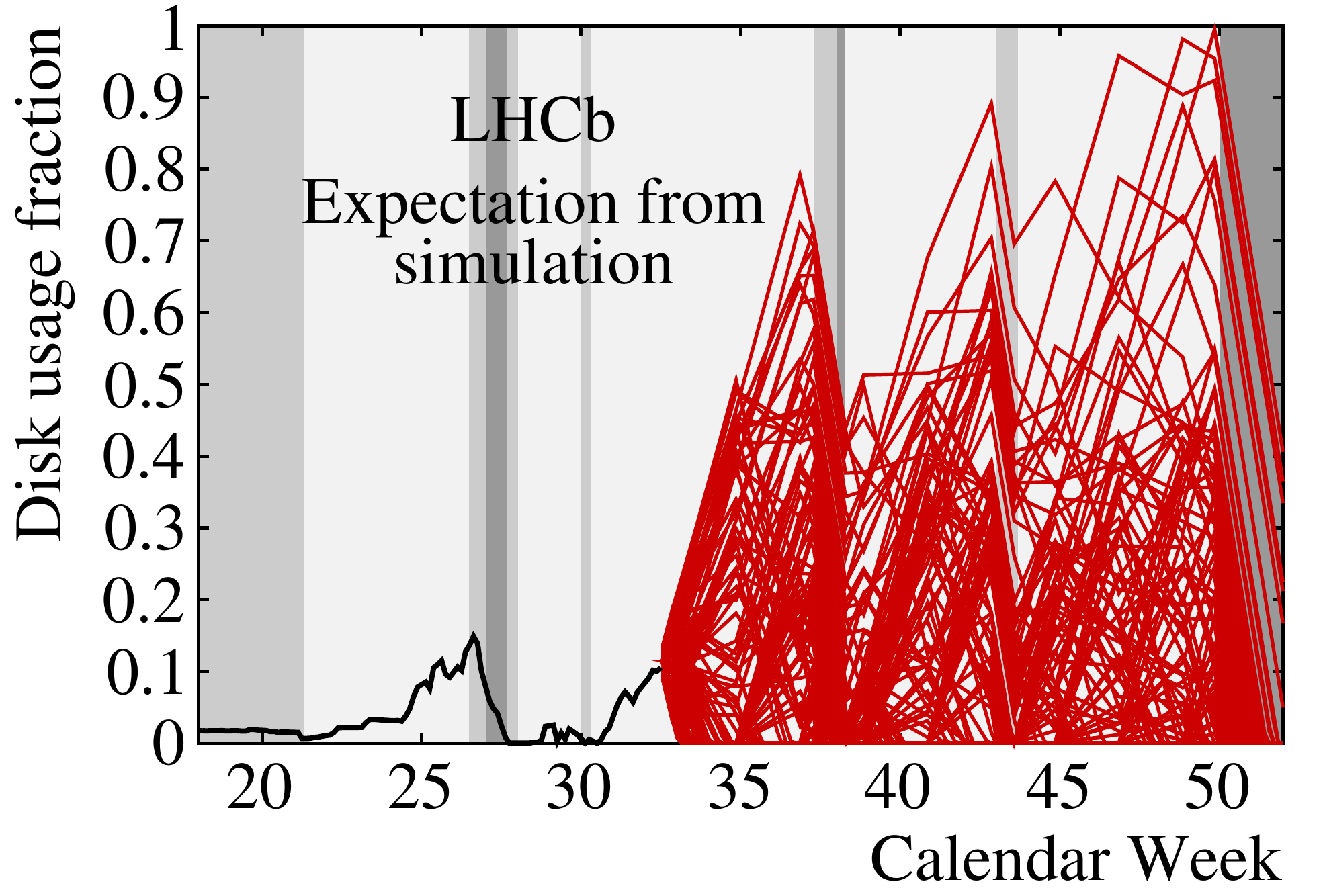}
  \includegraphics[width=0.49\textwidth]{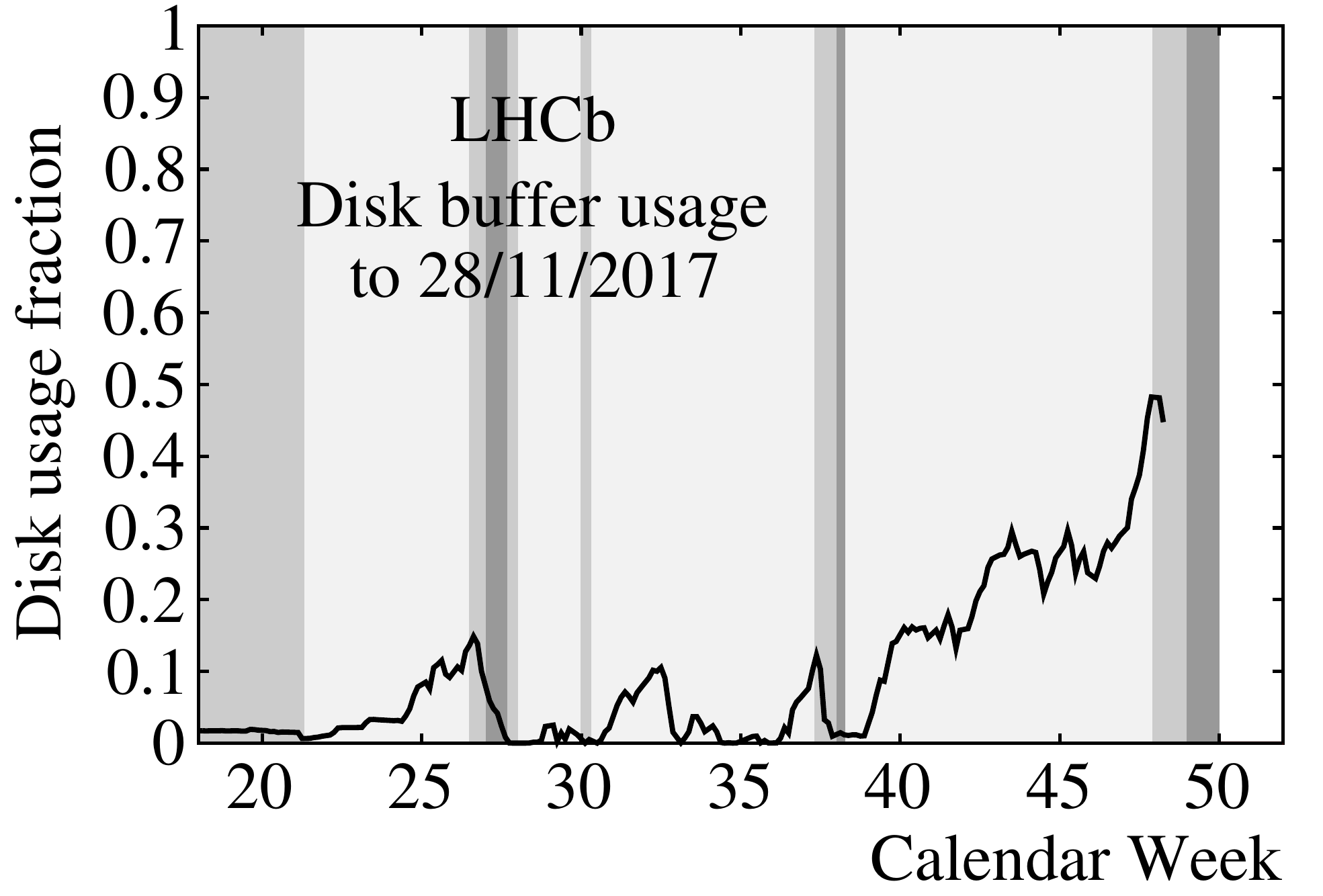}
	\caption{Disk buffer usage projections during (left) and at the end of (right) the 2017 data-taking period. During data taking, simulations (red, left) are used every two weeks to determine the probability of exceeding the 80\% usage threshold. In 2017, the loose \hltone configuration was used for the entire year leading to a maximum buffer capacity of 48\% (black, right). LHC Technical Stops and Machine Development (MD) periods are shown in dark and light grey, respectively. The schedule changed between when this simulation was run in week 32 and the end of the year. An MD period was removed and the duration of the data taking was reduced.}
  \label{fig:buffer}
\end{figure}

The LHC downtime is not uniformly distributed throughout the year. Most occurs
during a winter shutdown, lasting several months, and ``technical stops'' lasting approximately two weeks each and distributed throughout the year.
The runtime is consequently also concentrated, with a peak structure
of repeated 10--15~hour-long collision periods with inter-fill gaps of 2--3~hours between them. The timing budget is determined by simulating
the rate at which the disk buffer fills up and empties, using the processing speed measurements and the
most recent LHC fill structure as a guide.\footnote{In 2015 this optimization used the 2012 fill information.} The objective is to ensure
that the disk buffer will never exceed more than 80\% capacity at any point throughout the remaining data taking period, and is evaluated every two weeks using the actual disk occupancy at the time as the starting point. 
The output rate of \hltone is adjusted to keep the disk buffer usage within the desired limits. This output rate is controlled by switching
between two \hltone configurations, where the tighter configuration sacrifices some rate and efficiency
for the inclusive general purpose trigger lines while protecting the trigger lines used for specific areas of the physics
programme.
The buffer usage is monitored throughout the year, and biweekly simulations are made using the present buffer capacity, \hltone output rate and \hlttwo throughput to determine the projected disk usage until the end of the year. Should a significant fraction of these simulations exceed the 80\% usage threshold, the \hltone configuration is tightened. An example simulation and the disk usage throughout 2017 are shown in Fig.~\ref{fig:buffer}.

\subsection{Efficiency measurement method}
\label{sec:tistosmethod}
All efficiencies are measured on background-subtracted data using the so-called TISTOS method
described in the Run~1 performance paper~\cite{LHCb-DP-2012-004} and
briefly recapped here. Off\-line-selected signal events are divided into the following categories:
\begin{itemize}
\item{\bf TIS:} Events which are triggered independently of the presence of the signal decay. These are
unbiased by the trigger selection except for correlations between the signal decay and the rest of the event. (For example when triggering
on the ``other'' $B$ in the event and subsequently looking at the momentum distribution of the ``signal'' $B$,
the correlation in their momenta is caused by the fact that they both originate in the same fragmentation chain.)
\item{\bf TOS:} Events which are triggered on the signal decay independently of the presence of the rest of the event.
\end{itemize}
All efficiencies quoted in this paper are TOS efficiencies, given by
\begin{equation}
\epsilon = \frac{N(\textrm{TOS}\textrm{ and } \textrm{TIS})}{N(\textrm{TIS})},
\end{equation}
where $N(\textrm{TIS})$ is the number of signal TIS events in the sample, while $N(\textrm{TOS}\textrm{ and } \textrm{TIS})$ is the number of signal events which are both
TOS and TIS. The number of signal events passing and failing the TOS criterion is measured using a histogram sideband subtraction,
as described in Ref.~\cite{LHCb-DP-2013-001}.
In order to reduce the correlations between TOS and TIS events, 
the efficiency is plotted as a function of the \pt and, where appropriate, decay time of the signal particle.  

\subsection{Samples used for performance measurements}
\begin{figure}
  \centering
  \includegraphics[width=0.95\textwidth]{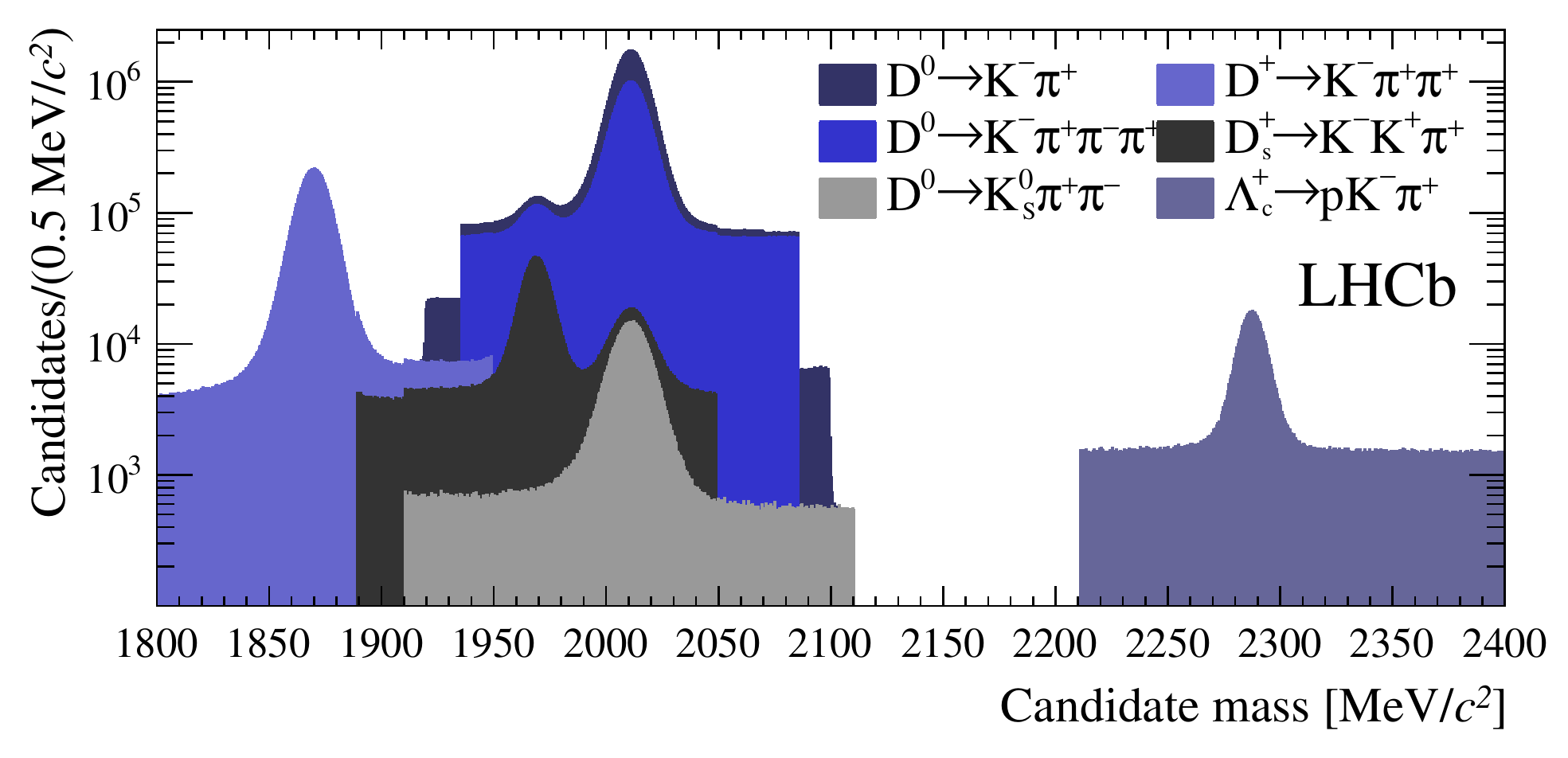}
  \caption{Charm candidates used for the evaluation of the trigger performance.}
  \label{fig:charmsignals}
\end{figure}

\begin{figure}
  \centering
  \includegraphics[width=0.95\textwidth]{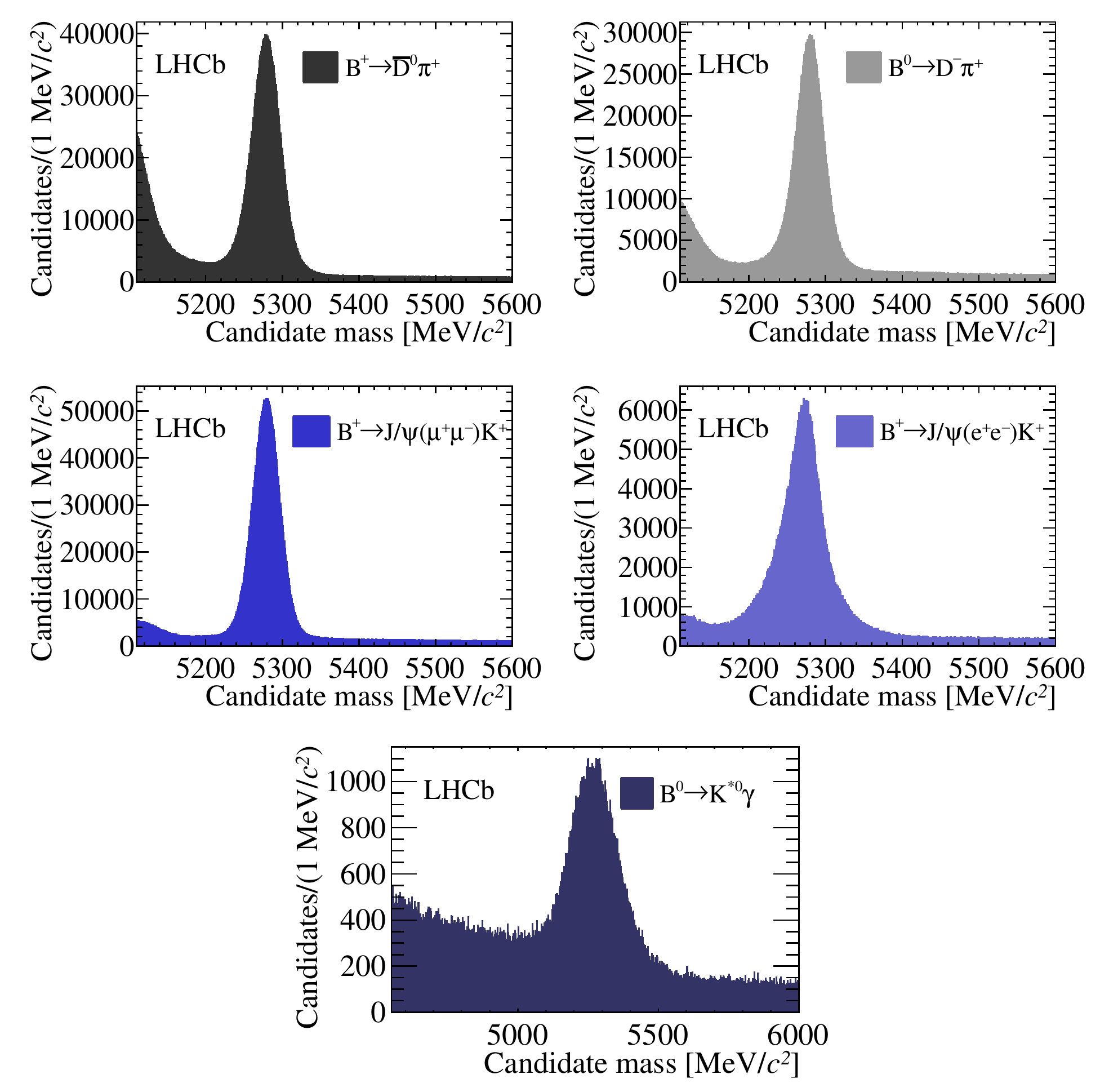}
  \caption{Beauty candidates used for the evaluation of the trigger performance.}
  \label{fig:beautysignals}
\end{figure}

The performance of the \lz and \hlt trigger selections is evaluated using samples of trigger-unbiased signals collected during Run~2,
shown in Figs.~\ref{fig:charmsignals}~and~\ref{fig:beautysignals}. The signal channels are representative of the LHCb physics programme, and
are selected using relatively loose criteria on the kinematics and displacement of the \Pb-hadron and the final-state particles.
As the \Pc-hadron signals are all fully selected by exclusive TURBO trigger lines, only their \lz and \hltone efficiencies
can be measured. The \Pb-hadron signals are selected by off\-line selections without imposing any trigger requirements,
and therefore, they can be used to measure the \lz, \hltone, and \hlttwo efficiencies. The exception is the $\Bz\to\Kstarz\gamma$ decay,
where requiring TIS at \hltone or \hlttwo results in a signal yield and purity which are too small to be usable.  Therefore, this mode is
only used to study the \lz photon and electron trigger performance.

\subsection{\lz performance}

\begin{figure}
  \centering
  \includegraphics[width=0.85\textwidth]{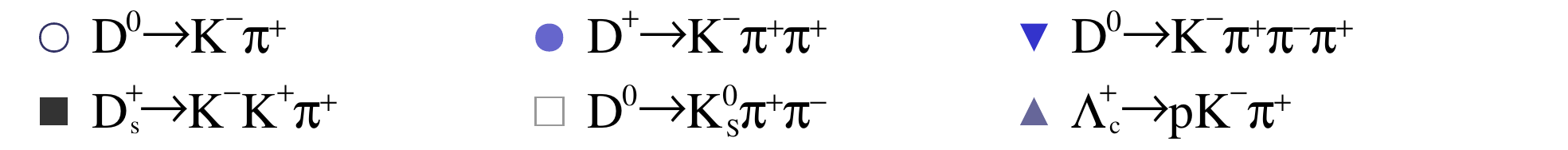}
  \includegraphics[width=0.44\textwidth]{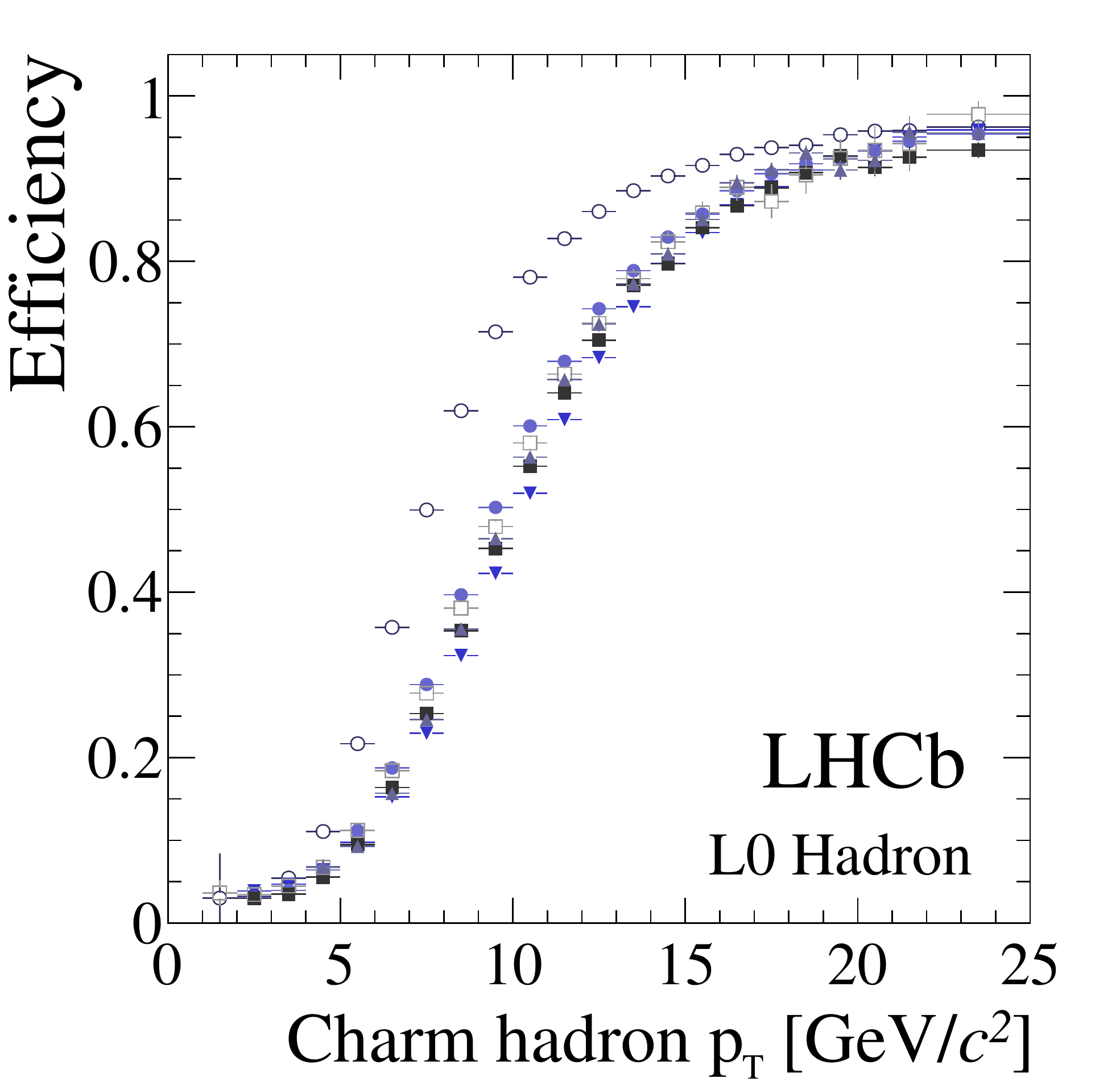}
  \includegraphics[width=0.44\textwidth]{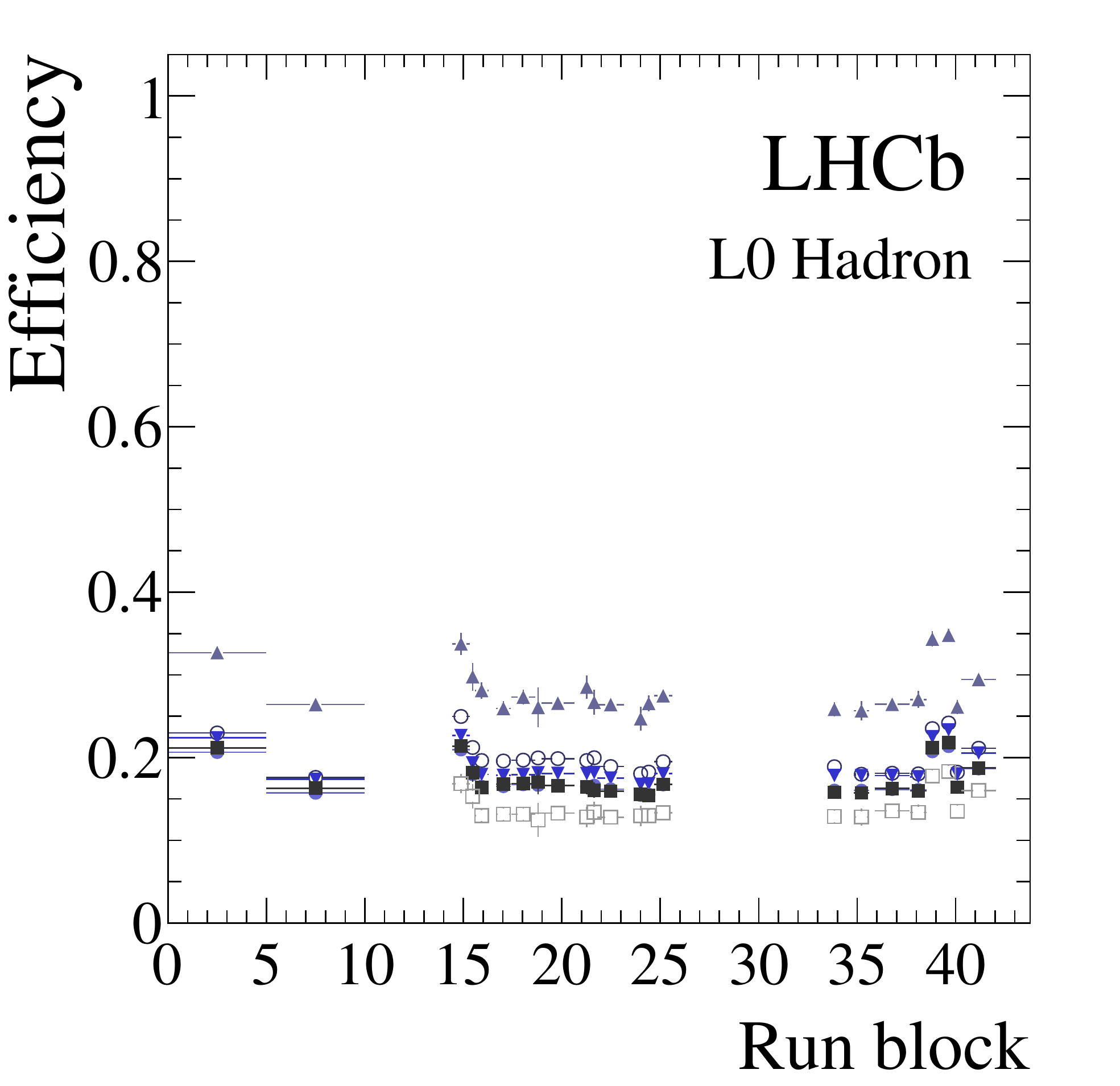}
  \caption{Efficiencies of the \lz trigger lines in Run~2 data for  \Pc-hadron decays. The left plot shows the efficiency
as a function of the hadron \pt, while the right plot shows the evolution of the efficiency as a function of the different trigger
configurations used during data taking. The three blocks visible in the plot, separated by vertical gaps, correspond to the three years of data taking (2015--2017). The \lz hadron
efficiency is shown.}
  \label{fig:L0performance_charm}
\end{figure}

\begin{figure}
  \centering
  \includegraphics[width=0.95\textwidth]{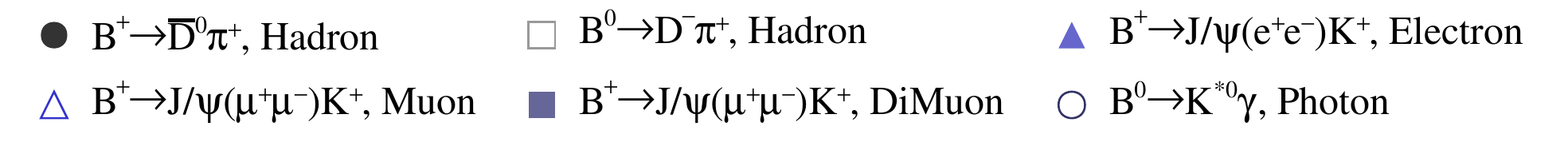}
  \includegraphics[width=0.44\textwidth]{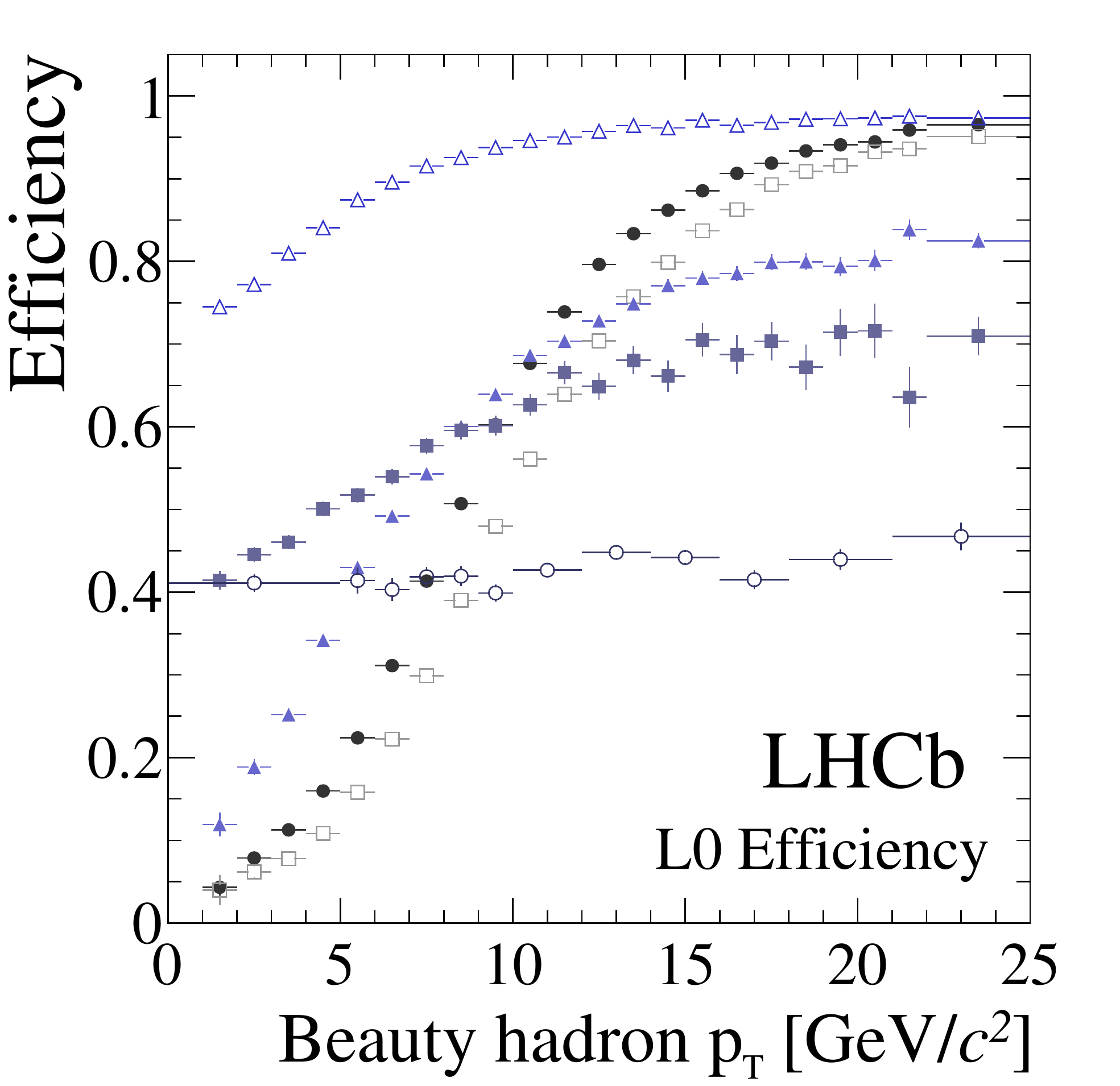}
  \includegraphics[width=0.44\textwidth]{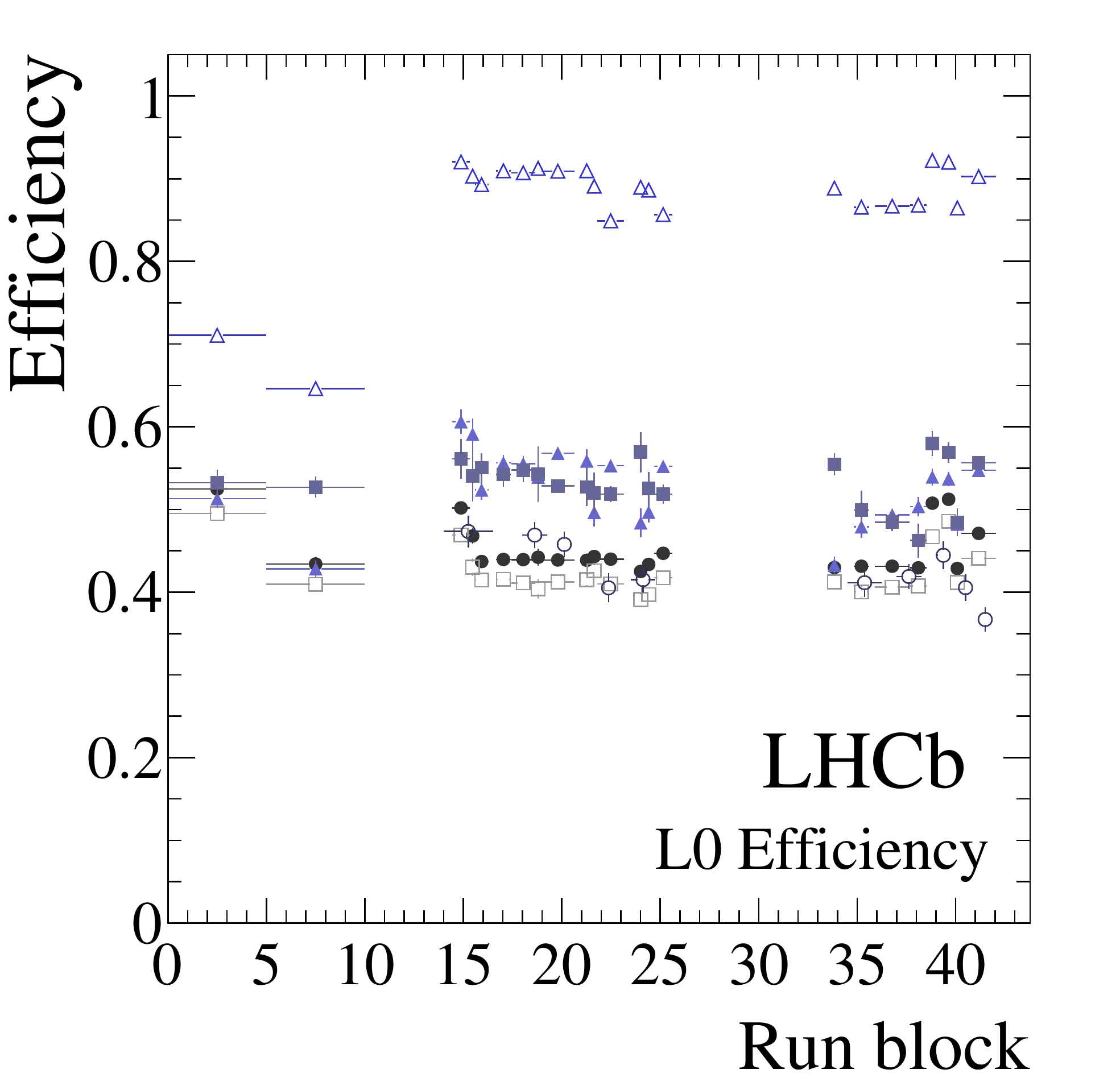}
  \caption{Efficiencies of the \lz trigger lines in Run~2 data for \Pb-hadron decays. The left plot shows the efficiency
as a function of the hadron \pt, while the right plot shows the evolution of the efficiency as a function of the different trigger
configurations used during data taking. The three blocks visible in the plot, separated by vertical gaps, correspond to the three years of data taking (2015--2017). 
	The plotted \lz efficiency for each \Pb-hadron is described in the legend above the plots.}
  \label{fig:L0performance_beauty}
\end{figure}

The efficiencies of the \lz trigger lines in Run~2 are shown in Fig.~\ref{fig:L0performance_charm} for \Pc-hadrons and Fig.~\ref{fig:L0performance_beauty}
for \Pb-hadrons, respectively, as a function of the \pt  and the data-taking period. 
The \lz is optimized to fill the available $\approx 1$~MHz bandwidth for a
given set of LHC running conditions, and so the \lz efficiency evolves as a function of those conditions. In particular,
if the LHC has to run with a reduced number of colliding bunches, the required rejection factor to reach 1~MHz output rate
is smaller and the \lz criteria can be correspondingly loosened, which is the cause of the jumps visible in the bottom plots.
In the case of the $\Bz\to\Kstarz\gamma$ decay, the low efficiency of the dedicated \lz photon trigger is due to the limited
information available to separate electrons and photons within the \lz system. This identification relies on the
amount of electromagnetic showering observed in the preshower detector, before the photons or electrons reach the \ecal, 
and whether there is a hit or not in the \spd detector.
The chosen working point is such that the \lz photon trigger has a high purity but relatively low efficiency. However, many genuine photons are selected by the \lz electron trigger, which is also efficient for photons.
In addiction, a significant amount of photons convert in the detector material between the magnet and the \spd plane. 
These converted photons are reconstructed as neutral clusters off\-line, but leave a hit in the \spd detector and are therefore triggered
as electrons.
In practice the $\Bz\to\Kstarz\gamma$ signal is selected using both electron and photon \lz trigger lines to account for these effects.

\begin{figure}
  \centering
  \includegraphics[width=0.55\textwidth]{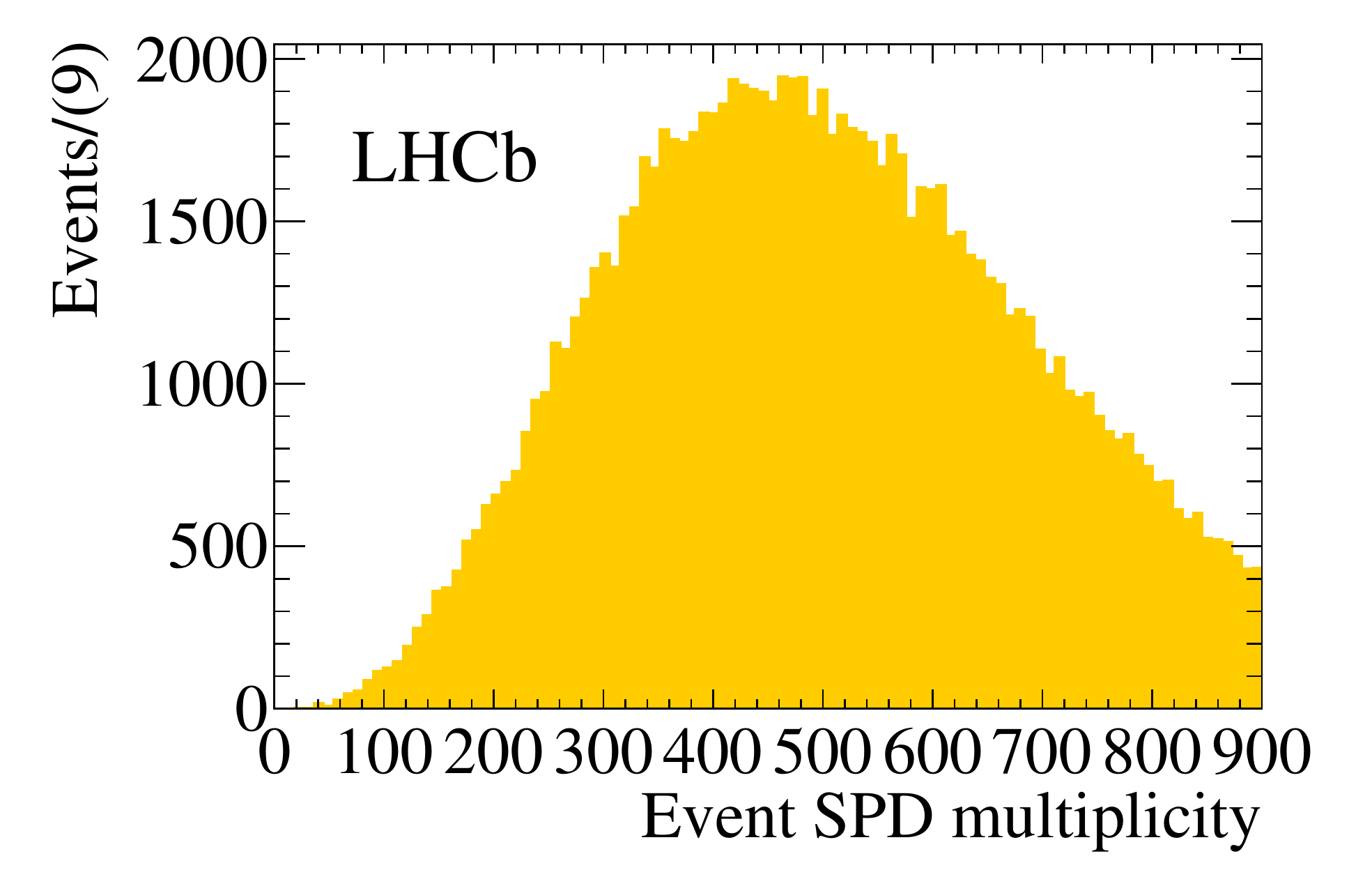}
  \caption{The \spd hit multiplicity of events containing $\Bu\to\Dzb\pip$ candidates in Run~2 data.}
  \label{fig:L0SPD}
\end{figure}

The efficiency of each \lz trigger is measured with respect to events where the corresponding \spd criterion from Table~\ref{tab:LzThresholds} has
already been applied. 
The distribution of \spd hits for $\Bu\to\Dzb\pip$ signal candidates in Run~2 data is shown in Fig.~\ref{fig:L0SPD},
and is representative of typical
heavy-flavour signals in LHCb. The efficiency of the \spd thresholds is  generally around $90\%$ for the \lzdimuon and $50\%$ for
the other heavy-flavour \lz trigger lines.
The advantage of these SPD requirements is that they allow looser \lz kinematic thresholds.

\begin{figure}
  \centering
  \includegraphics[width=0.44\textwidth]{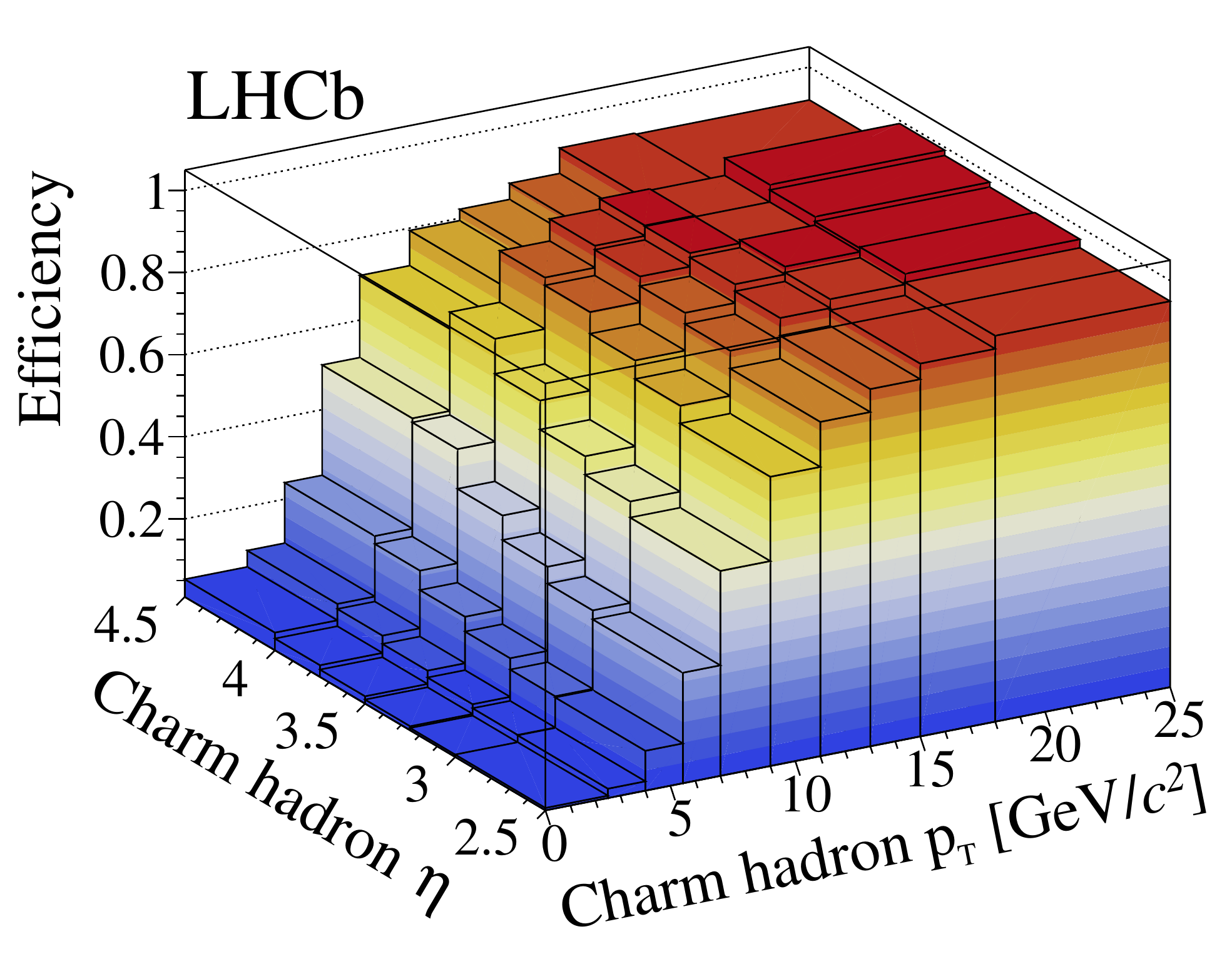}
  \includegraphics[width=0.44\textwidth]{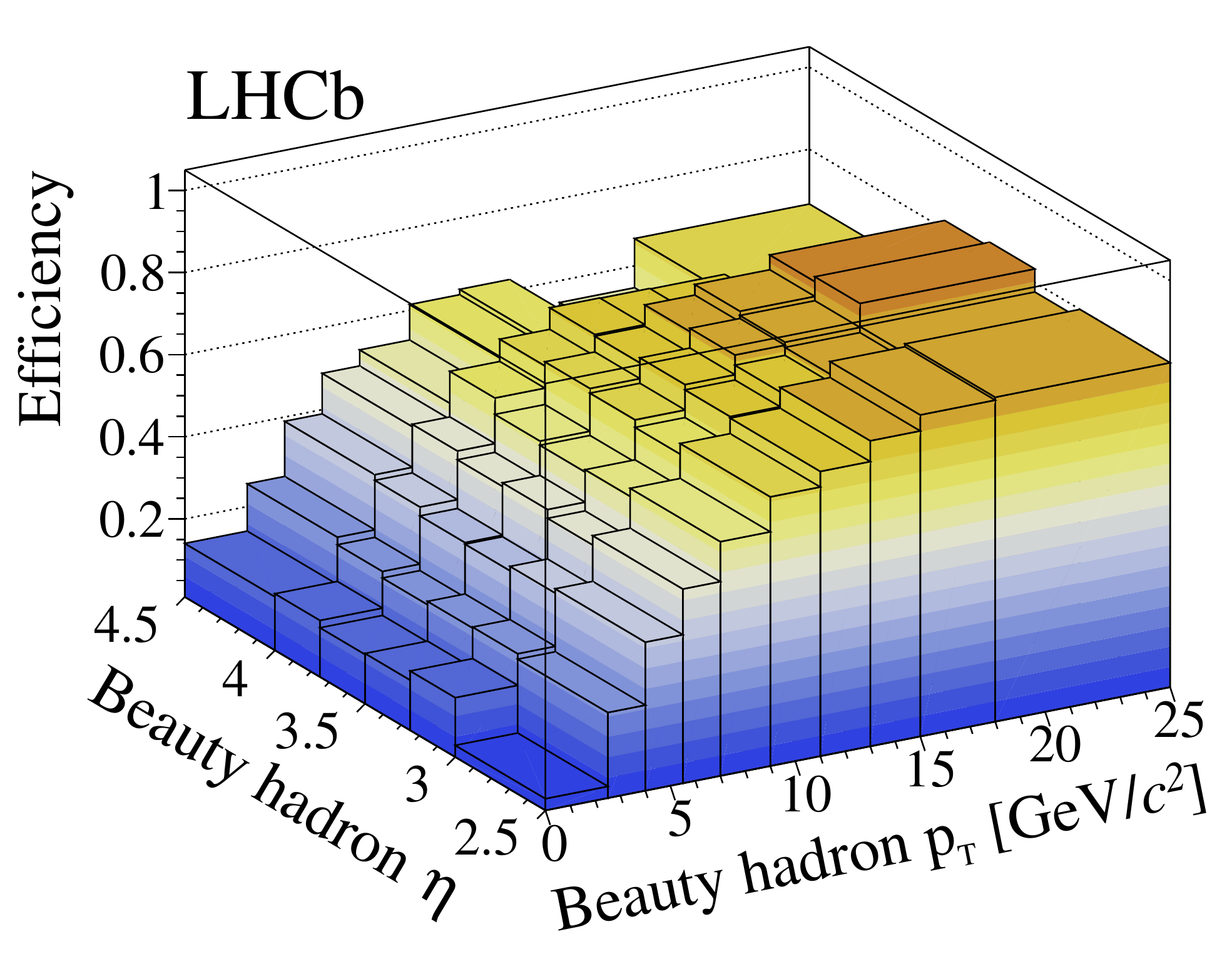}
  \includegraphics[width=0.44\textwidth]{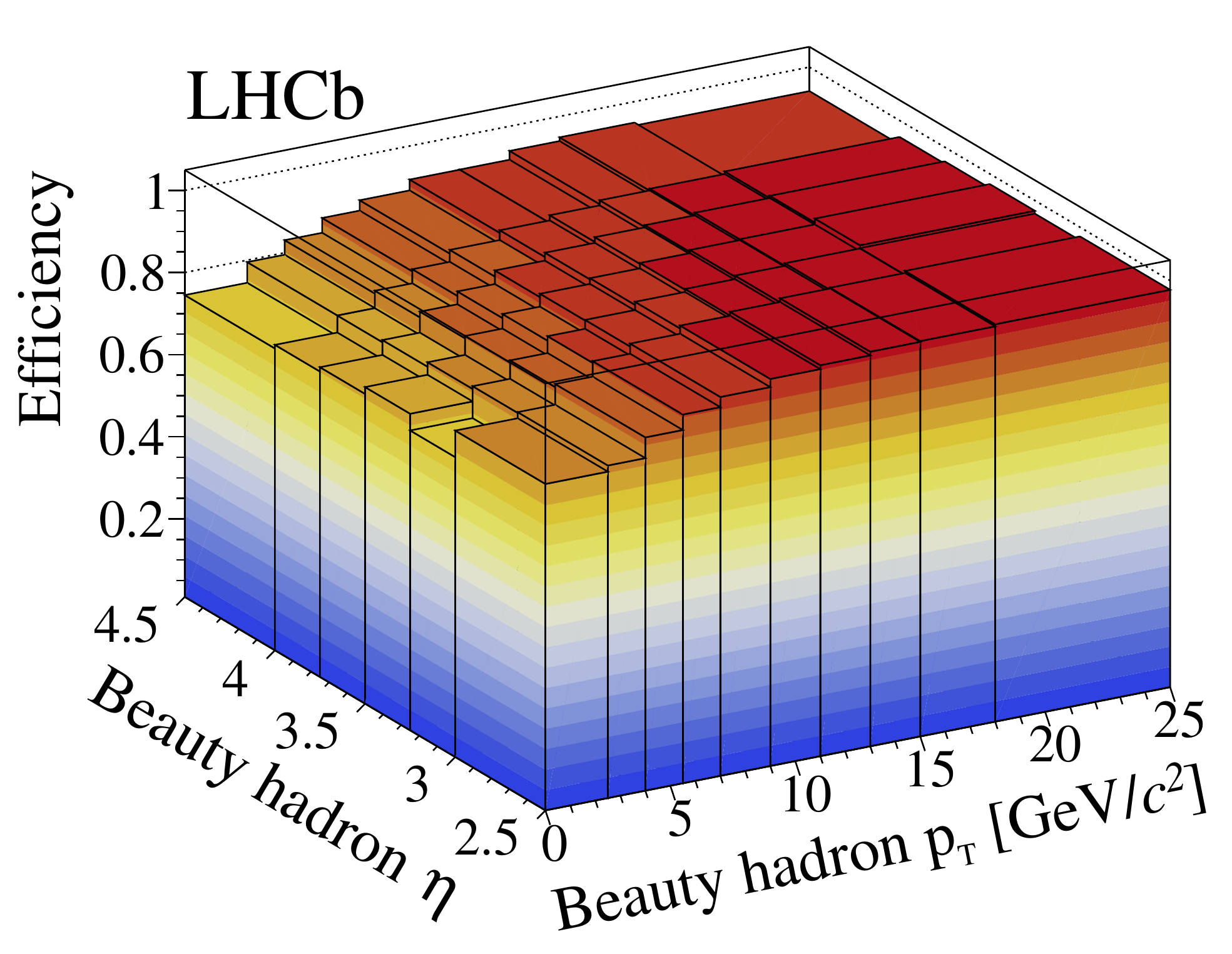}
  \includegraphics[width=0.44\textwidth]{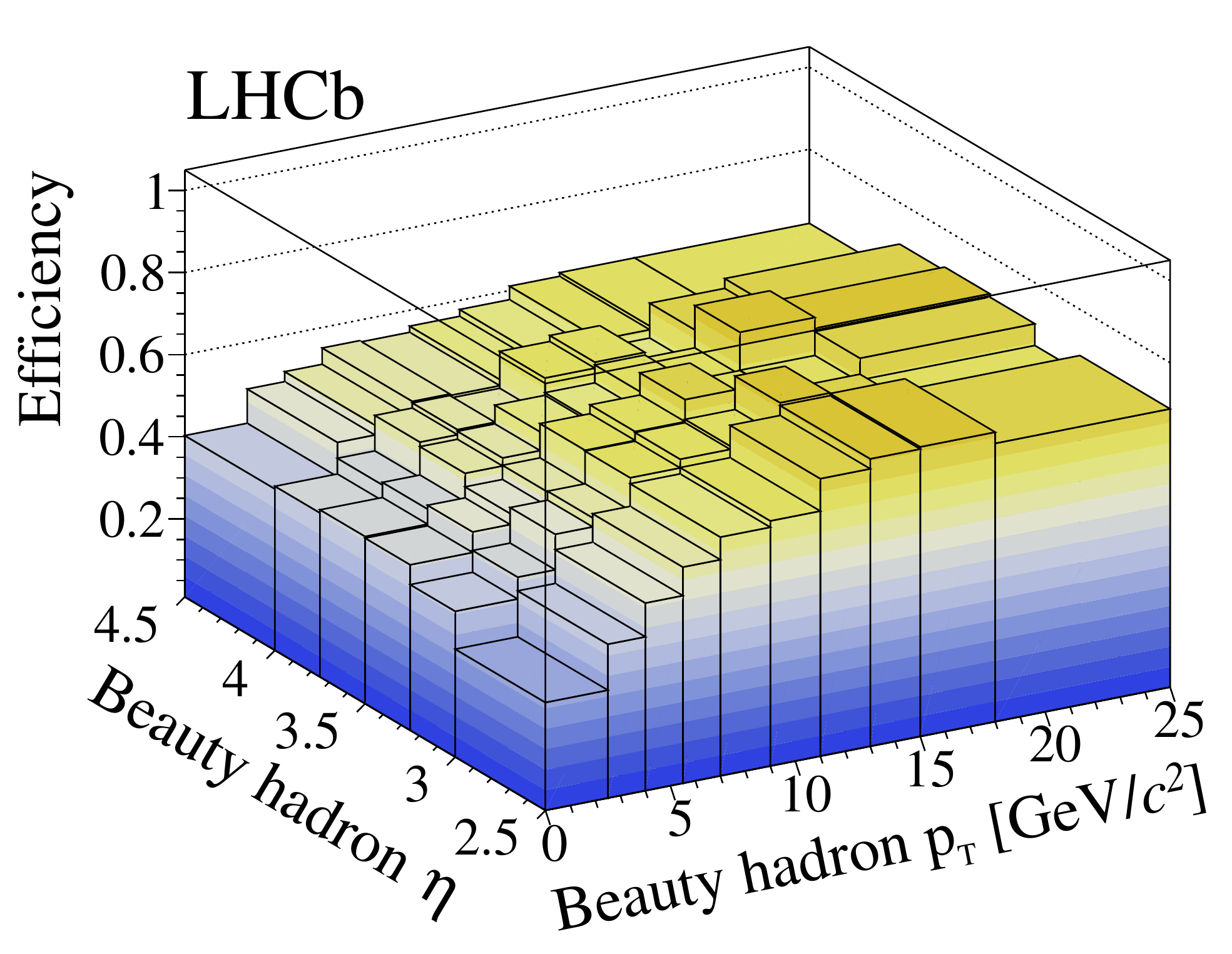}
  \caption{Two-dimensional efficiencies of the \lz trigger lines in Run~2 data: (top left) \lz hadron;
(top right) \lz electron; (bottom left) \lz muon; and
(bottom right) \lz dimuon. The \lz hadron efficiency is evaluated using $\Dz\to\Km\pip$ decays,
whereas the others are evaluated using the relevant signals listed in Fig.~\ref{fig:L0performance_charm} and Fig.~\ref{fig:L0performance_beauty}.}
  \label{fig:L0performance2D}
\end{figure}

The \lz trigger efficiencies as functions of the hadron \pt and $\eta$ are shown in Fig.~\ref{fig:L0performance2D},
except for the photon trigger where the signal yields are too small.
The efficiency is relatively flat in $\eta$ for any given \pt bin, although the calorimeter-based trigger lines do have
a slightly better efficiency at high pseudorapidities.

\subsection{\hltone performance}
The \hltone trigger stage processes approximately 1~MHz of events that pass the \lz trigger, and reduces the event rate to around 110~kHz, which
are further processed by \hlttwo. The \hltone reconstruction sequence was described in Sec.~\ref{sec:hlt1}, while this section describes
the performance of the \hltone trigger lines.

\subsubsection{Inclusive lines}

\hltone has two inclusive trigger lines which select events containing a particle whose decay vertex is displaced from the PV:
a line which selects a single displaced track with high \pt, and a line which selects
a displaced two-track vertex with high \pt. The single track line is a reoptimization of the Run~1 inclusive single track
trigger~\cite{LHCb-PUB-2011-003}, while the displaced two-track vertex trigger is a new development for Run~2.
Both lines start by selecting good quality tracks that are inconsistent with originating from the PV.
The single-track trigger then selects events based on a hyperbolic requirement in the 2D plane of the track displacement
and \pt.\footnote{More complicated multivariate selection criteria, for example boosted decision trees using track quality
information in addition to the displacement and \pt, were tried but gave no significant increase in performance.} The
two-track displaced vertex trigger selects events based on a MatrixNet classifier~\cite{gulin11a} whose input variables are the
vertex-fit quality, the vertex displacement, the scalar sum of the \pt of the two tracks, and the displacement of the
tracks making up the vertex.

\begin{figure}
  \centering
  \includegraphics[width=0.85\textwidth]{figs/L0_Charm_Legend.pdf}\\
  \includegraphics[width=0.40\textwidth]{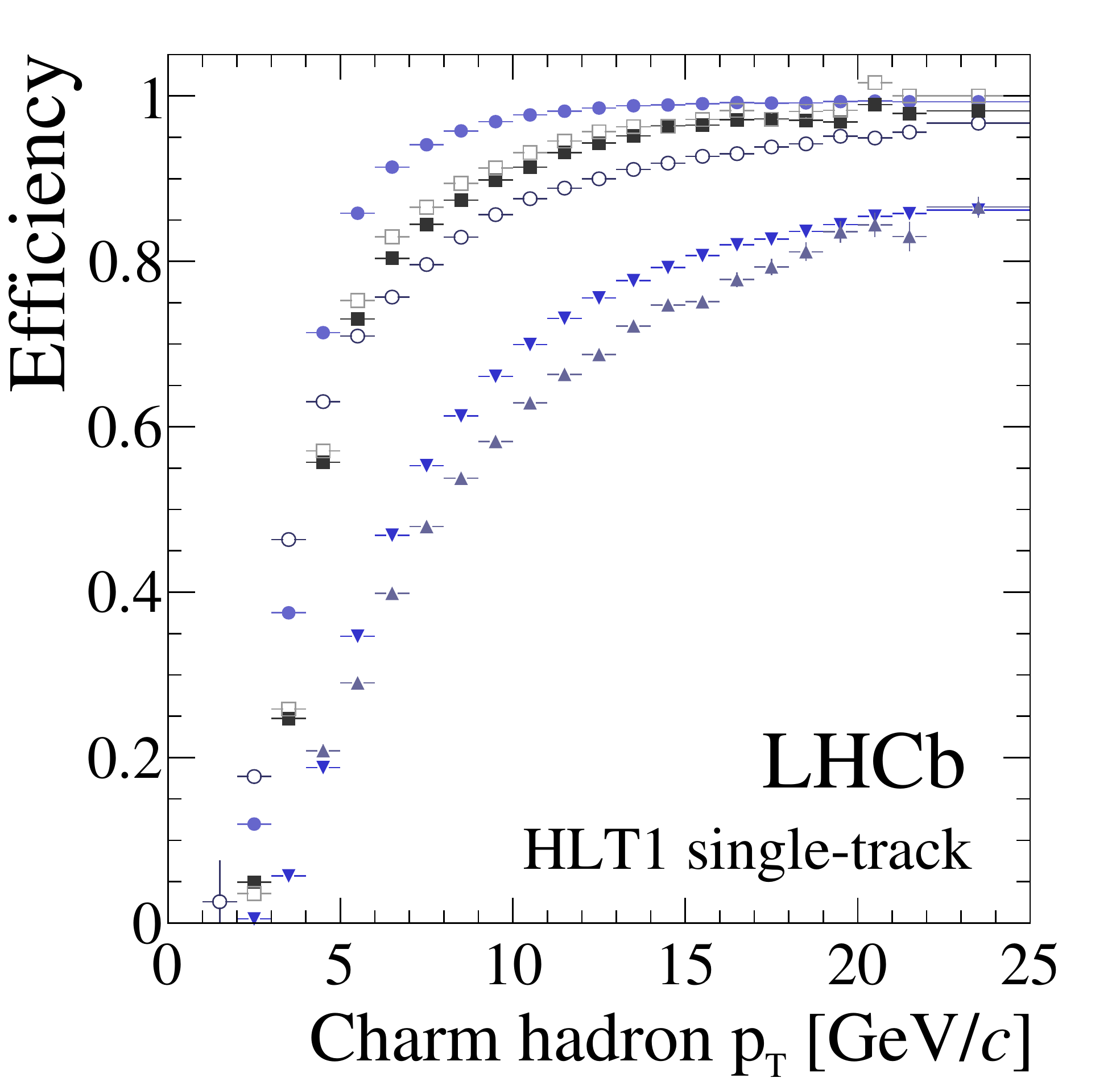}
  \includegraphics[width=0.40\textwidth]{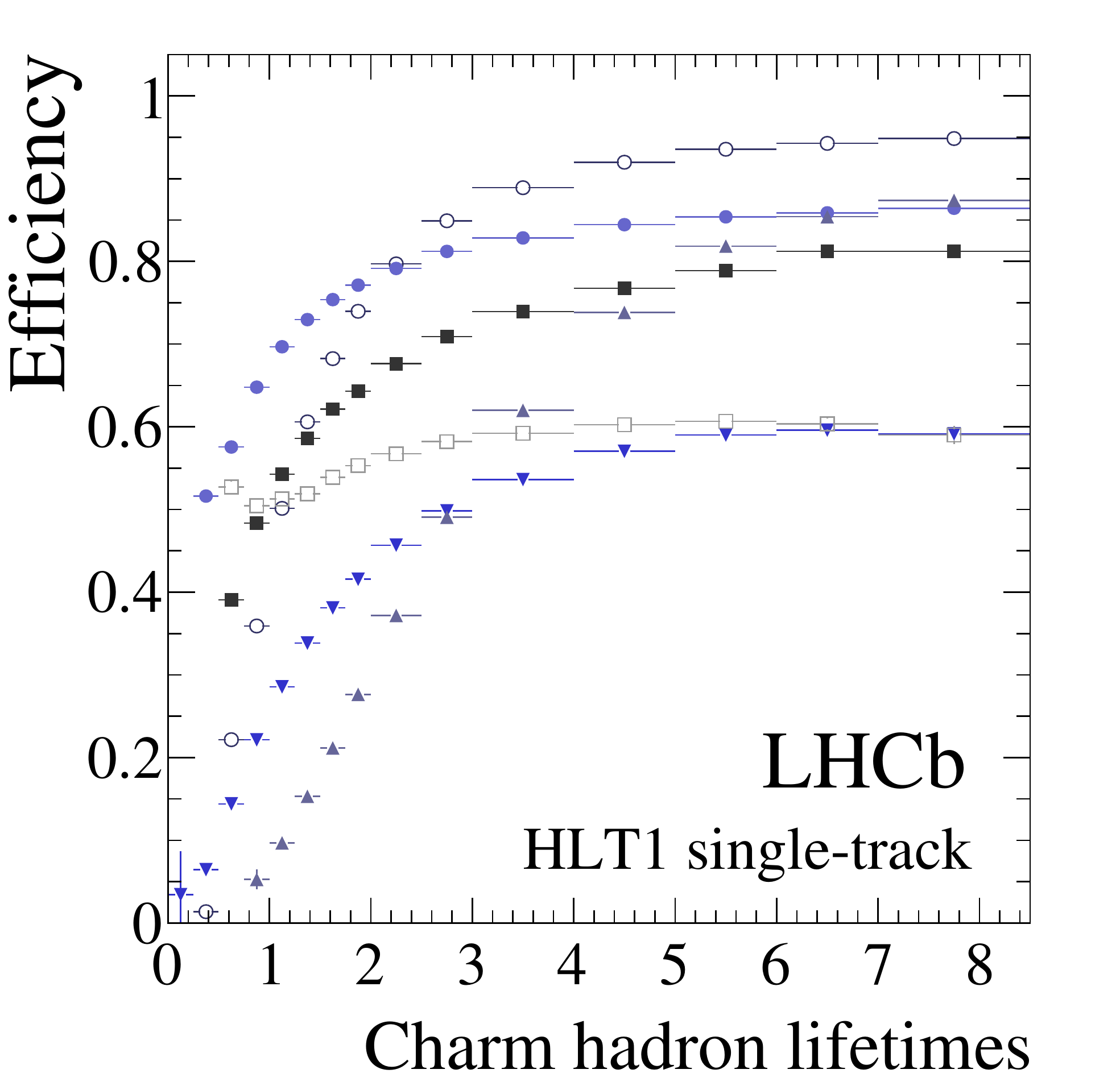}
  \includegraphics[width=0.40\textwidth]{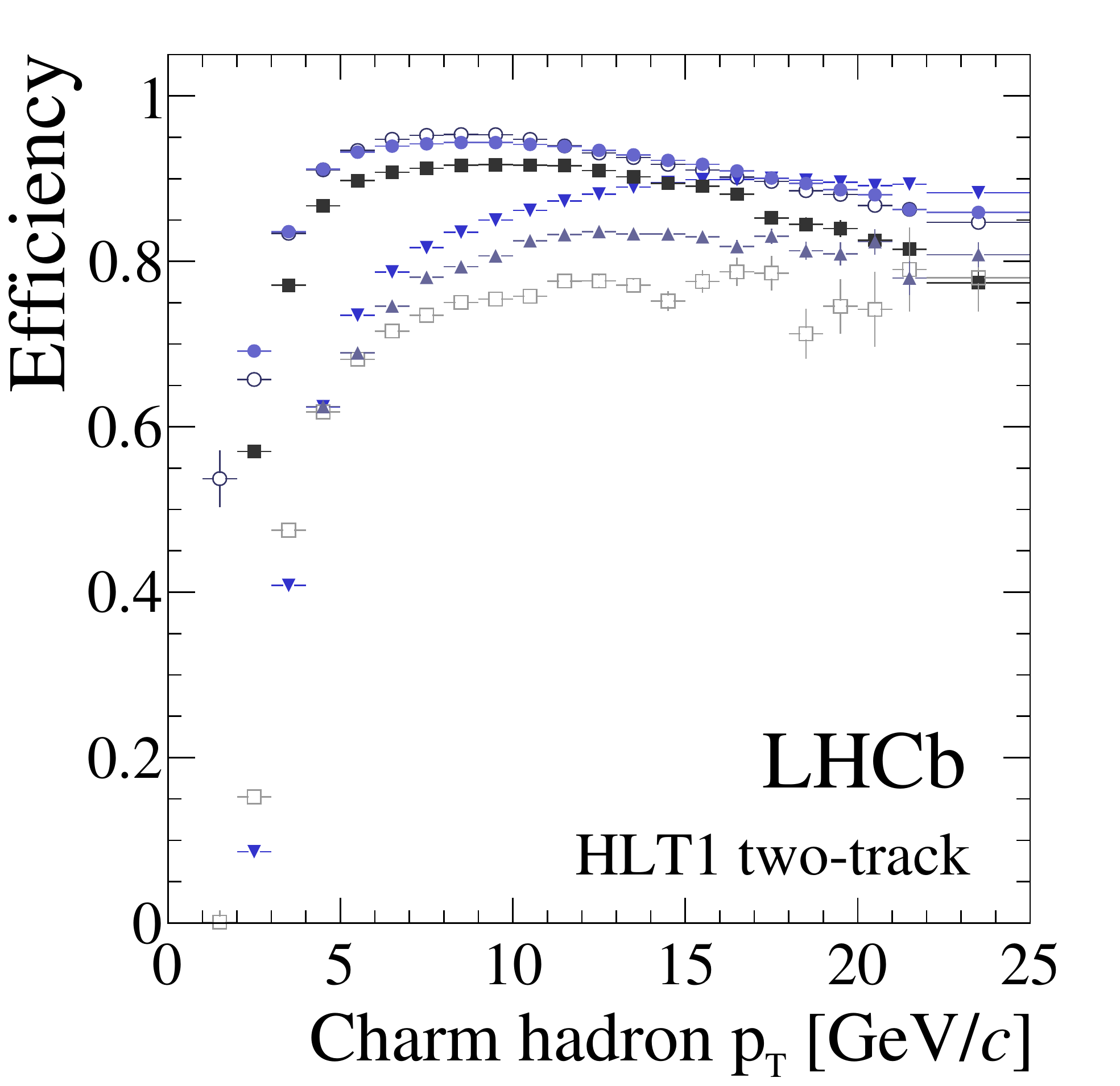}
  \includegraphics[width=0.40\textwidth]{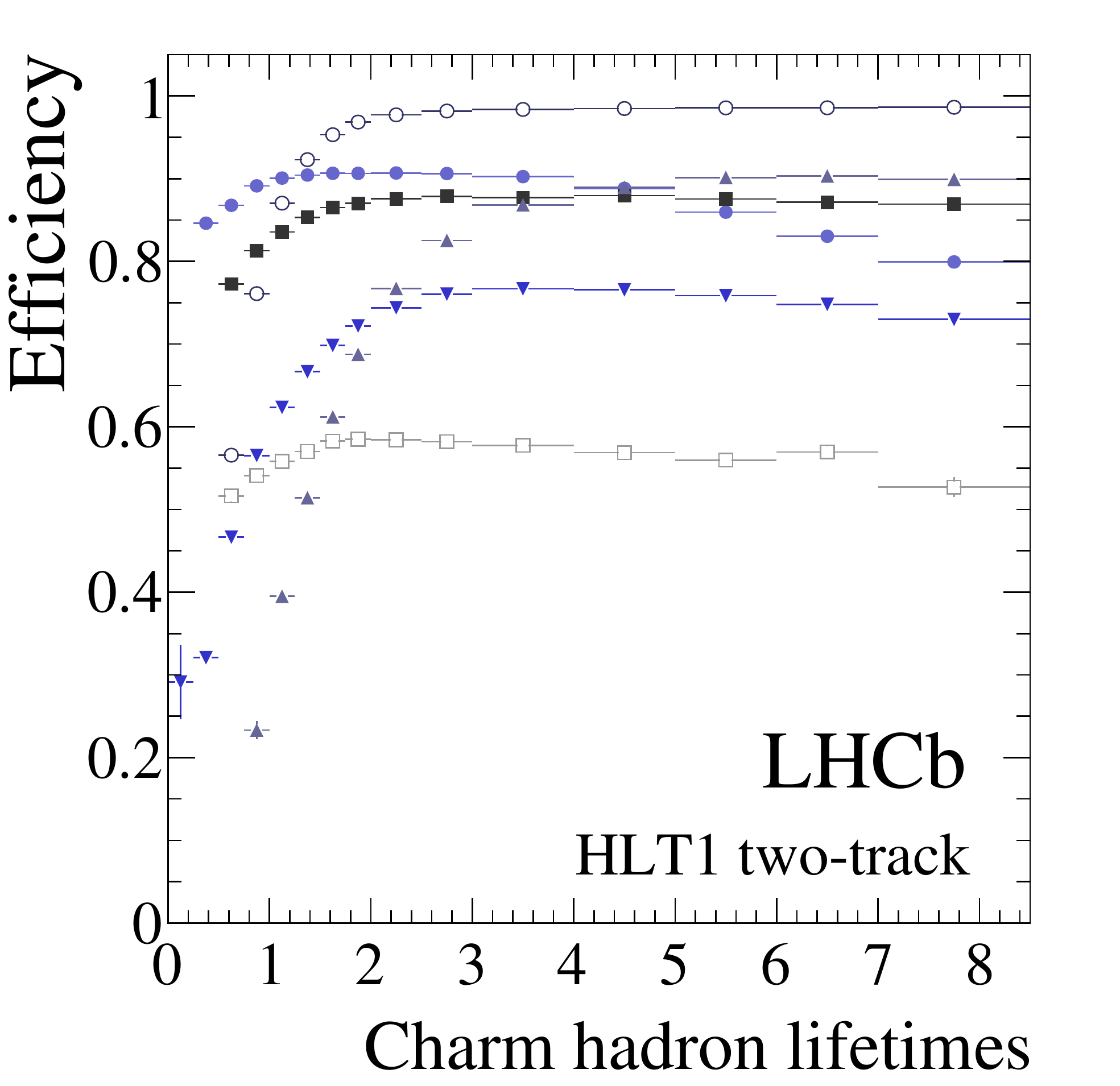}
  \includegraphics[width=0.40\textwidth]{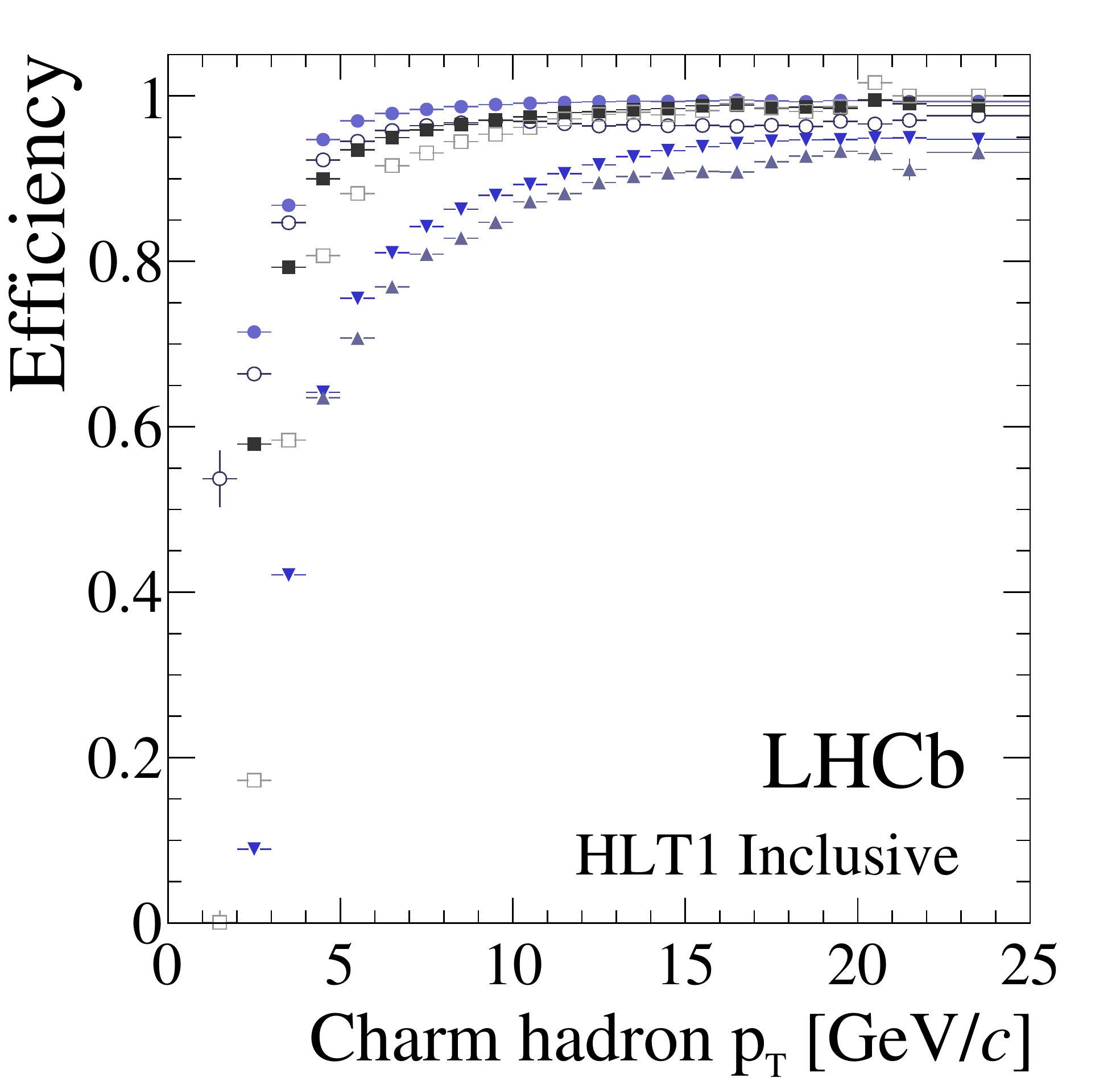}
  \includegraphics[width=0.40\textwidth]{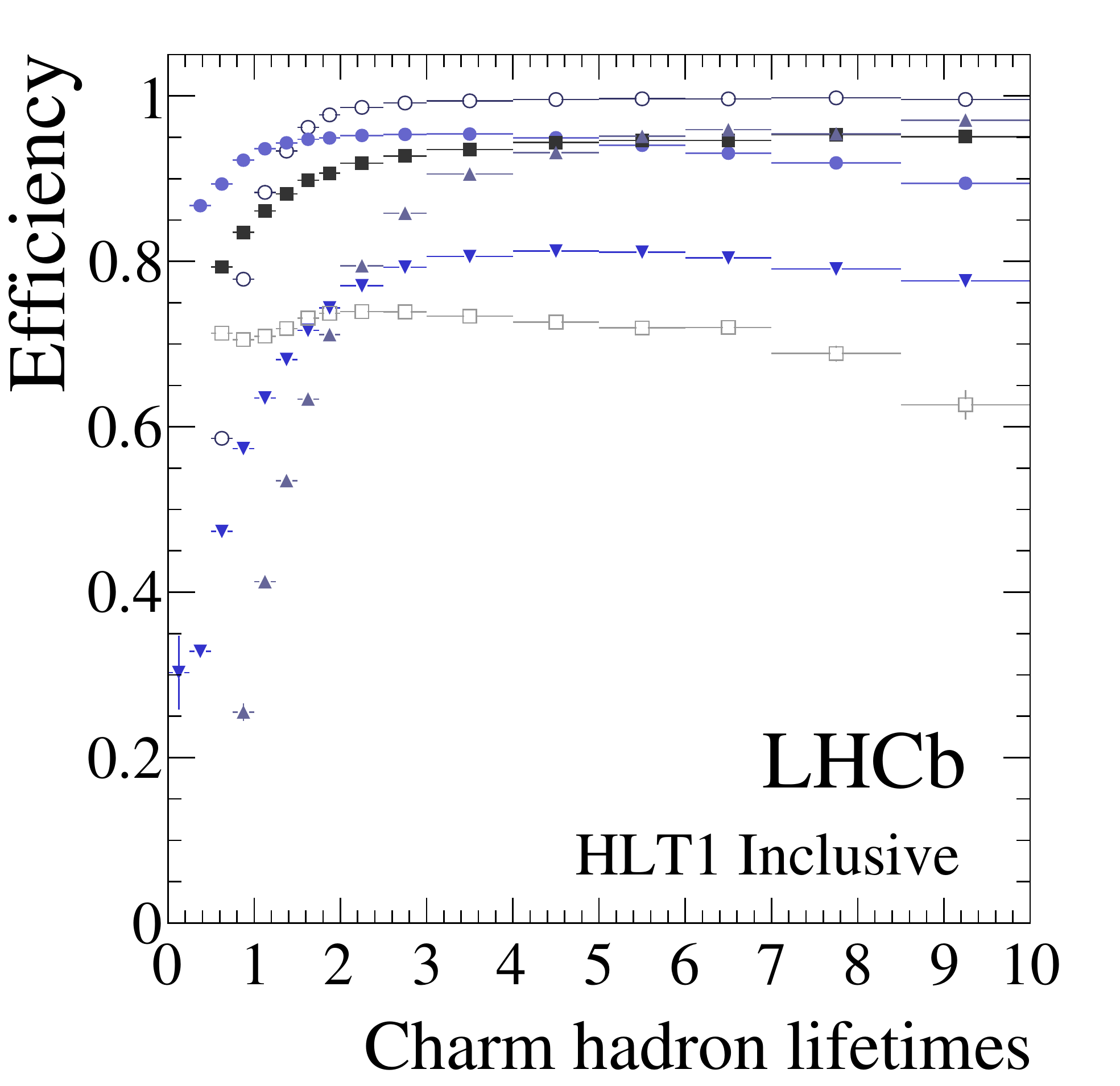}
	\caption{\small{Efficiency of the \hltone inclusive trigger lines as a function of (left) \Pc-hadron \pt and (right) decay time. The decay time plots
are drawn such that the x-axis is binned in units of the lifetime for each hadron in its rest frame. The plots in each column show,
	from top to bottom, the single-track, two-track, and combined \hltone inclusive performance.}}
  \label{fig:hlt1_track_tistos_charm}
\end{figure}

\begin{figure}
  \centering
  \includegraphics[width=0.85\textwidth]{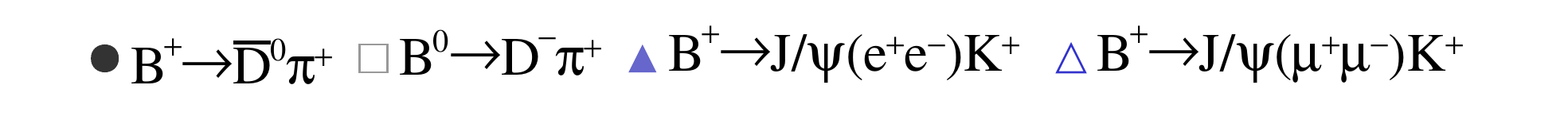}
  \includegraphics[width=0.40\textwidth]{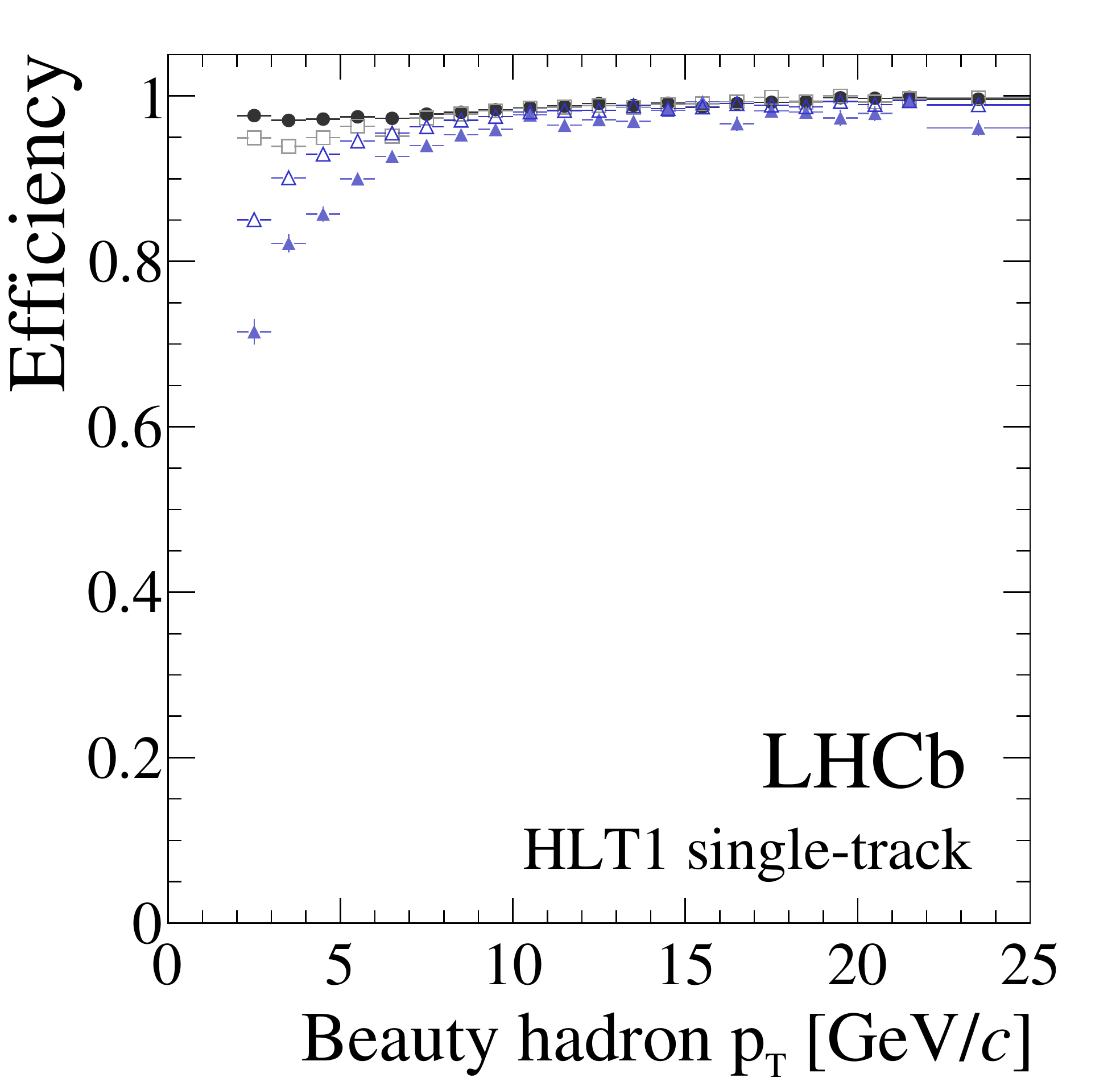}
  \includegraphics[width=0.40\textwidth]{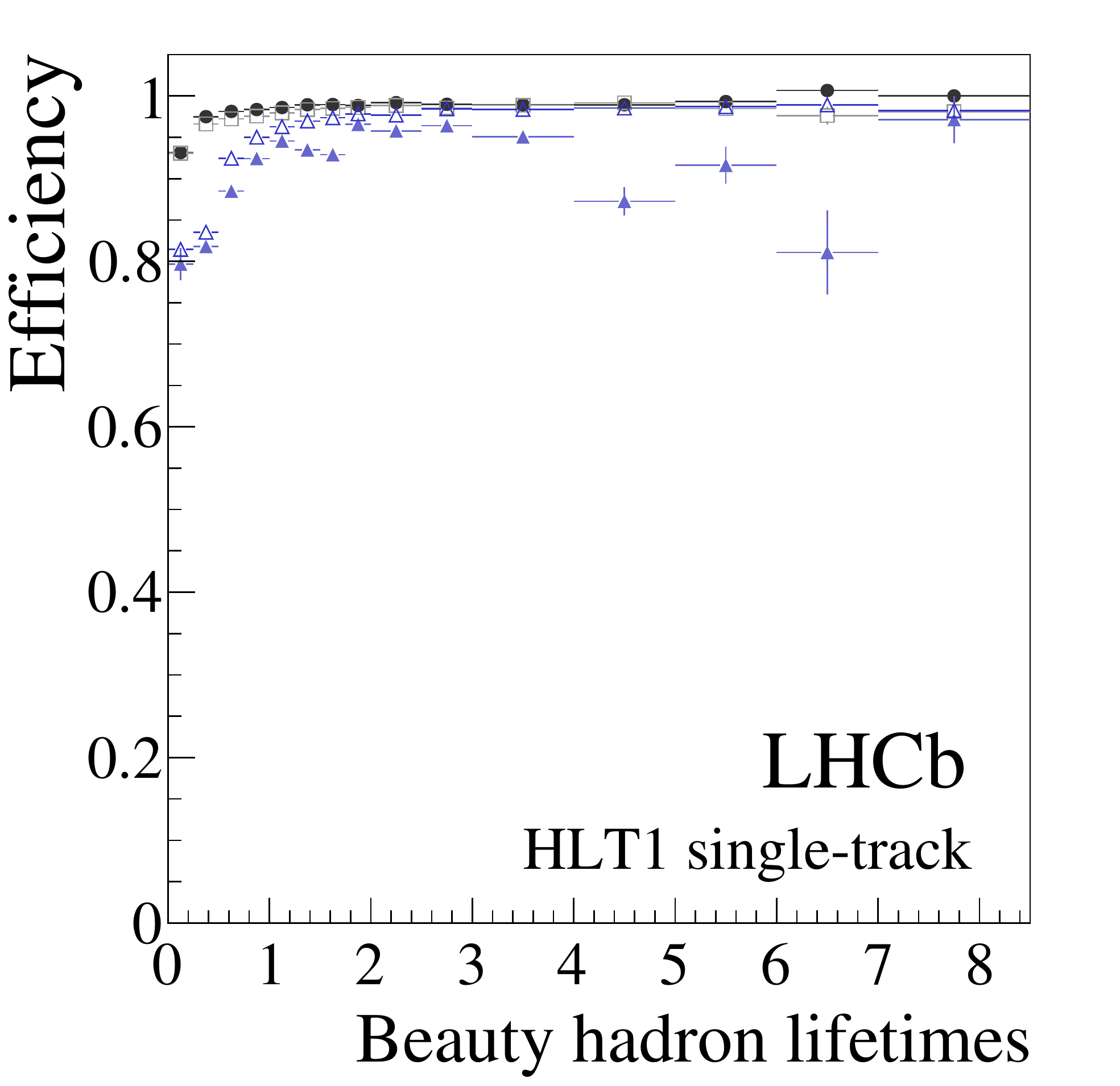}
  \includegraphics[width=0.40\textwidth]{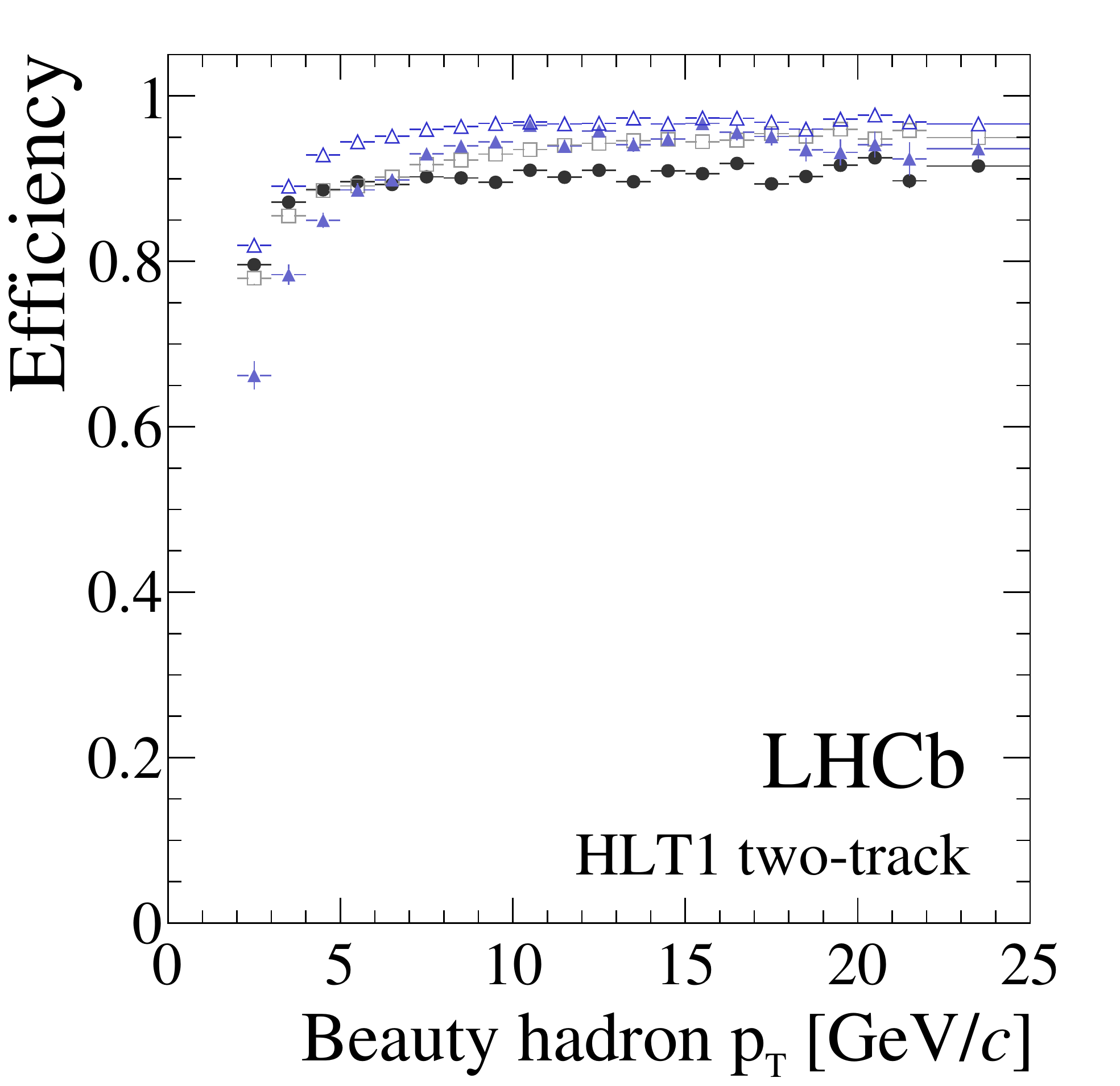}
  \includegraphics[width=0.40\textwidth]{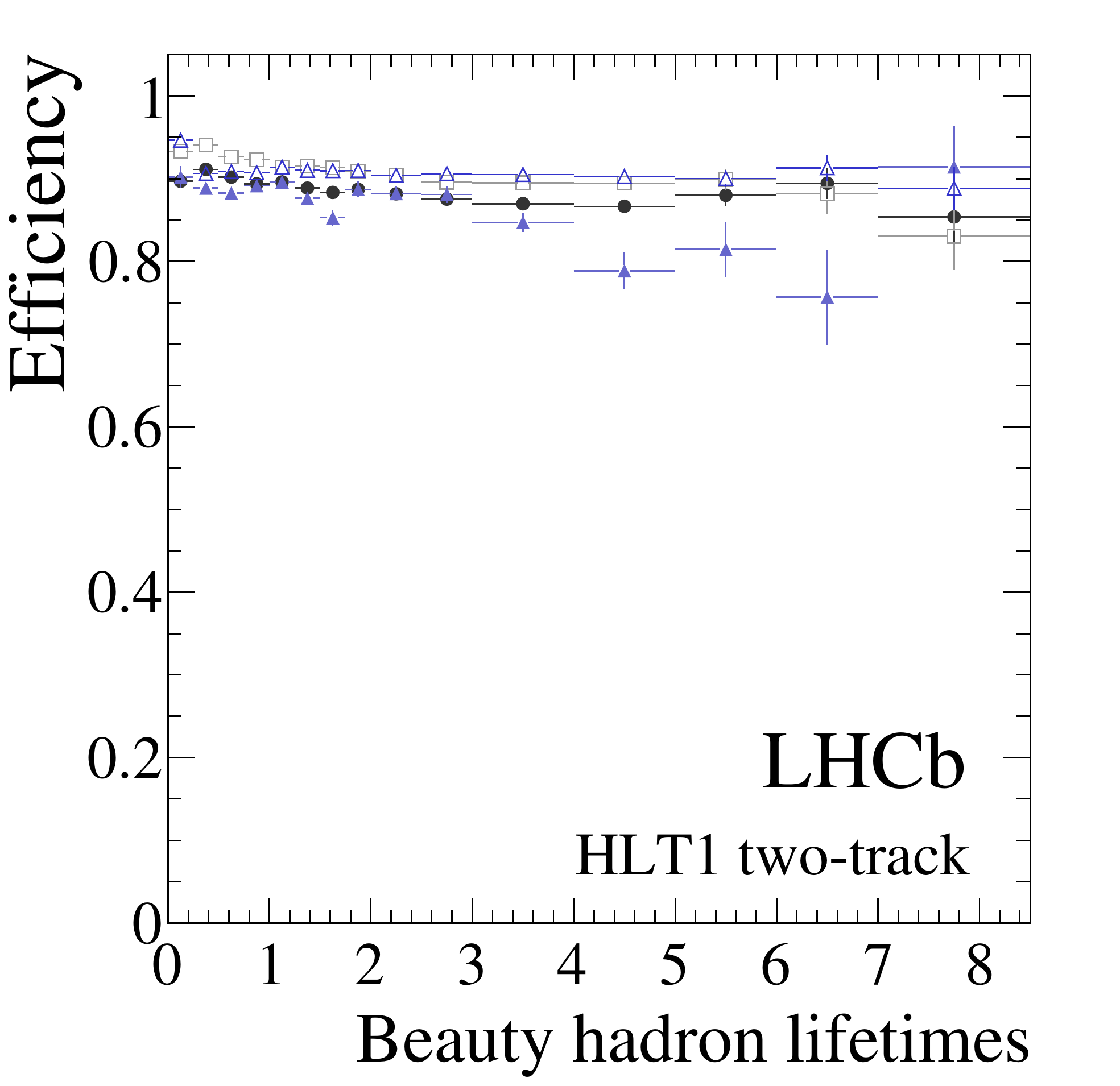}
  \includegraphics[width=0.40\textwidth]{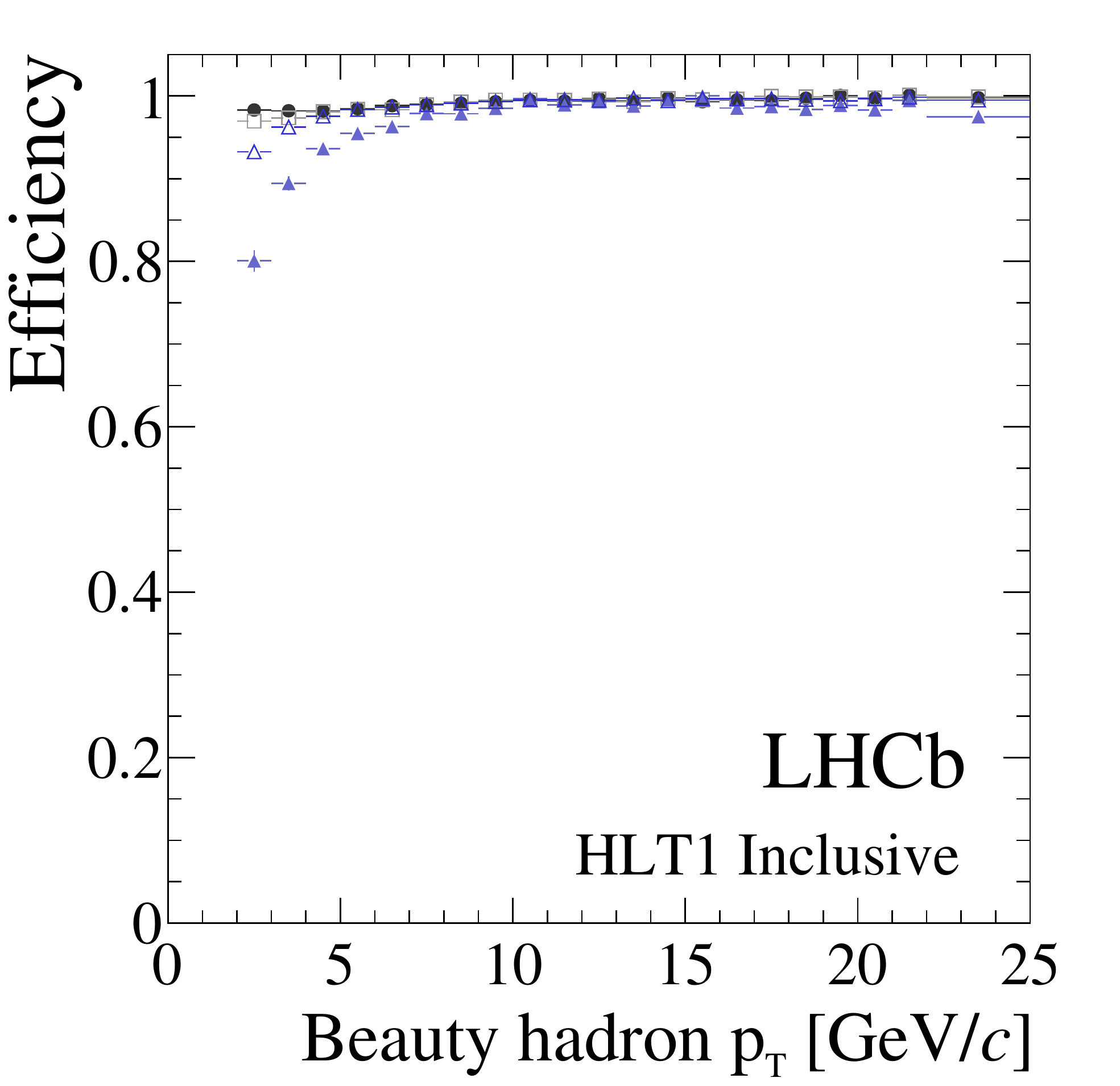}
  \includegraphics[width=0.40\textwidth]{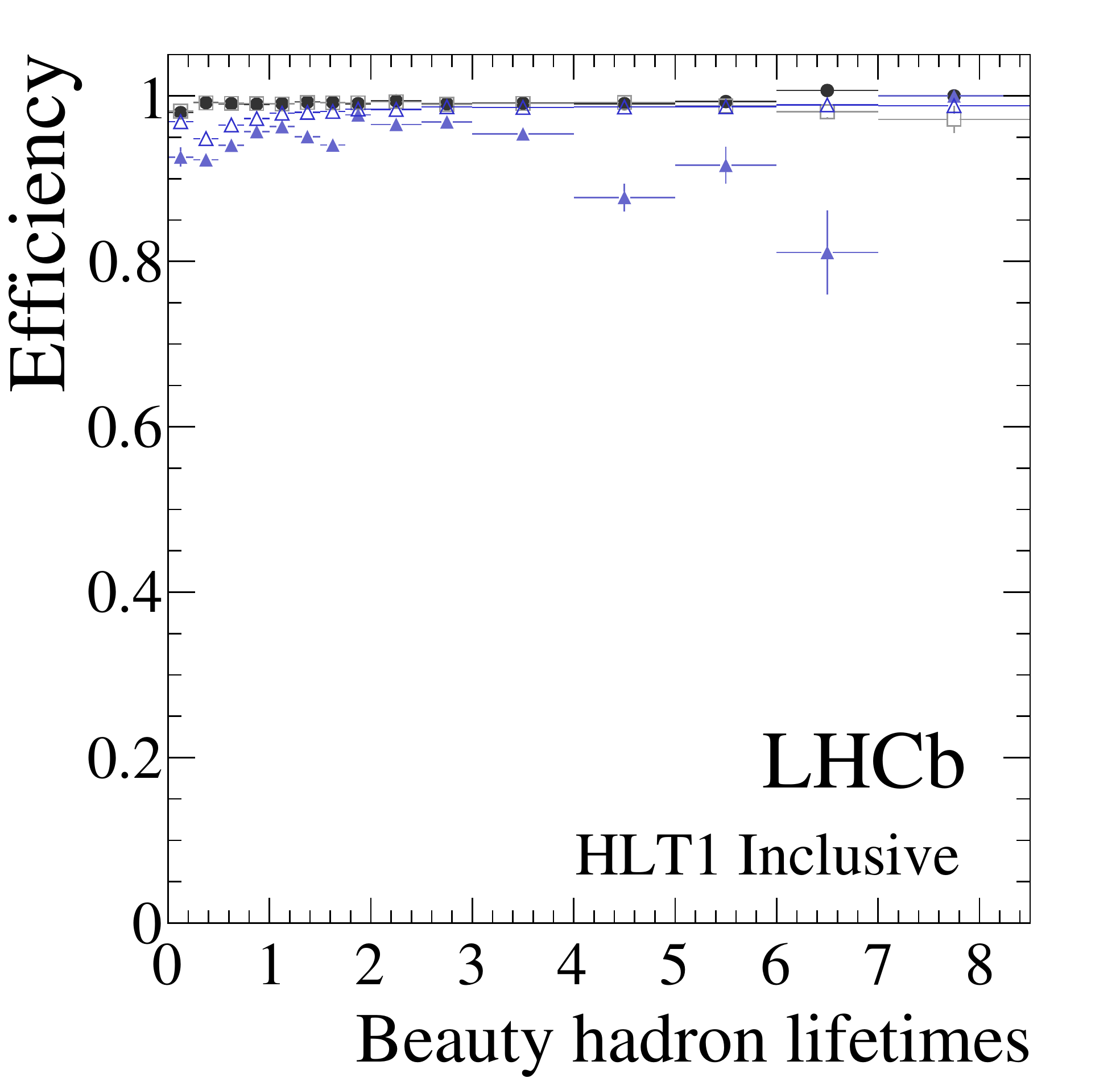}
	\caption{\small{Efficiency of the \hltone inclusive trigger lines as a function of (left) \Pb-hadron \pt and (right) decay time. The decay time plots
are drawn such that the x-axis is binned in units of the lifetime for each hadron in its rest frame. The plots in each column show,
	from top to bottom, the single-track, two-track, and combined \hltone inclusive performance.}}
  \label{fig:hlt1_track_tistos_beauty}
\end{figure}

These trigger lines were primarily optimized for inclusively selecting the decays of \Pb~and~\Pc~hadrons, and were trained using 26 different
\Pb-~and~\Pc-hadron decays in order to make them as efficient as possible on the full spectrum of possible decay topologies. Care was
taken, however, to make sure that these trigger lines would also be efficient for more exotic displaced signatures, for example hypothetical
supersymmetric particles.
The performance of these trigger lines is shown in Figs.~\ref{fig:hlt1_track_tistos_charm}~and~\ref{fig:hlt1_track_tistos_beauty}.
The two-track line is more efficienct at low \pt, whereas the single track line performs best at high \pt,
such that combined they provide high efficiency over the full \pt range.

\begin{figure}
  \centering
  \includegraphics[width=0.44\textwidth]{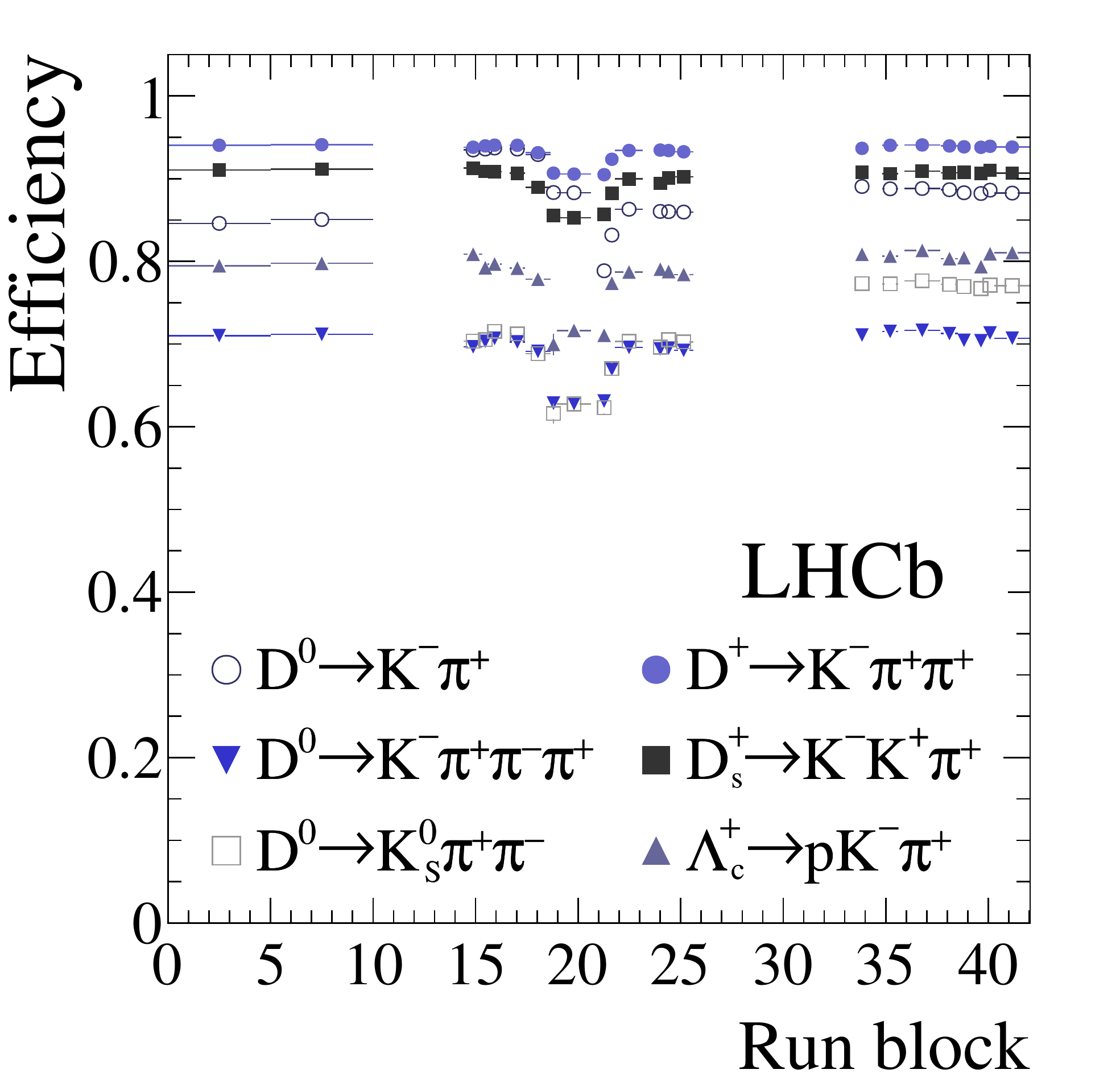}
  \includegraphics[width=0.44\textwidth]{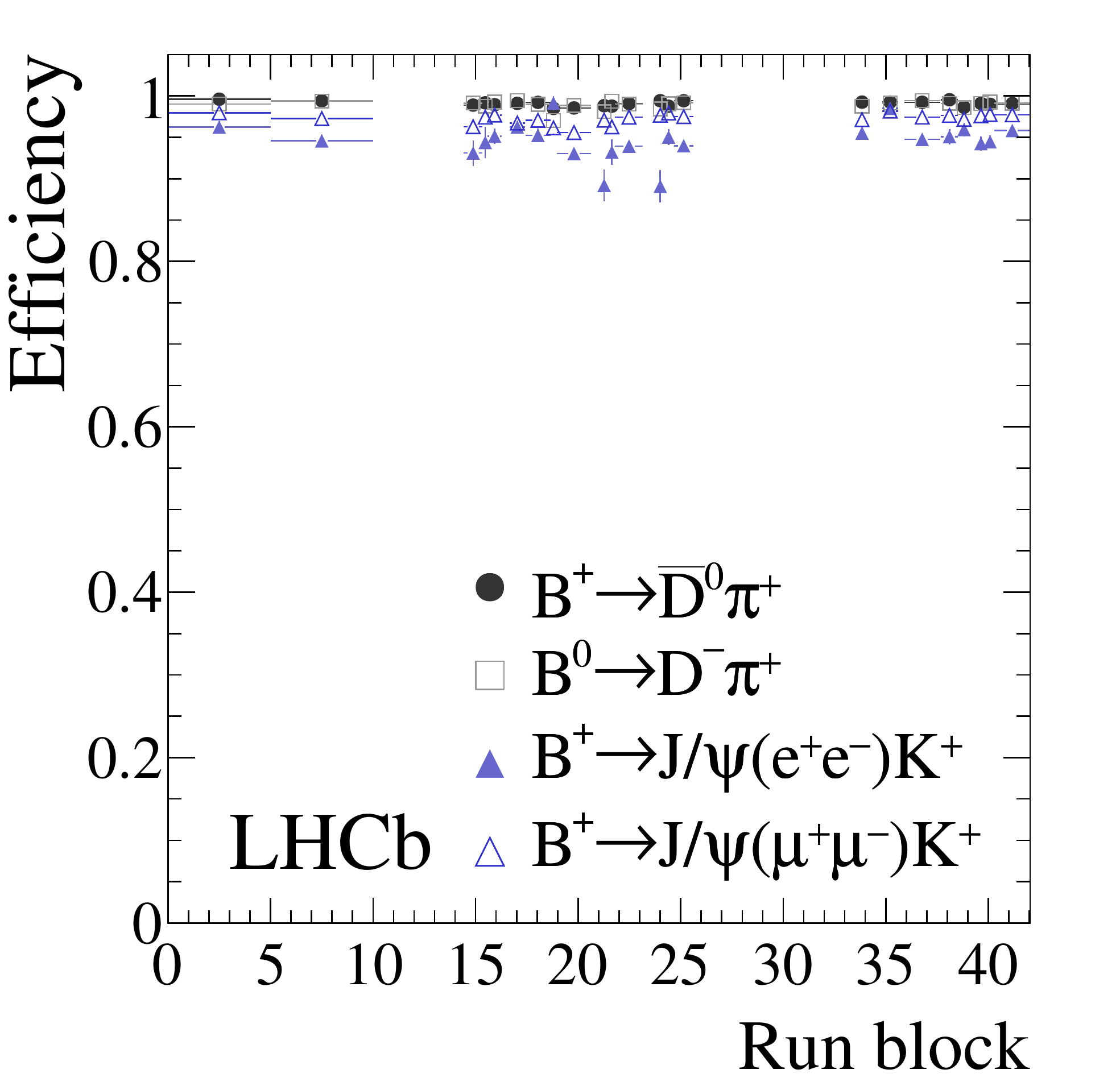}
  \caption{The \hltone efficiency as a function of the different trigger
configurations used during data taking for (left) \Pc-hadrons and (right) \Pb-hadrons.
The three blocks visible in the plot, separated by vertical gaps, correspond to the three years of data taking (2015--2017).}
  \label{fig:hlt1_totaleff_tistos_run}
\end{figure}

Unlike the \lz trigger configurations, which changed frequently in response to varying LHC conditions, the \hltone trigger configuration
was kept largely stable, with some updates at the end of each data-taking year.
The variation in the total \hltone efficiency as a function of the data-taking
period is shown in Fig.~\ref{fig:hlt1_totaleff_tistos_run}. The \Pb-hadron efficiencies have been stable throughout the Run~2 data taking.
The \Pc-hadron efficiencies decreased midway through 2016, when a tighter \hltone configuration was used to prevent the disk buffer from
overflowing due to unexpectedly high LHC efficiency and availability. The improvements seen for
some of the \Pc-hadron channels in 2017 with respect to 2016 are caused by changes in the corresponding reference
off\-line selections leading to a different average \pt
and displacement of the \Pc-hadron, not any intrinsic variation in the \hltone performance or thresholds.

\subsubsection{Muon lines}
\label{sec:hlt1_muon}

\begin{figure}[t]
  \centering
  \includegraphics[width=0.85\textwidth]{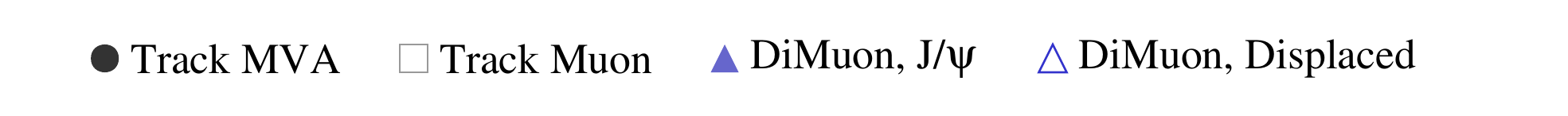}
  \includegraphics[width=0.44\textwidth]{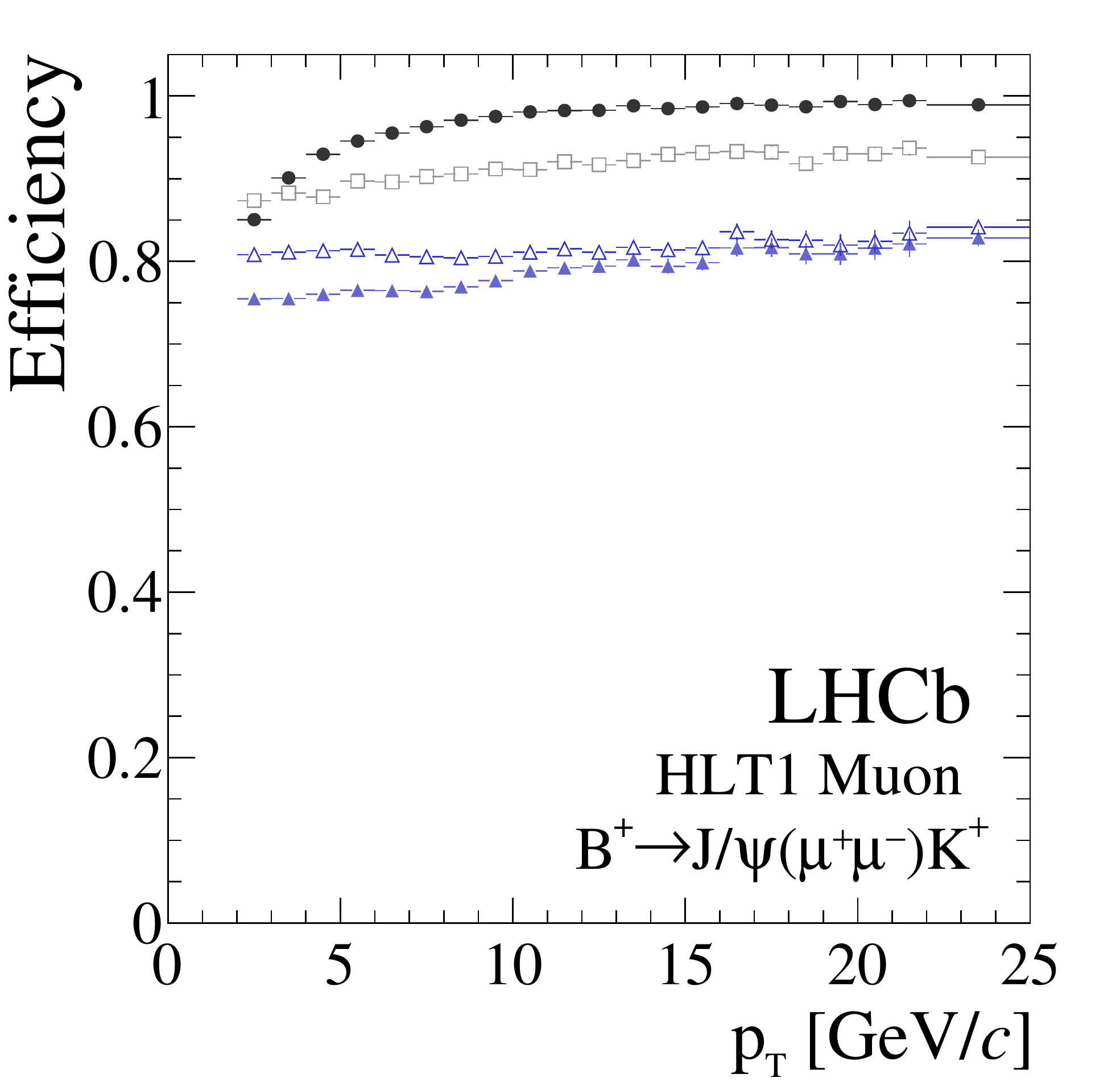}
  \includegraphics[width=0.44\textwidth]{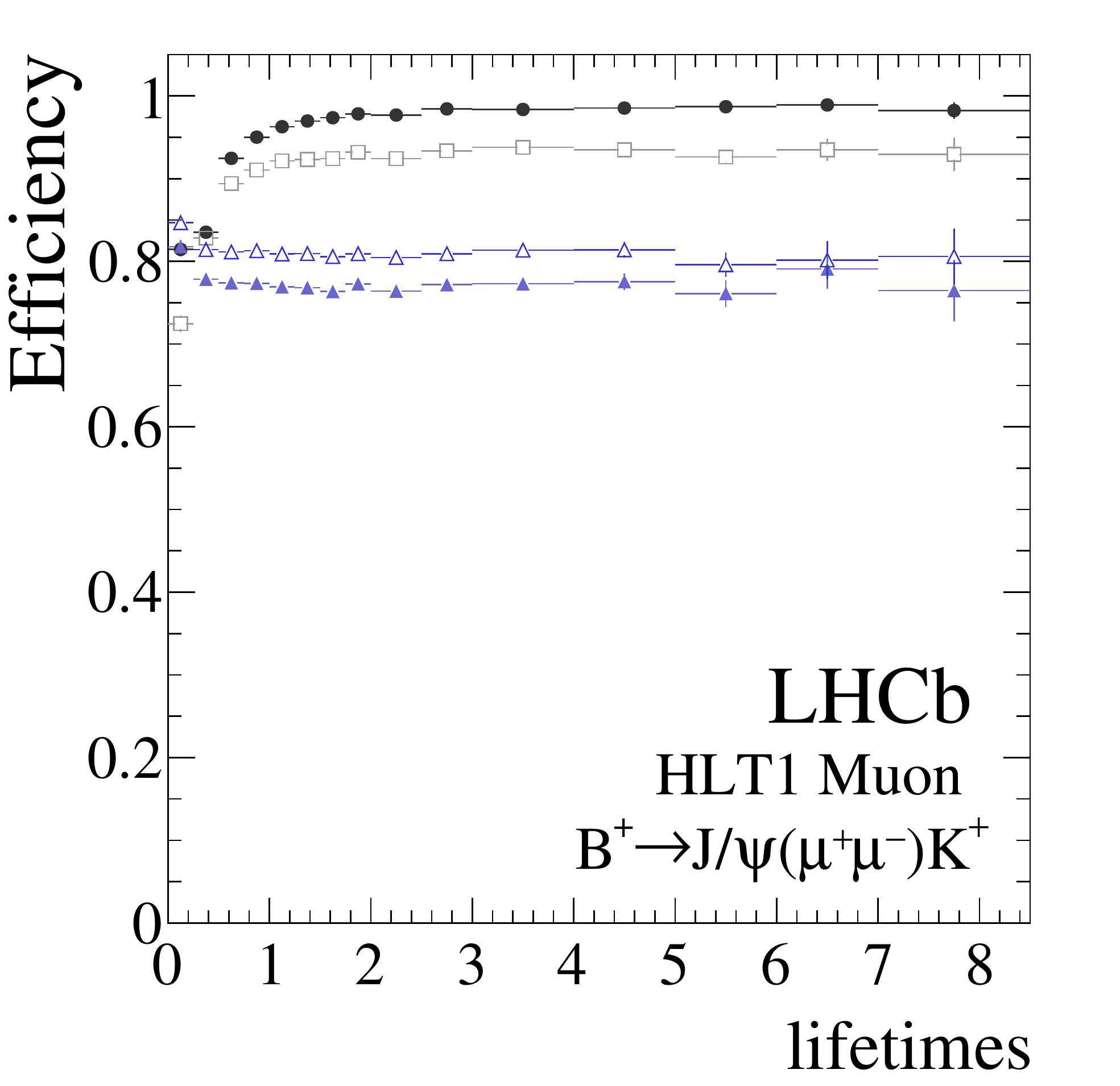}
  \caption{The efficiency of the \hltone muon trigger lines as a function of the (left) \Pb-hadron \pt and (right) units of the average $\Bp$ decay time. The decay time plot is drawn such that the x-axis is binned in units of the $\Bp$ lifetime in its rest frame. The
efficiency of the inclusive single-track \hltone trigger is plotted for reference.}
  \label{fig:hlt1_muon_tistos}
\end{figure}

The \hltone muon lines select muonic decays of \Pb~and~\Pc~hadrons, as well muons originating from decays of \W and \Z bosons.
As muons have an intrinsically cleaner signature than hadrons, the muon lines make use of simple rectangular
selection criteria as opposed to the multivariate inclusive lines.
There are four primary lines: one line that selects a single displaced muon with high \pt for flavour physics;
a second single muon line that selects very high \pt muons, without displacement criteria, for electroweak physics;
a third line that selects a dimuon pair compatible with originating from the decay of a charmonium or bottomonium resonance, or from Drell-Yan production; and a fourth line
that selects displaced dimuons with no requirement on the dimuon mass.
The efficiencies of the lines relevant for \Pb-hadron decays are shown in Fig.~\ref{fig:hlt1_muon_tistos}
as obtained from data with the TISTOS method.
Note that because these \hltone muon trigger lines only run on events selected by \lzmuon and \lzdimuon trigger lines, their absolute efficiency is
lower than that of the inclusive single-track \hltone trigger, which runs on all \lz-selected events.
In addition to these lines, for Run 2 a new line dedicated to lower-\pt dimuons
has been developed which has tighter criteria on the displacement of the dimuon but runs on all \lz-selected events,
rather than just the muon ones. This line is particularly important for selecting rare decays of strange hadrons, that are not triggered by the \lz muon lines,
increasing their \hltone efficiency up to a factor three~\cite{Dettori:2297352}.

\subsubsection{Calibration trigger lines}
\label{sec:hlt1_calib}

\begin{figure}
  \centering
  \includegraphics[width=0.99\textwidth]{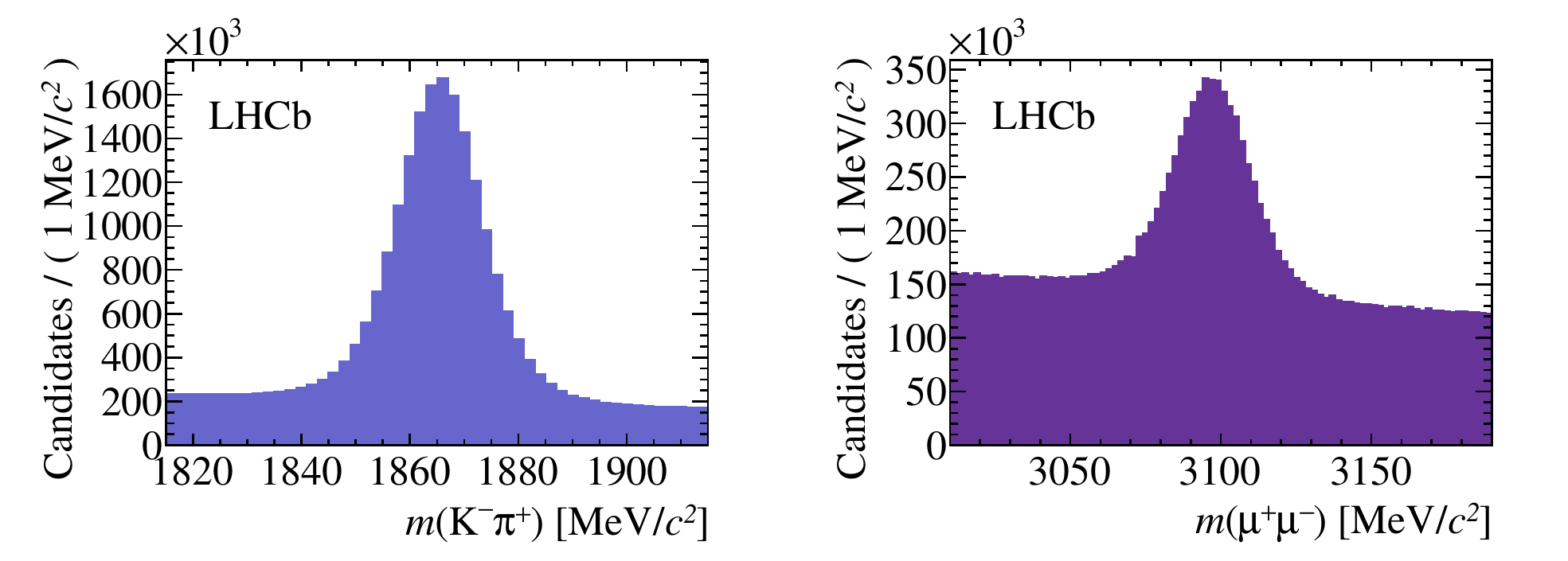}
  \caption{The $D^0$ (left) and $J/\psi$ (right) candidates selected by the \hltone calibration lines. Both plots show candidates reconstructed online.} 
  \label{fig:hlt1_calibtrigs}
\end{figure}

\hltone contains two primary types of calibration trigger lines: a line which selects $D^0 \to K^- \pi^+$ candidates with significant displacement
from the PV, and a line which selects $J/\psi \to \mu^+ \mu^-$ candidates. The former is used for providing a pure sample of
good tracks (the $D^0$ decay products) for the alignment of the tracking system, while the latter is used to provide
a pure sample of muons for the alignment
of the muon chambers. In addition, other trigger lines select events enriched in off-axis \velo tracks or tracks which populate the lower-occupancy
regions of the \rich detectors, for use in the \velo and \rich alignment, respectively. The purity and yield of the calibration trigger lines is illustrated
in Fig.~\ref{fig:hlt1_calibtrigs}, which shows the $D^0$ and $J/\psi$ candidates reconstructed online in their respective lines for a specific fill
corresponding to approximately 18.5~pb$^{-1}$ of integrated luminosity.

\subsubsection{Low multiplicity event and exclusive trigger lines}
Special trigger lines for low-multiplicity events are needed to enable the study of central exclusive production (CEP). This kind of process takes place by
colourless, low-\pt $t$-channel exchange between protons and can result in particle production in the central rapidity region. The protons remain intact
and are deflected only slightly, so such production is typically accompanied by large ranges of rapidity with little detector activity, known as ``rapidity gaps''.
The trigger development initially focussed  on acquiring large samples of exclusively-produced dimuon candidates, but evolved to cover final states
involving hadrons and calorimeter objects.

Since low levels of activity are anticipated for CEP, events with more than 30 hits in the \spd are rejected at the
hardware level. Lower-bounds are also placed on relevant detector activity measurements as appropriate for each final state. 
These criteria indirectly favour the selection of events with exactly one $pp$ interaction, as opposed to either multiple- or zero-interaction events.
At the \hltone stage, the low-multiplicity events containing muons or electromagnetic calorimeter objects occur at a low enough rate
that can be selected with no additional requirements. 
but low-multiplicity events containing hadrons are required to have at least two tracks reconstructed in the \velo.

In addition, the low \pt thresholds implemented in the Run 2 \hltone tracking allowed several special exclusive \hltone trigger selections to
be implemented for the first time, notably trigger lines that select two-body beauty and charm hadron decays without biasing their decay times~\cite{Kenzie:2110638}.
In 2018 the HeRSCheL detector~\cite{LHCb-DP-2016-003} is employed in the \lz selection of CEP events,
allowing for a reduction of the \pt thresholds.

\subsubsection{\hltone bandwidth division}
\begin{figure}
  \centering
  \includegraphics[width=0.95\textwidth]{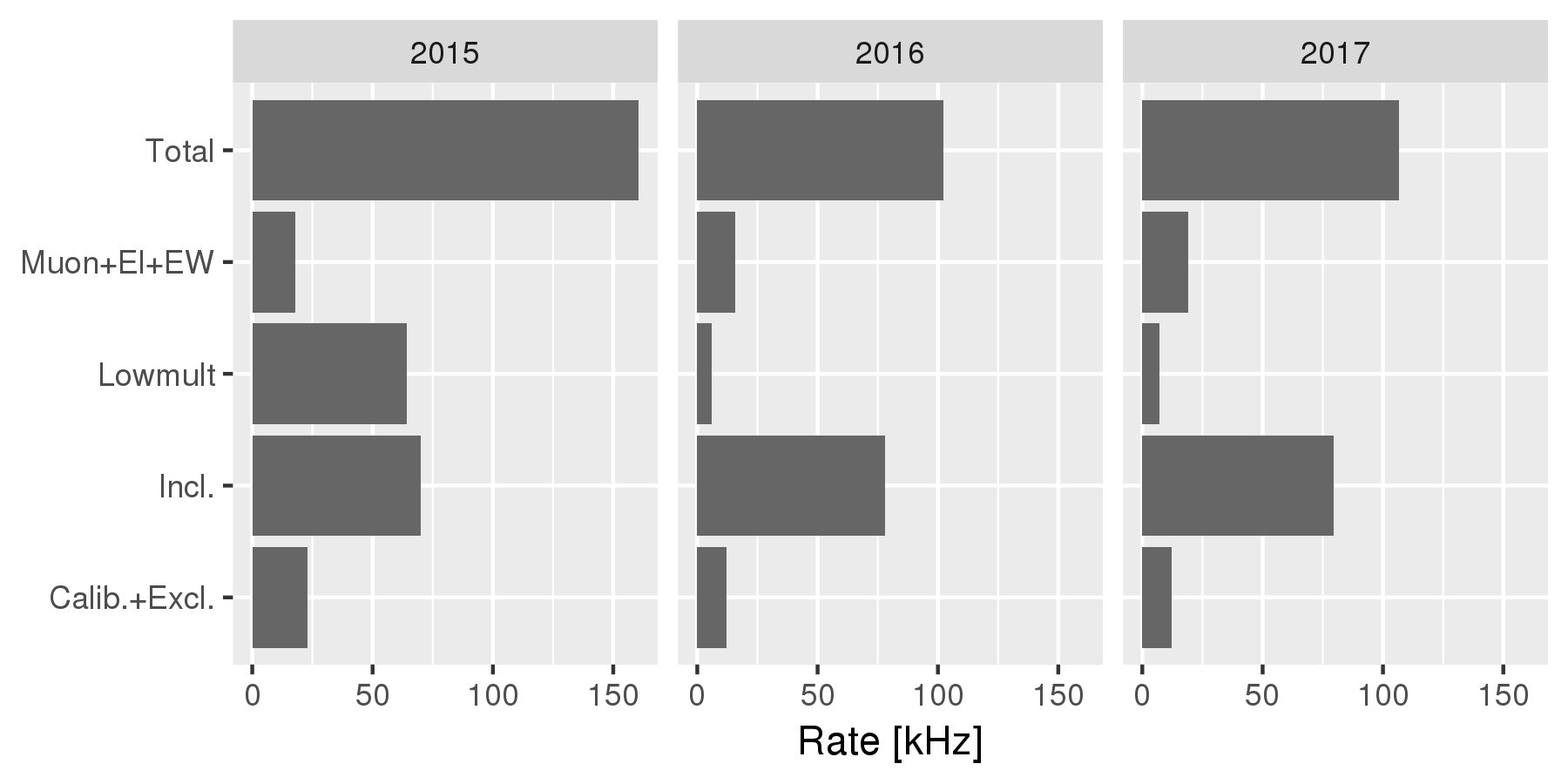}
  \caption{Rates of the main groups of \hltone trigger lines and the total \hltone rate as a function of the year of data taking, shown for the trigger configuration used to take most of the luminosity in each year.}
  \label{fig:hlt1_rates}
\end{figure}

The \hltone bandwidth is preferentially allocated to the inclusive and muon trigger lines which, by selecting \Pb-~and~\Pc-~hadron decays, cover most
of the \lhcb physics programme. The large disk buffer available in Run~2 also makes it possible to allow generous rates for other
trigger lines, however, with a total \hltone output rate of 150~kHz which is around two times the Run~1 average. The \hltone rates
and the overlaps in the events selected by the different \hltone trigger lines are shown in Fig.~\ref{fig:hlt1_rates}.

\subsection{\hlttwo performance}
The \hlttwo trigger stage reduces the event rate to around 12.5~kHz, at which point the remaining events
are saved to permanent storage for further analysis. The \hlttwo reconstruction sequence was described in Sec.~\ref{sec:hlt2},
while this section describes the performance of a representative set of \hlttwo trigger lines.

\subsubsection{Inclusive \Pb-hadron trigger lines}
The \hlttwo inclusive \Pb-hadron trigger lines look for a two-, three-, or four-track vertex with sizable \pt, significant displacement from
the PV, and a topology compatible with the decay of a \Pb~hadron. As in Run~1~\cite{LHCb-PUB-2011-016,BBDT},
these ``topological'' trigger lines rely on a multivariate selection of the displaced vertex. This selection is implemented in a
MatrixNet classifier whose inputs have been discretized~\cite{BBDT} in order to minimize the variation in selection performance
with varying detector conditions and speed up the evaluation time. The efficiency of the topological trigger lines is increased
for decays involving muons by relaxing the requirement on the multivariate discriminant whenever one or more of the tracks associated
with the topological vertex is positively identified as a muon or electron.

\begin{figure}
  \centering
  \includegraphics[width=0.80\textwidth]{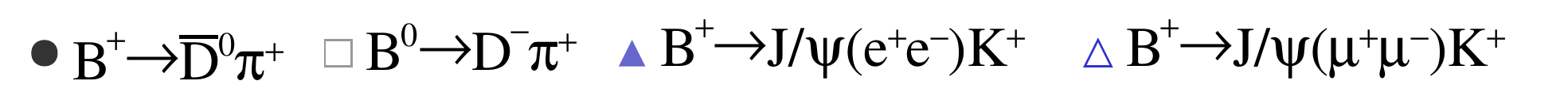}
  \includegraphics[width=0.44\textwidth]{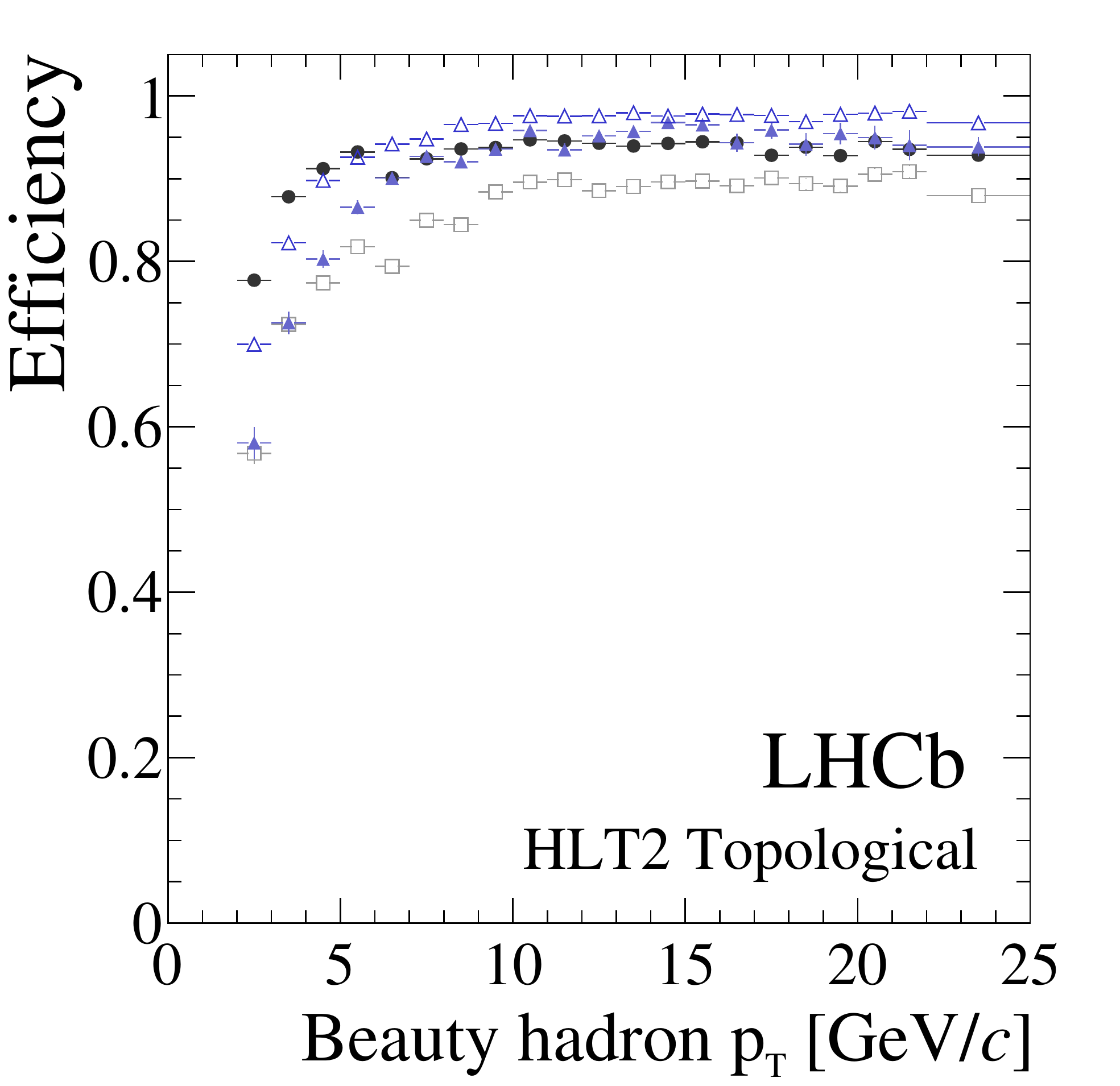}
  \includegraphics[width=0.44\textwidth]{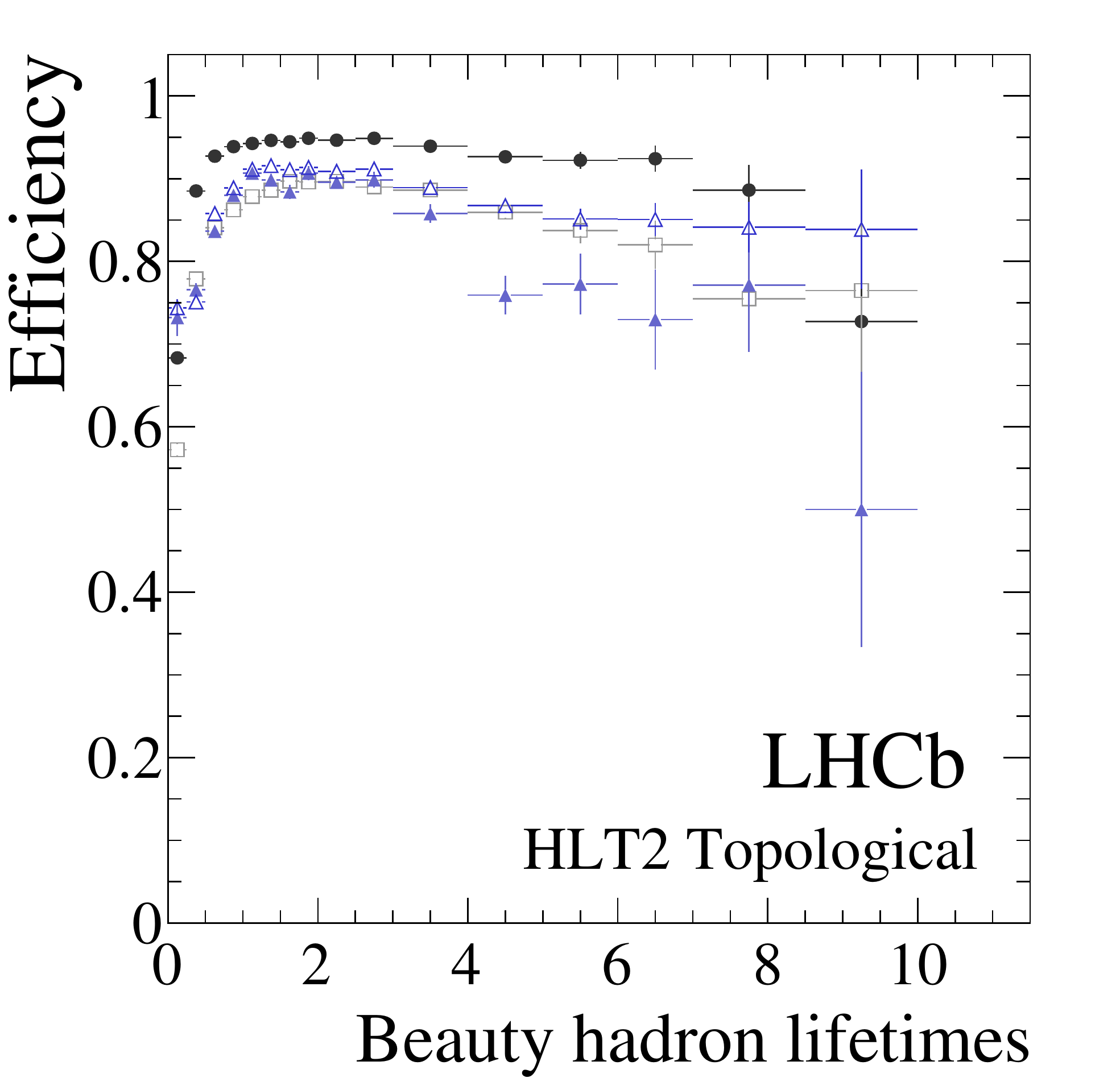}
  \caption{Efficiency of the \hlttwo topological trigger lines as a function of the (left) \Pb-hadron \pt and (right) in units of the average \Pb-hadron decay time. The decay time plots
are drawn such that the x-axis is binned in units of the lifetime for each hadron in its rest frame.
The plots show the combined efficiency of the topological trigger lines for each \Pb-hadron decay mode.}
  \label{fig:hlt2_topoeff}
\end{figure}

\begin{figure}
  \centering
  \includegraphics[width=0.90\textwidth]{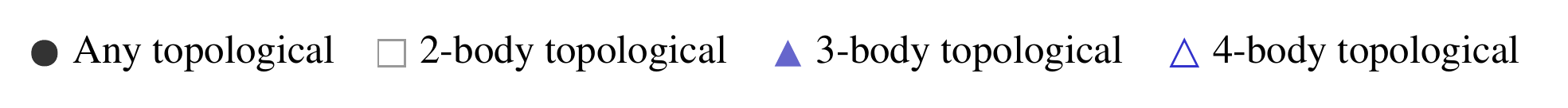}
  \includegraphics[width=0.44\textwidth]{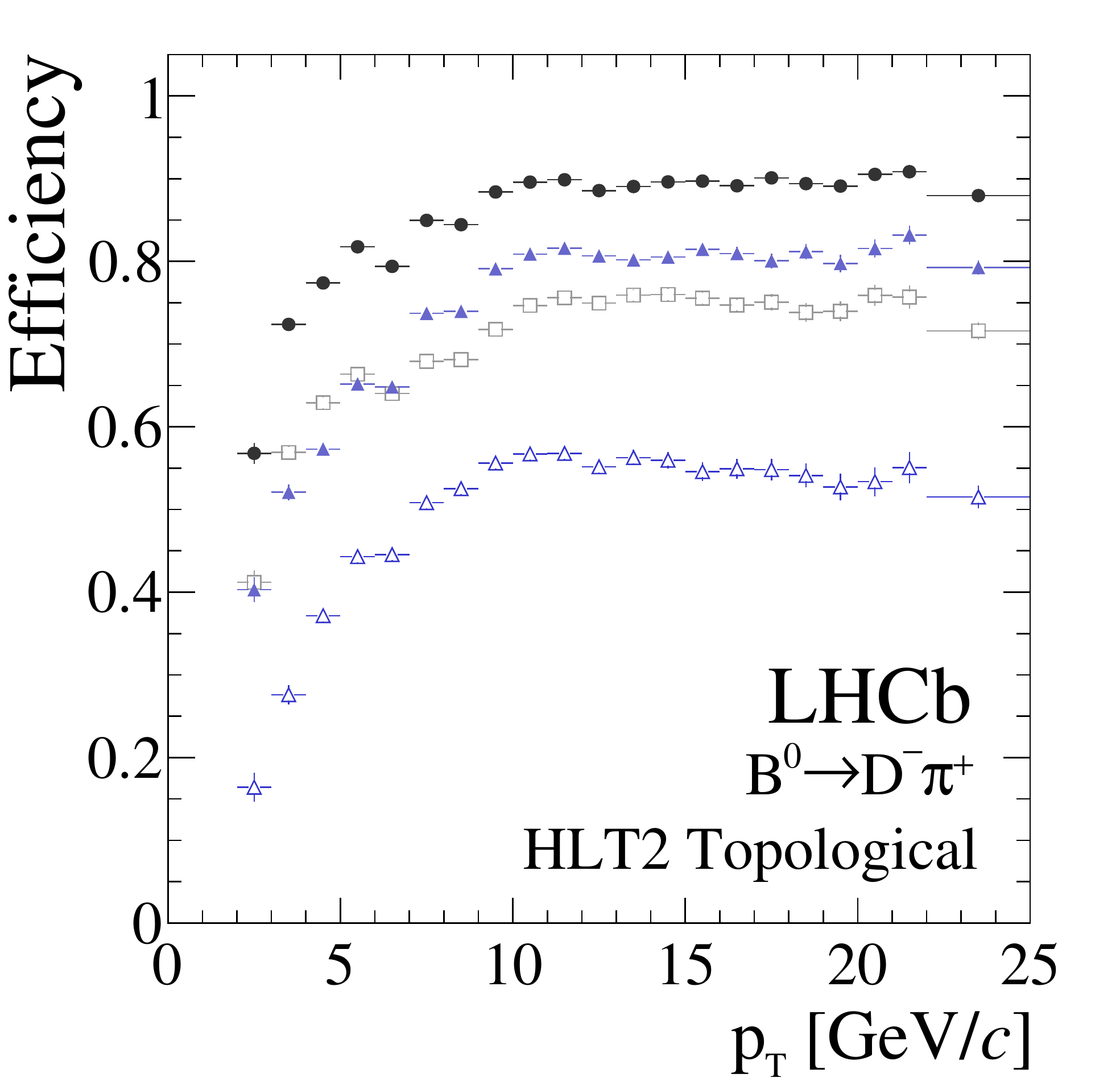}
  \includegraphics[width=0.44\textwidth]{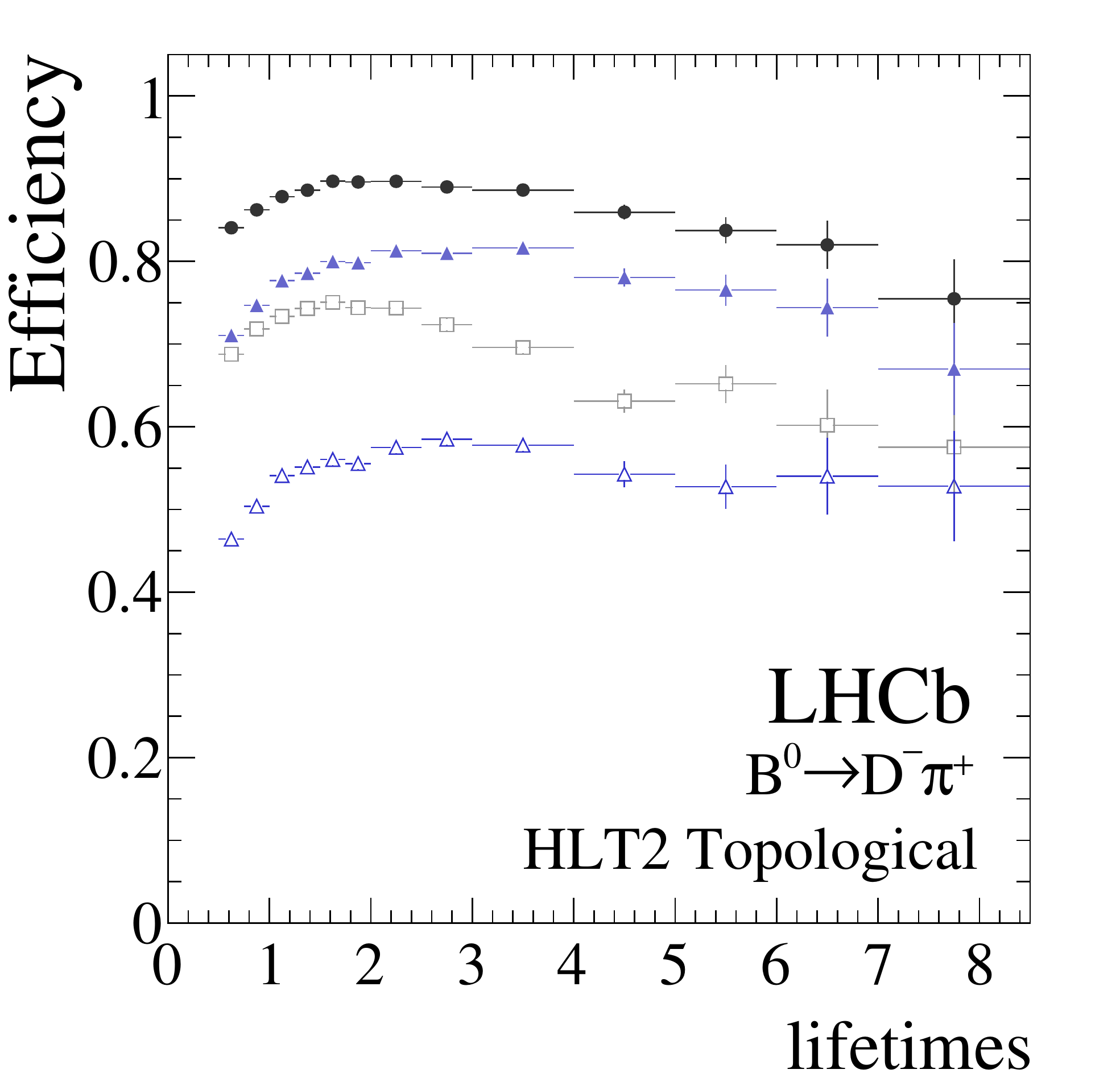}
  \caption{Efficiency of the \hlttwo topological trigger lines as a function of the (left) \Pb-hadron \pt and (right) in units of the average \Pb-hadron decay time. The decay time plots
are drawn such that the x-axis is binned in units of the lifetime for each hadron in its rest frame.
The plots show the different contributions
of the 2-, 3-, and 4-body topological trigger lines to a specific decay.}
  \label{fig:hlt2_topoeff_bdpi}
\end{figure}

\begin{figure}[t]
  \centering
  \includegraphics[width=0.44\textwidth]{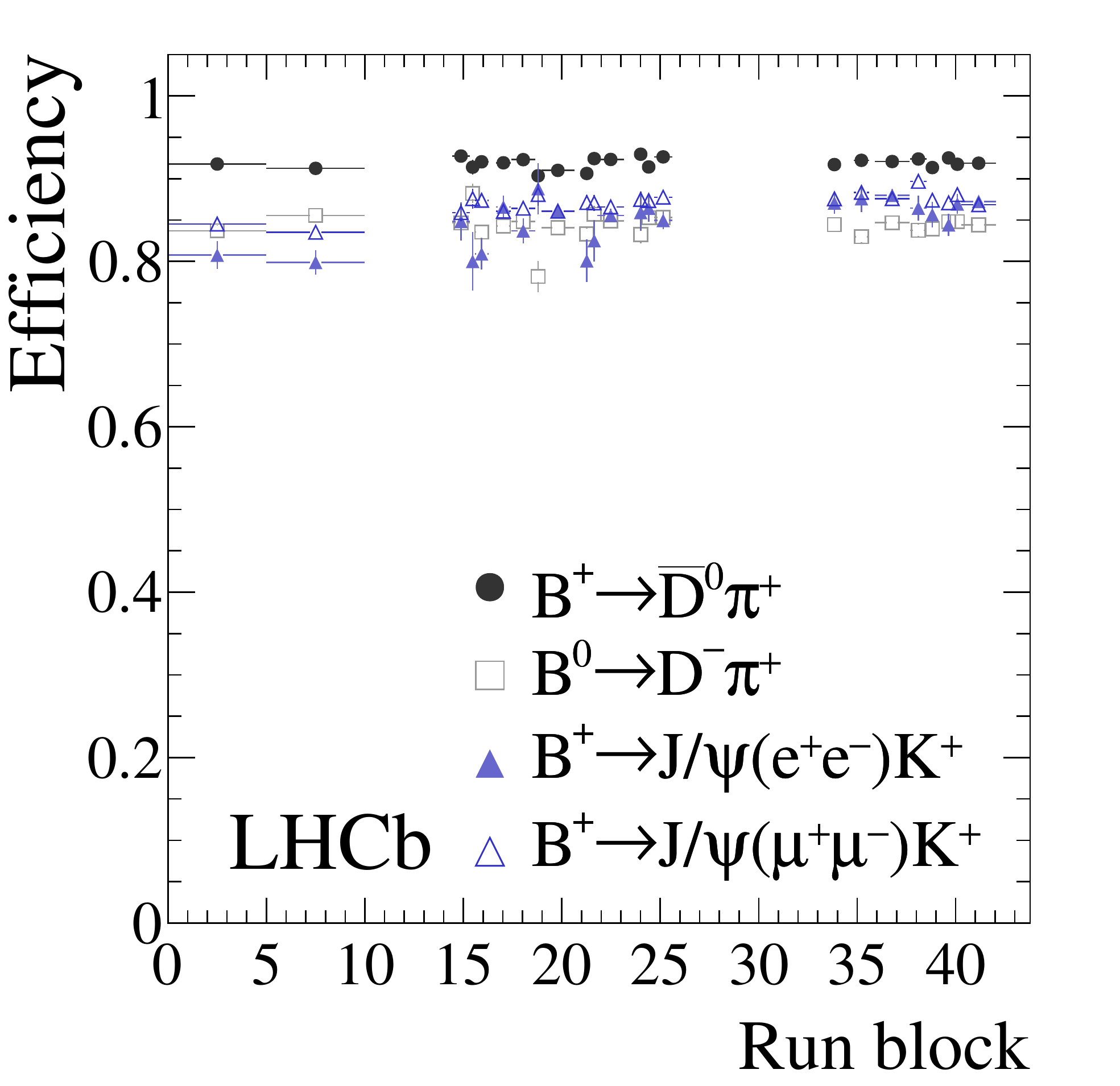}
  \caption{Evolution of the \hlttwo efficiency as a function of the different trigger
configurations used during data taking.}
  \label{fig:hlt2_topoeff_run}
\end{figure}

The topological trigger lines are trained to separate signal \Pb-hadron decays
which can be fully reconstructed inside the detector acceptance from those which cannot, as well as from
displaced vertices formed from the decays of \Pc~hadrons originating from the PV. The displaced
vertices from \Pc~hadrons are the most numerous background.
Harder to discriminate against, however, are the backgrounds from \Pb-hadron decays that are only 
partially contained in the detector acceptance, or \Pb-hadron
decays in which much of the energy is taken by neutral particles. 
The selection has been reoptimized~\cite{LHCb-PROC-2015-018} for Run~2, taking advantage of the full off\-line reconstruction
now available in \hlttwo to
loosen the selection criteria when building vertex candidates, and fully relying on the multivariate algorithm to discriminate
between signal and background. The resulting efficiencies
are shown in Fig.~\ref{fig:hlt2_topoeff} and 
Fig.~\ref{fig:hlt2_topoeff_bdpi} for a specific decay mode, 
while the evolution of the efficiency as a function of the data-taking conditions is shown in Fig.~\ref{fig:hlt2_topoeff_run}.



\subsubsection{Muon and dimuon trigger lines}

\begin{figure}
  \centering
  \includegraphics[width=0.90\textwidth]{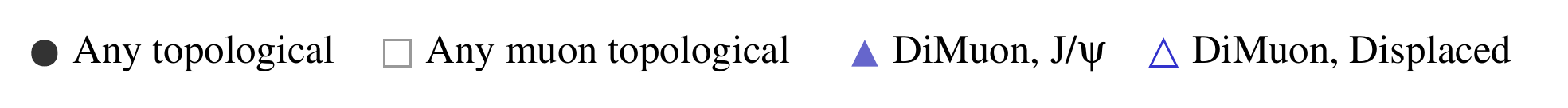}
  \includegraphics[width=0.44\textwidth]{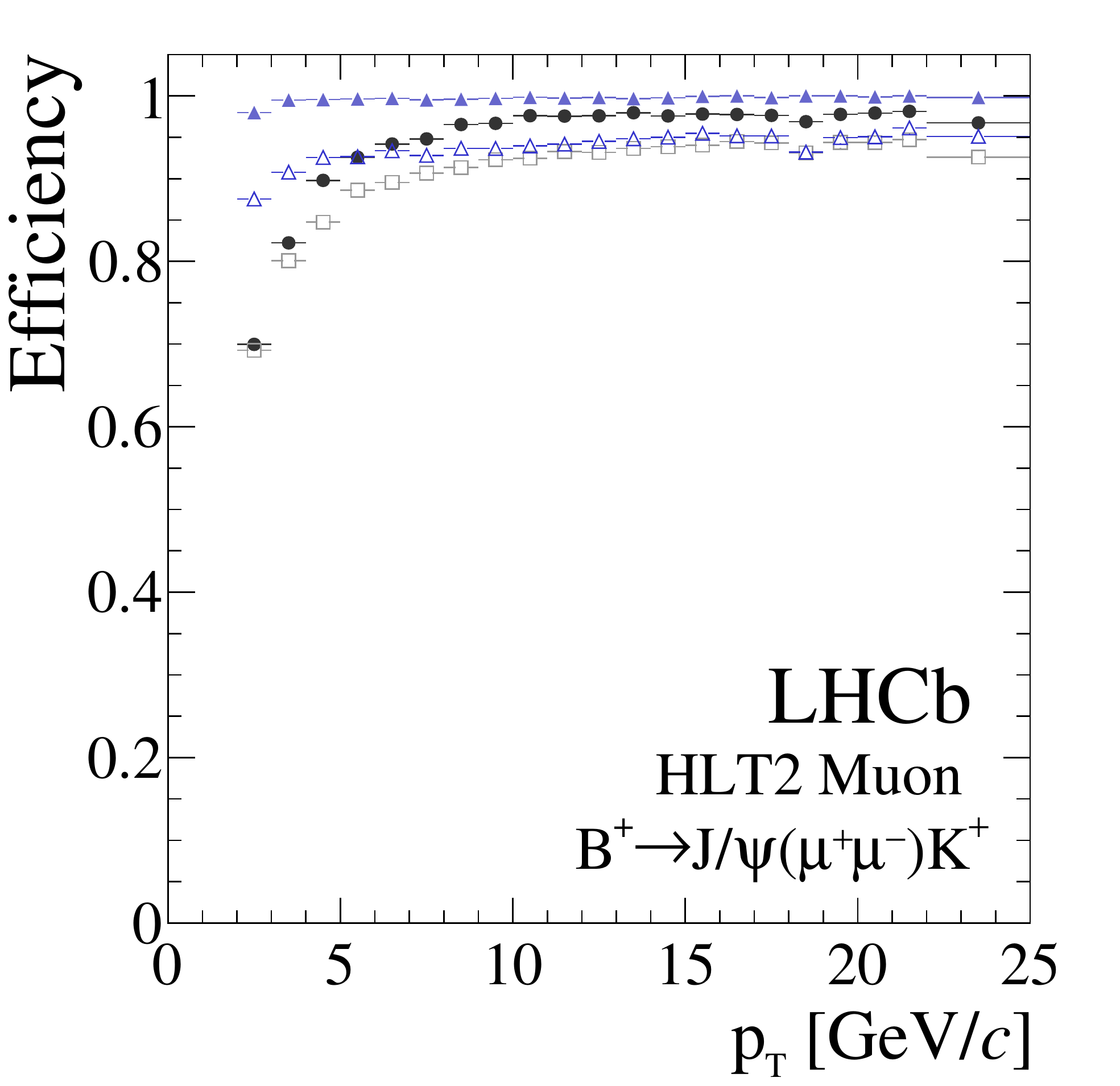}
  \includegraphics[width=0.44\textwidth]{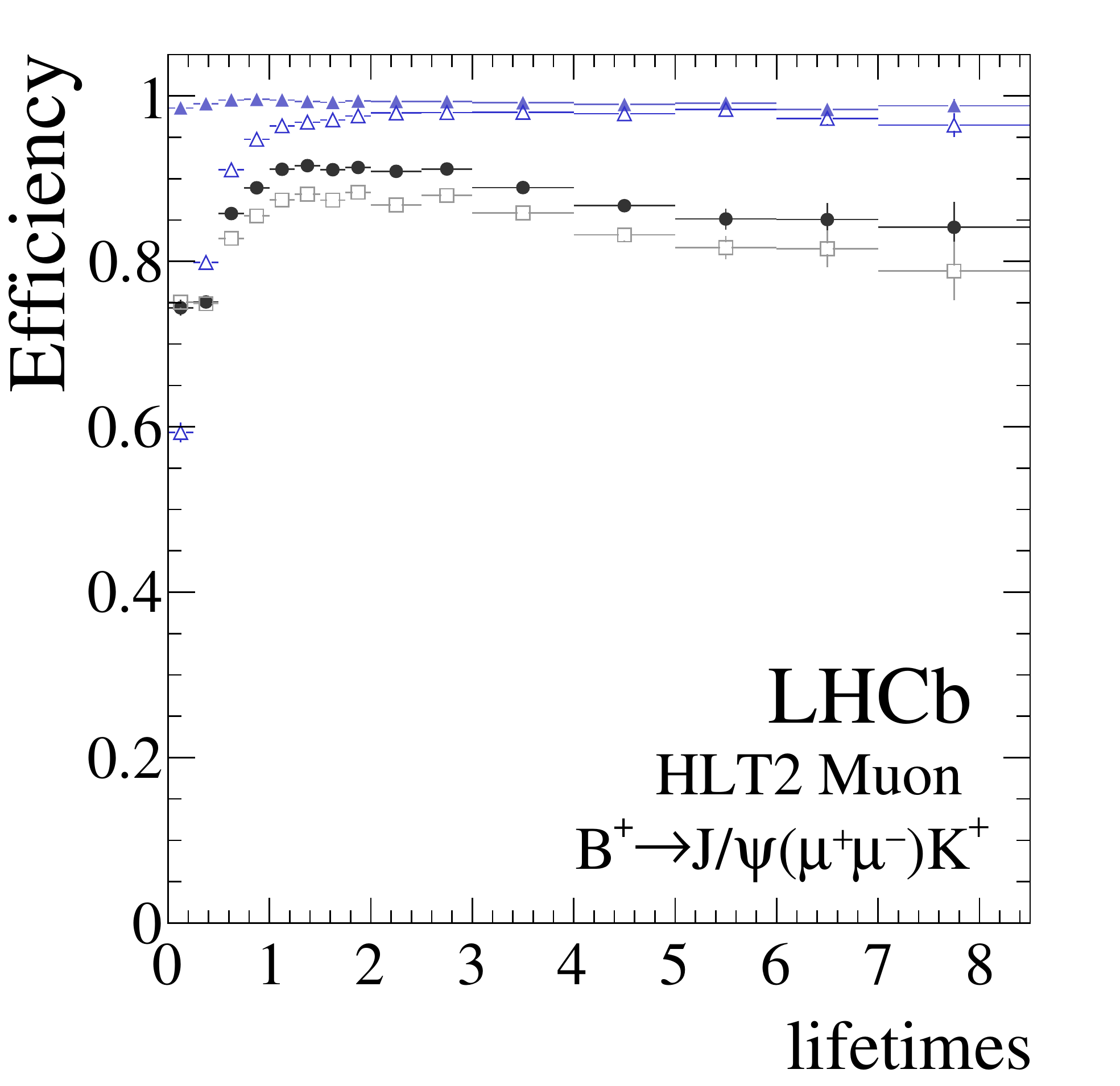}
  \caption{The TOS efficiency of the \hlttwo muon trigger lines as a function of the (left) $\Bp$ \pt and (right) units of the average $\Bp$ decay time. The decay time plot is drawn such that the x-axis is binned in units of the $\Bp$ lifetime in its rest frame. The
efficiency of the inclusive topological (``any topological'') trigger lines and topological trigger lines requiring one track identified as a muon (``any muon topological'')
are plotted for reference.}
  \label{fig:hlt2_muon_tistos}
\end{figure}

The \hlttwo muon and dimuon trigger lines select a wide spectrum of signals: low-mass Drell--Yan dimuons for electroweak physics,
dimuons originating from the PV for production measurements, dimuons with displacement from the PV for the study
of \Pb-, \Pc-, and \Ps-hadron decays and heavy dimuons for exotic particle searches and electroweak physics.
As mentioned in Sec.~\ref{sec:muonid}, in Run 2 the \hlttwo and off\-line muon-identification procedures are identical.
Owing to this improvement and because muons provide a relatively rare and clean event signature,
the dimuon trigger lines generally have a  high efficiency which is only limited in some cases
by the rate of the selected signal, most notably for production measurements.
This is illustrated in Fig.~\ref{fig:hlt2_muon_tistos} where the efficiency of the \hlttwo muon trigger lines is shown for $B^+\to\jpsi\Kp$ decays.
Note that the muon topological trigger lines have a lower absolute efficiency compared to the hadron topological trigger lines because they only
process events passing the \hltone single-muon selection.
In addition to the standard inclusive muon lines used in Run 1, for Run 2 new lines have been developed in particular for
dimuons with lower \pt for exotic particle searches (\eg dark photons) and for rare strange-hadron decays~\cite{Dettori:2297352}.

\subsubsection{Exclusive and calibration trigger lines}
\label{sec:hlt2_calib}

In addition to the inclusive trigger lines, the full off\-line reconstruction performed at the start of \hlttwo
means that it is possible to fully reconstruct certain decays of interest and select them using dedicated trigger lines without
any loss in efficiency compared to the off\-line analysis. This is especially important for \Pc-hadron trigger lines because around $10\%$ of all
13~TeV proton-proton collisions produce a $c\bar{c}$ pair, and it is not possible to write all \Pc-hadron signals to off\-line storage.
In order to reduce the necessary disk space, the \lhcb exclusive \Pc-hadron trigger lines make extensive use of the TURBO stream. 
All events selected by those trigger lines, except those containing neutral particles, are sent to the TURBO stream.
The selection criteria of these trigger lines are usually a slightly looser version of those used
in the off\-line analysis, enabling the candidates saved in the TURBO stream to be directly used by the analysts.
In total, over 200 different
exclusive trigger lines which select the decays of \Pc~hadrons are deployed in Run~2. They are generally tuned to have $S/B$ ratios well in excess of 1 already at
the output of the trigger, with the final selection performed off\-line using information reconstructed in the trigger and tuned to minimize systematic uncertainties.
The purity achievable using the trigger-level information has already been illustrated in Fig.~\ref{fig:charmsignals} for a representative sample of \Pc-hadron decays.

In addition, \hlttwo contains a suite of calibration trigger lines, which are used to measure the performance of
the track-finding and particle-identification algorithms in a data-driven way. These trigger lines select
high-yield charm, charmonium, and $\KS$ decays using a tag-and-probe
approach, where the probe particle is kept unbiased with respect to either the tracking or particle-identification information. There are around 50 such lines
in total, and they select around 500~Hz of calibration signals.


\subsubsection{Low multiplicity event trigger lines}
At the \hlttwo stage there are dedicated selections for each relevant final state with a low track multiplicity.
There are 32 lines: two to select exclusive dimuon production, three to select exclusive production of photons or electrons, and the remainder
to select various hadronic final states, dominated by lines that select low-\pt hadrons.

\begin{figure}[t]
\includegraphics[width=.45\textwidth]{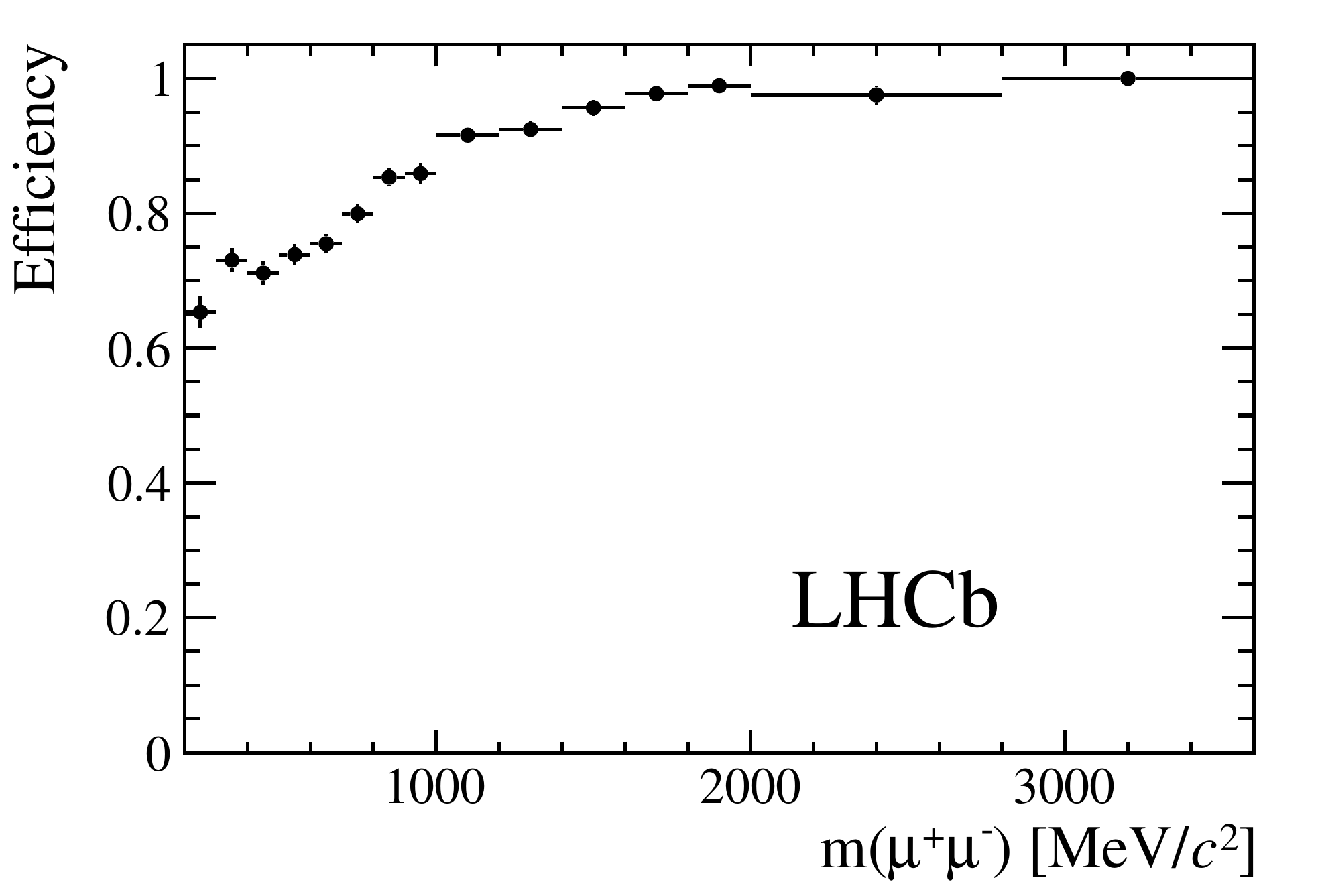}
\includegraphics[width=.45\textwidth]{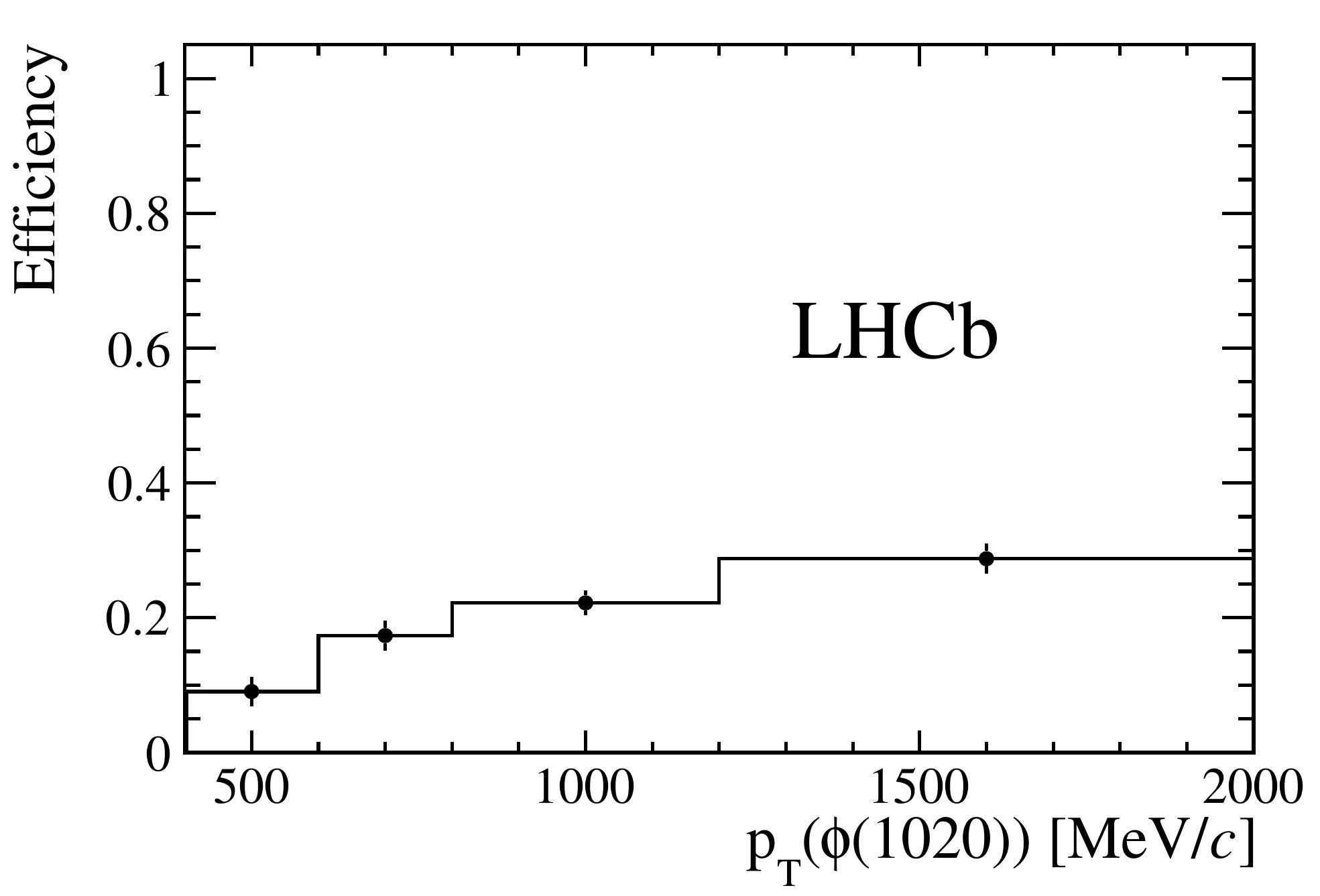}
\caption{\hlttwo trigger efficiencies of the dedicated selections for low-multiplicity events: (left) for dimuon candidates as a function of dimuon mass, and (right) for $\phi(1020)$ candidates as a function of candidate \pt.\label{im:LowMultTriggerEff}}
\end{figure}

The \hlttwo trigger efficiencies have been determined in data and are shown in Fig.~\ref{im:LowMultTriggerEff} for two channels of particular interest: dimuon and dihadron. The dimuon \hlttwo trigger efficiency is determined using a sample of independently triggered candidates reconstructed in events containing exactly two muon tracks inside the detector acceptance. The dimuon candidate is required to have satisfied the relevant low-multiplicity \lz trigger. The efficiency is shown as a function of dimuon mass, where the rise at 800\mevcc results from the 400\mevc \pt requirement for each muon. In the case of exclusive production, where the candidate is expected to be produced with low \pt, this leads to an implicit lower bound on the mass of the exclusively-produced object at $m(\mup\mun)\approx 800\mevcc$. The non-zero efficiency for candidates with $m(\mup\mun) \lesssim 800\mevcc$ arises from candidates with higher \pt.

The dihadron \hlttwo trigger efficiency, which includes the effect of a 50\% prescale, is determined using $\phi(1020)\to\Kp\Km$ candidates reconstructed in low-multiplicity events and triggered independently of the signal candidate. The $\phi(1020)$ candidate is required to pass the relevant low-multiplicity \lz and \hltone trigger lines, and the background from misidentified pions is reduced using information from the RICH sub-detectors. The efficiency is shown as a function of the \pt of the $\phi(1020)$ meson.

\subsubsection{\hlttwo bandwidth division}
\begin{figure}
  \centering
  \includegraphics[width=0.95\textwidth]{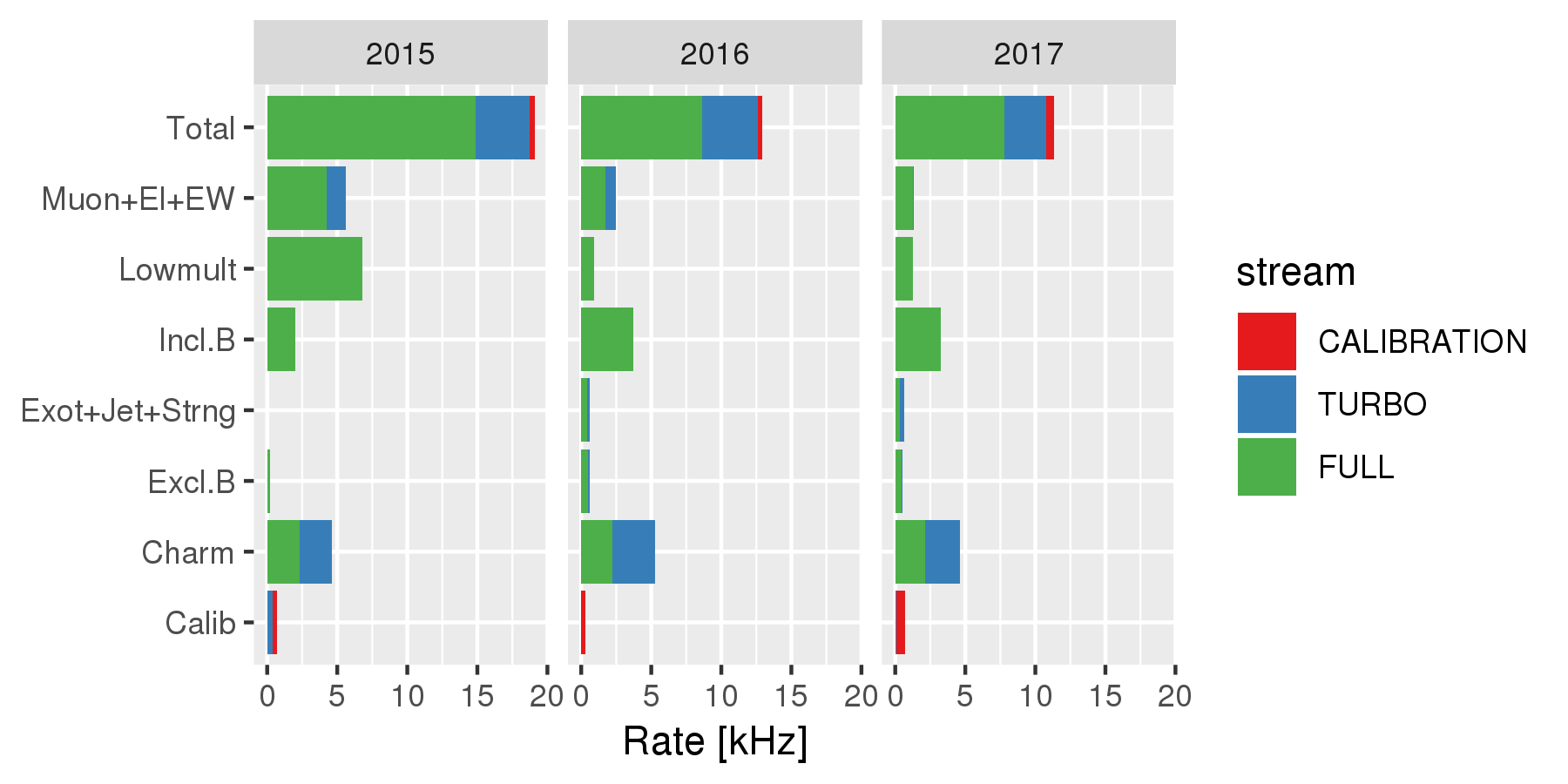}
  \caption{Rates of the main categories of \hlttwo trigger lines and the total \hlttwo rate for each year of data taking, shown for the trigger configuration used to take most of the luminosity in the given year. TURBO, CALIBRATION, and FULL refer to different output data streams as discussed in Ref.~\cite{LHCb-DP-2016-001}.}
  \label{fig:hlt2_rates}
\end{figure}

The \hlttwo bandwidth is divided into the full stream, containing inclusive trigger lines, and the TURBO stream, which contains exclusive
trigger lines that fully reconstruct relevant decays. Most of the full stream rate is taken up by the topological \Pb-hadron, inclusive \Pc-hadron, and
dimuon trigger lines, while the TURBO stream rate is divided among several hundred exclusive \Pc-hadron trigger lines. As the TURBO stream
trigger lines perform a full selection of high-purity signals, their rates are generally proportional to the signal abundance. The \hlttwo rates
and the overlaps in the events selected by the different \hlttwo trigger lines are shown in Fig.~\ref{fig:hlt2_rates}, where the exclusive trigger lines are
counted as one item for brevity.

\section{Conclusions}
The design and performance of the LHCb Run~2 reconstruction and High Level Trigger have been presented.
The use of real-time alignment and calibration and improvements in the reconstruction software allows for events to be fully reconstructed in the High Level Trigger with equivalent quality to the Run 1 off\-line performance, and enables signals
to be selected with a purity close to that achievable off\-line.
This in turn enables physics analysis to be performed
directly with the output of the reconstruction in the trigger. To this end, a significant fraction of triggered
events is saved in a reduced ``real-time analysis'' format, 
saving only higher-level reconstructed objects relevant to physics analysis and not the full raw detector data.
The successful deployment of this full real-time reconstruction and analysis during Run~2 is a critical
stepping stone towards the LHCb upgrade, whose software trigger will have to deal with roughly 100 times greater
data rates while maintaining a high acceptance over the same broad range of physics channels.


\section*{Acknowledgements}
\noindent We express our gratitude to our colleagues in the CERN
accelerator departments for the excellent performance of the LHC. We
thank the technical and administrative staff at the LHCb
institutes.
We acknowledge support from CERN and from the national agencies:
CAPES, CNPq, FAPERJ and FINEP (Brazil); 
MOST and NSFC (China); 
CNRS/IN2P3 (France); 
BMBF, DFG and MPG (Germany); 
INFN (Italy); 
NWO (Netherlands); 
MNiSW and NCN (Poland); 
MEN/IFA (Romania); 
MSHE (Russia); 
MinECo (Spain); 
SNSF and SER (Switzerland); 
NASU (Ukraine); 
STFC (United Kingdom); 
NSF (USA).
We acknowledge the computing resources that are provided by CERN, IN2P3
(France), KIT and DESY (Germany), INFN (Italy), SURF (Netherlands),
PIC (Spain), GridPP (United Kingdom), RRCKI and Yandex
LLC (Russia), CSCS (Switzerland), IFIN-HH (Romania), CBPF (Brazil),
PL-GRID (Poland) and OSC (USA).
We are indebted to the communities behind the multiple open-source
software packages on which we depend.
Individual groups or members have received support from
AvH Foundation (Germany);
EPLANET, Marie Sk\l{}odowska-Curie Actions and ERC (European Union);
ANR, Labex P2IO and OCEVU, and R\'{e}gion Auvergne-Rh\^{o}ne-Alpes (France);
Key Research Program of Frontier Sciences of CAS, CAS PIFI, and the Thousand Talents Program (China);
RFBR, RSF and Yandex LLC (Russia);
GVA, XuntaGal and GENCAT (Spain);
the Royal Society
and the Leverhulme Trust (United Kingdom);
Laboratory Directed Research and Development program of LANL (USA).


\addcontentsline{toc}{section}{References}
\setboolean{inbibliography}{true}
\bibliographystyle{LHCb}
\bibliography{standard,LHCb-PAPER,LHCb-CONF,LHCb-DP,LHCb-TDR}

\ifx\mcitethebibliography\mciteundefinedmacro
\PackageError{LHCb.bst}{mciteplus.sty has not been loaded}
{This bibstyle requires the use of the mciteplus package.}\fi
\providecommand{\href}[2]{#2}
\begin{mcitethebibliography}{10}
\mciteSetBstSublistMode{n}
\mciteSetBstMaxWidthForm{subitem}{\alph{mcitesubitemcount})}
\mciteSetBstSublistLabelBeginEnd{\mcitemaxwidthsubitemform\space}
{\relax}{\relax}

\bibitem{LHCb-DP-2012-004}
R.~Aaij {\em et~al.}, \ifthenelse{\boolean{articletitles}}{\emph{{The \lhcb
  trigger and its performance in 2011}},
  }{}\href{https://doi.org/10.1088/1748-0221/8/04/P04022}{JINST \textbf{8}
  (2013) P04022}, \href{http://arxiv.org/abs/1211.3055}{{\normalfont\ttfamily
  arXiv:1211.3055}}\relax
\mciteBstWouldAddEndPuncttrue
\mciteSetBstMidEndSepPunct{\mcitedefaultmidpunct}
{\mcitedefaultendpunct}{\mcitedefaultseppunct}\relax
\EndOfBibitem
\bibitem{LHCb-DP-2016-001}
R.~Aaij {\em et~al.}, \ifthenelse{\boolean{articletitles}}{\emph{{Tesla: an
  application for real-time data analysis in High Energy Physics}},
  }{}\href{https://doi.org/10.1016/j.cpc.2016.07.022}{Comput.\ Phys.\ Commun.\
  \textbf{208} (2016) 35},
  \href{http://arxiv.org/abs/1604.05596}{{\normalfont\ttfamily
  arXiv:1604.05596}}\relax
\mciteBstWouldAddEndPuncttrue
\mciteSetBstMidEndSepPunct{\mcitedefaultmidpunct}
{\mcitedefaultendpunct}{\mcitedefaultseppunct}\relax
\EndOfBibitem
\bibitem{Aaij:2019uij}
R.~Aaij {\em et~al.}, \ifthenelse{\boolean{articletitles}}{\emph{{A
  comprehensive real-time analysis model at the LHCb experiment}},
  }{}\href{http://arxiv.org/abs/1903.01360}{{\normalfont\ttfamily
  arXiv:1903.01360}}\relax
\mciteBstWouldAddEndPuncttrue
\mciteSetBstMidEndSepPunct{\mcitedefaultmidpunct}
{\mcitedefaultendpunct}{\mcitedefaultseppunct}\relax
\EndOfBibitem
\bibitem{LHCbCollaboration:2319756}
C.~M. LHCb~Collaboration, \ifthenelse{\boolean{articletitles}}{\emph{{Computing
  Model of the Upgrade LHCb experiment}}, }{} Tech. Rep. CERN-LHCC-2018-014.
  LHCB-TDR-018, CERN, Geneva, May, 2018\relax
\mciteBstWouldAddEndPuncttrue
\mciteSetBstMidEndSepPunct{\mcitedefaultmidpunct}
{\mcitedefaultendpunct}{\mcitedefaultseppunct}\relax
\EndOfBibitem
\bibitem{LHCb}
LHCb collaboration, A.~A. Alves~Jr.\ {\em et~al.},
  \ifthenelse{\boolean{articletitles}}{\emph{{The \lhcb detector at the LHC}},
  }{}\href{https://doi.org/10.1088/1748-0221/3/08/S08005}{JINST \textbf{3}
  (2008) S08005}\relax
\mciteBstWouldAddEndPuncttrue
\mciteSetBstMidEndSepPunct{\mcitedefaultmidpunct}
{\mcitedefaultendpunct}{\mcitedefaultseppunct}\relax
\EndOfBibitem
\bibitem{LHCb-DP-2014-002}
LHCb collaboration, R.~Aaij {\em et~al.},
  \ifthenelse{\boolean{articletitles}}{\emph{{LHCb detector performance}},
  }{}\href{https://doi.org/10.1142/S0217751X15300227}{Int.\ J.\ Mod.\ Phys.\
  \textbf{A30} (2015) 1530022},
  \href{http://arxiv.org/abs/1412.6352}{{\normalfont\ttfamily
  arXiv:1412.6352}}\relax
\mciteBstWouldAddEndPuncttrue
\mciteSetBstMidEndSepPunct{\mcitedefaultmidpunct}
{\mcitedefaultendpunct}{\mcitedefaultseppunct}\relax
\EndOfBibitem
\bibitem{LHCb-DP-2014-001}
R.~Aaij {\em et~al.}, \ifthenelse{\boolean{articletitles}}{\emph{{Performance
  of the LHCb Vertex Locator}},
  }{}\href{https://doi.org/10.1088/1748-0221/9/09/P09007}{JINST \textbf{9}
  (2014) P09007}, \href{http://arxiv.org/abs/1405.7808}{{\normalfont\ttfamily
  arXiv:1405.7808}}\relax
\mciteBstWouldAddEndPuncttrue
\mciteSetBstMidEndSepPunct{\mcitedefaultmidpunct}
{\mcitedefaultendpunct}{\mcitedefaultseppunct}\relax
\EndOfBibitem
\bibitem{LHCb-DP-2013-003}
R.~Arink {\em et~al.}, \ifthenelse{\boolean{articletitles}}{\emph{{Performance
  of the LHCb Outer Tracker}},
  }{}\href{https://doi.org/10.1088/1748-0221/9/01/P01002}{JINST \textbf{9}
  (2014) P01002}, \href{http://arxiv.org/abs/1311.3893}{{\normalfont\ttfamily
  arXiv:1311.3893}}\relax
\mciteBstWouldAddEndPuncttrue
\mciteSetBstMidEndSepPunct{\mcitedefaultmidpunct}
{\mcitedefaultendpunct}{\mcitedefaultseppunct}\relax
\EndOfBibitem
\bibitem{LHCb-DP-2012-003}
M.~Adinolfi {\em et~al.},
  \ifthenelse{\boolean{articletitles}}{\emph{{Performance of the \lhcb RICH
  detector at the LHC}},
  }{}\href{https://doi.org/10.1140/epjc/s10052-013-2431-9}{Eur.\ Phys.\ J.\
  \textbf{C73} (2013) 2431},
  \href{http://arxiv.org/abs/1211.6759}{{\normalfont\ttfamily
  arXiv:1211.6759}}\relax
\mciteBstWouldAddEndPuncttrue
\mciteSetBstMidEndSepPunct{\mcitedefaultmidpunct}
{\mcitedefaultendpunct}{\mcitedefaultseppunct}\relax
\EndOfBibitem
\bibitem{LHCb-DP-2012-002}
A.~A. Alves~Jr.\ {\em et~al.},
  \ifthenelse{\boolean{articletitles}}{\emph{{Performance of the LHCb muon
  system}}, }{}\href{https://doi.org/10.1088/1748-0221/8/02/P02022}{JINST
  \textbf{8} (2013) P02022},
  \href{http://arxiv.org/abs/1211.1346}{{\normalfont\ttfamily
  arXiv:1211.1346}}\relax
\mciteBstWouldAddEndPuncttrue
\mciteSetBstMidEndSepPunct{\mcitedefaultmidpunct}
{\mcitedefaultendpunct}{\mcitedefaultseppunct}\relax
\EndOfBibitem
\bibitem{Sjostrand:2006za}
T.~Sj\"{o}strand, S.~Mrenna, and P.~Skands,
  \ifthenelse{\boolean{articletitles}}{\emph{{PYTHIA 6.4 physics and manual}},
  }{}\href{https://doi.org/10.1088/1126-6708/2006/05/026}{JHEP \textbf{05}
  (2006) 026}, \href{http://arxiv.org/abs/hep-ph/0603175}{{\normalfont\ttfamily
  arXiv:hep-ph/0603175}}\relax
\mciteBstWouldAddEndPuncttrue
\mciteSetBstMidEndSepPunct{\mcitedefaultmidpunct}
{\mcitedefaultendpunct}{\mcitedefaultseppunct}\relax
\EndOfBibitem
\bibitem{Sjostrand:2007gs}
T.~Sj\"{o}strand, S.~Mrenna, and P.~Skands,
  \ifthenelse{\boolean{articletitles}}{\emph{{A brief introduction to PYTHIA
  8.1}}, }{}\href{https://doi.org/10.1016/j.cpc.2008.01.036}{Comput.\ Phys.\
  Commun.\  \textbf{178} (2008) 852},
  \href{http://arxiv.org/abs/0710.3820}{{\normalfont\ttfamily
  arXiv:0710.3820}}\relax
\mciteBstWouldAddEndPuncttrue
\mciteSetBstMidEndSepPunct{\mcitedefaultmidpunct}
{\mcitedefaultendpunct}{\mcitedefaultseppunct}\relax
\EndOfBibitem
\bibitem{LHCb-PROC-2010-056}
I.~Belyaev {\em et~al.}, \ifthenelse{\boolean{articletitles}}{\emph{{Handling
  of the generation of primary events in Gauss, the LHCb simulation
  framework}}, }{}\href{https://doi.org/10.1088/1742-6596/331/3/032047}{J.\
  Phys.\ Conf.\ Ser.\  \textbf{331} (2011) 032047}\relax
\mciteBstWouldAddEndPuncttrue
\mciteSetBstMidEndSepPunct{\mcitedefaultmidpunct}
{\mcitedefaultendpunct}{\mcitedefaultseppunct}\relax
\EndOfBibitem
\bibitem{Lange:2001uf}
D.~J. Lange, \ifthenelse{\boolean{articletitles}}{\emph{{The EvtGen particle
  decay simulation package}},
  }{}\href{https://doi.org/10.1016/S0168-9002(01)00089-4}{Nucl.\ Instrum.\
  Meth.\  \textbf{A462} (2001) 152}\relax
\mciteBstWouldAddEndPuncttrue
\mciteSetBstMidEndSepPunct{\mcitedefaultmidpunct}
{\mcitedefaultendpunct}{\mcitedefaultseppunct}\relax
\EndOfBibitem
\bibitem{Golonka:2005pn}
P.~Golonka and Z.~Was, \ifthenelse{\boolean{articletitles}}{\emph{{PHOTOS Monte
  Carlo: A precision tool for QED corrections in $Z$ and $W$ decays}},
  }{}\href{https://doi.org/10.1140/epjc/s2005-02396-4}{Eur.\ Phys.\ J.\
  \textbf{C45} (2006) 97},
  \href{http://arxiv.org/abs/hep-ph/0506026}{{\normalfont\ttfamily
  arXiv:hep-ph/0506026}}\relax
\mciteBstWouldAddEndPuncttrue
\mciteSetBstMidEndSepPunct{\mcitedefaultmidpunct}
{\mcitedefaultendpunct}{\mcitedefaultseppunct}\relax
\EndOfBibitem
\bibitem{Allison:2006ve}
Geant4 collaboration, J.~Allison {\em et~al.},
  \ifthenelse{\boolean{articletitles}}{\emph{{Geant4 developments and
  applications}}, }{}\href{https://doi.org/10.1109/TNS.2006.869826}{IEEE
  Trans.\ Nucl.\ Sci.\  \textbf{53} (2006) 270}\relax
\mciteBstWouldAddEndPuncttrue
\mciteSetBstMidEndSepPunct{\mcitedefaultmidpunct}
{\mcitedefaultendpunct}{\mcitedefaultseppunct}\relax
\EndOfBibitem
\bibitem{Agostinelli:2002hh}
Geant4 collaboration, S.~Agostinelli {\em et~al.},
  \ifthenelse{\boolean{articletitles}}{\emph{{Geant4: A simulation toolkit}},
  }{}\href{https://doi.org/10.1016/S0168-9002(03)01368-8}{Nucl.\ Instrum.\
  Meth.\  \textbf{A506} (2003) 250}\relax
\mciteBstWouldAddEndPuncttrue
\mciteSetBstMidEndSepPunct{\mcitedefaultmidpunct}
{\mcitedefaultendpunct}{\mcitedefaultseppunct}\relax
\EndOfBibitem
\bibitem{LHCb-PROC-2011-006}
M.~Clemencic {\em et~al.}, \ifthenelse{\boolean{articletitles}}{\emph{{The
  \lhcb simulation application, Gauss: Design, evolution and experience}},
  }{}\href{https://doi.org/10.1088/1742-6596/331/3/032023}{J.\ Phys.\ Conf.\
  Ser.\  \textbf{331} (2011) 032023}\relax
\mciteBstWouldAddEndPuncttrue
\mciteSetBstMidEndSepPunct{\mcitedefaultmidpunct}
{\mcitedefaultendpunct}{\mcitedefaultseppunct}\relax
\EndOfBibitem
\bibitem{LHCb-DP-2017-001}
P.~d'Argent {\em et~al.}, \ifthenelse{\boolean{articletitles}}{\emph{{Improved
  performance of the LHCb Outer Tracker in LHC Run 2}},
  }{}\href{https://doi.org/10.1088/1748-0221/12/11/P11016}{JINST \textbf{9}
  (2017) P11016}, \href{http://arxiv.org/abs/1708.00819}{{\normalfont\ttfamily
  arXiv:1708.00819}}\relax
\mciteBstWouldAddEndPuncttrue
\mciteSetBstMidEndSepPunct{\mcitedefaultmidpunct}
{\mcitedefaultendpunct}{\mcitedefaultseppunct}\relax
\EndOfBibitem
\bibitem{Dujany}
G.~Dujany and B.~Storaci, \ifthenelse{\boolean{articletitles}}{\emph{{Real-time
  alignment and calibration of the LHCb Detector in Run II}}, }{}
  \href{http://cdsweb.cern.ch/search?p=LHCb-PROC-2015-011&f=reportnumber&action_search=Search&c=LHCb+Conference+Proceedings}
  {LHCb-PROC-2015-011}\relax
\mciteBstWouldAddEndPuncttrue
\mciteSetBstMidEndSepPunct{\mcitedefaultmidpunct}
{\mcitedefaultendpunct}{\mcitedefaultseppunct}\relax
\EndOfBibitem
\bibitem{Borghi:2017hfp}
LHCb, S.~Borghi, \ifthenelse{\boolean{articletitles}}{\emph{{Novel real-time
  alignment and calibration of the LHCb detector and its performance}},
  }{}\href{https://doi.org/10.1016/j.nima.2016.06.050}{Nucl.\ Instrum.\ Meth.\
  \textbf{A845} (2017) 560}\relax
\mciteBstWouldAddEndPuncttrue
\mciteSetBstMidEndSepPunct{\mcitedefaultmidpunct}
{\mcitedefaultendpunct}{\mcitedefaultseppunct}\relax
\EndOfBibitem
\bibitem{FastVelo}
O.~Callot, \ifthenelse{\boolean{articletitles}}{\emph{{FastVelo, a fast and
  efficient pattern recognition package for the Velo}}, }{}
  \href{http://cdsweb.cern.ch/search?p=LHCb-PUB-2011-001.
  CERN-LHCb-PUB-2011-001&f=reportnumber&action_search=Search&c=LHCb+Notes}
  {LHCb-PUB-2011-001. CERN-LHCb-PUB-2011-001}, LHCb\relax
\mciteBstWouldAddEndPuncttrue
\mciteSetBstMidEndSepPunct{\mcitedefaultmidpunct}
{\mcitedefaultendpunct}{\mcitedefaultseppunct}\relax
\EndOfBibitem
\bibitem{VeloTT}
E.~E. Bowen, B.~Storaci, and M.~Tresch,
  \ifthenelse{\boolean{articletitles}}{\emph{{VeloTT tracking for LHCb Run
  II}}, }{} \href{http://cdsweb.cern.ch/search?p=LHCb-PUB-2015-024.
  CERN-LHCb-PUB-2015-024.
  LHCb-INT-2014-040&f=reportnumber&action_search=Search&c=LHCb+Notes}
  {LHCb-PUB-2015-024. CERN-LHCb-PUB-2015-024. LHCb-INT-2014-040}\relax
\mciteBstWouldAddEndPuncttrue
\mciteSetBstMidEndSepPunct{\mcitedefaultmidpunct}
{\mcitedefaultendpunct}{\mcitedefaultseppunct}\relax
\EndOfBibitem
\bibitem{forward}
O.~Callot and S.~Hansmann-Menzemer,
  \ifthenelse{\boolean{articletitles}}{\emph{{The Forward Tracking: Algorithm
  and Performance Studies}}, }{}
  \href{http://cdsweb.cern.ch/search?p=LHCb-2007-015.
  CERN-LHCb-2007-015&f=reportnumber&action_search=Search&c=LHCb+Reports}
  {LHCb-2007-015. CERN-LHCb-2007-015}\relax
\mciteBstWouldAddEndPuncttrue
\mciteSetBstMidEndSepPunct{\mcitedefaultmidpunct}
{\mcitedefaultendpunct}{\mcitedefaultseppunct}\relax
\EndOfBibitem
\bibitem{Stahl:2260684}
LHCb Collaboration, M.~Stahl,
  \ifthenelse{\boolean{articletitles}}{\emph{{Machine learning and parallelism
  in the reconstruction of LHCb and its upgrade. Machine learning and
  parallelism in the reconstruction of LHCb and its upgrade}}, }{}J.\ Phys.\ :
  Conf.\ Ser.\  \textbf{898} (2017) 042042. 8 p\relax
\mciteBstWouldAddEndPuncttrue
\mciteSetBstMidEndSepPunct{\mcitedefaultmidpunct}
{\mcitedefaultendpunct}{\mcitedefaultseppunct}\relax
\EndOfBibitem
\bibitem{Dziurda:2115353}
A.~Dziurda, T.~Lesiak, and V.~Gligorov,
  \ifthenelse{\boolean{articletitles}}{\emph{{Studies of time-dependent \CP
  violation in charm decays of $B_s^0$ mesons}}, }{} Apr, 2015.
\newblock Presented 19 Jun 2015\relax
\mciteBstWouldAddEndPuncttrue
\mciteSetBstMidEndSepPunct{\mcitedefaultmidpunct}
{\mcitedefaultendpunct}{\mcitedefaultseppunct}\relax
\EndOfBibitem
\bibitem{Aaij:2253050}
R.~Aaij {\em et~al.}, \ifthenelse{\boolean{articletitles}}{\emph{{Optimization
  of the muon reconstruction algorithms for LHCb Run 2}}, }{}
  \href{http://cdsweb.cern.ch/search?p=LHCb-PUB-2017-007.
  CERN-LHCb-PUB-2017-007&f=reportnumber&action_search=Search&c=LHCb+Notes}
  {LHCb-PUB-2017-007. CERN-LHCb-PUB-2017-007}\relax
\mciteBstWouldAddEndPuncttrue
\mciteSetBstMidEndSepPunct{\mcitedefaultmidpunct}
{\mcitedefaultendpunct}{\mcitedefaultseppunct}\relax
\EndOfBibitem
\bibitem{seeding}
O.~Callot and M.~Schiller,
  \ifthenelse{\boolean{articletitles}}{\emph{{PatSeeding: a standalone track
  reconstruction algorithm}}, }{}
  \href{http://cdsweb.cern.ch/search?p=LHCb-2008-042.
  CERN-LHCb-2008-042&f=reportnumber&action_search=Search&c=LHCb+Reports}
  {LHCb-2008-042. CERN-LHCb-2008-042}\relax
\mciteBstWouldAddEndPuncttrue
\mciteSetBstMidEndSepPunct{\mcitedefaultmidpunct}
{\mcitedefaultendpunct}{\mcitedefaultseppunct}\relax
\EndOfBibitem
\bibitem{matching1}
M.~Needham and J.~Van~Tilburg,
  \ifthenelse{\boolean{articletitles}}{\emph{{Performance of the track
  matching}}, }{} \href{http://cdsweb.cern.ch/search?p=LHCb-2007-020.
  CERN-LHCb-2007-020&f=reportnumber&action_search=Search&c=LHCb+Reports}
  {LHCb-2007-020. CERN-LHCb-2007-020}\relax
\mciteBstWouldAddEndPuncttrue
\mciteSetBstMidEndSepPunct{\mcitedefaultmidpunct}
{\mcitedefaultendpunct}{\mcitedefaultseppunct}\relax
\EndOfBibitem
\bibitem{matching2}
M.~Needham, \ifthenelse{\boolean{articletitles}}{\emph{{Performance of the
  Track Matching}}, }{} \href{http://cdsweb.cern.ch/search?p=LHCb-2007-129.
  CERN-LHCb-2007-129&f=reportnumber&action_search=Search&c=LHCb+Reports}
  {LHCb-2007-129. CERN-LHCb-2007-129}\relax
\mciteBstWouldAddEndPuncttrue
\mciteSetBstMidEndSepPunct{\mcitedefaultmidpunct}
{\mcitedefaultendpunct}{\mcitedefaultseppunct}\relax
\EndOfBibitem
\bibitem{patllt}
A.~Davis, M.~De~Cian, A.~M. Dendek, and T.~Szumlak,
  \ifthenelse{\boolean{articletitles}}{\emph{{PatLongLivedTracking: A tracking
  algorithm for the reconstruction of the daughters of long-lived particles in
  LHCb}}, }{} \href{http://cdsweb.cern.ch/search?p=LHCb-PUB-2017-001.
  CERN-LHCb-PUB-2017-001&f=reportnumber&action_search=Search&c=LHCb+Notes}
  {LHCb-PUB-2017-001. CERN-LHCb-PUB-2017-001}\relax
\mciteBstWouldAddEndPuncttrue
\mciteSetBstMidEndSepPunct{\mcitedefaultmidpunct}
{\mcitedefaultendpunct}{\mcitedefaultseppunct}\relax
\EndOfBibitem
\bibitem{TMVA4}
A.~Hoecker {\em et~al.}, \ifthenelse{\boolean{articletitles}}{\emph{{TMVA 4 ---
  Toolkit for Multivariate Data Analysis. Users Guide.}},
  }{}\href{http://arxiv.org/abs/physics/0703039}{{\normalfont\ttfamily
  arXiv:physics/0703039}}\relax
\mciteBstWouldAddEndPuncttrue
\mciteSetBstMidEndSepPunct{\mcitedefaultmidpunct}
{\mcitedefaultendpunct}{\mcitedefaultseppunct}\relax
\EndOfBibitem
\bibitem{Hocker:2007ht}
H.~Voss, A.~Hoecker, J.~Stelzer, and F.~Tegenfeldt,
  \ifthenelse{\boolean{articletitles}}{\emph{{TMVA - Toolkit for Multivariate
  Data Analysis}}, }{}\href{https://doi.org/10.22323/1.050.0040}{PoS
  \textbf{ACAT} (2007) 040}\relax
\mciteBstWouldAddEndPuncttrue
\mciteSetBstMidEndSepPunct{\mcitedefaultmidpunct}
{\mcitedefaultendpunct}{\mcitedefaultseppunct}\relax
\EndOfBibitem
\bibitem{ghostprob}
M.~De~Cian, S.~Farry, P.~Seyfert, and S.~Stahl,
  \ifthenelse{\boolean{articletitles}}{\emph{{Fast neural-net based fake track
  rejection in the LHCb reconstruction}}, }{}
  \href{http://cdsweb.cern.ch/search?p=LHCb-PUB-2017-011.
  CERN-LHCb-PUB-2017-011&f=reportnumber&action_search=Search&c=LHCb+Notes}
  {LHCb-PUB-2017-011. CERN-LHCb-PUB-2017-011}\relax
\mciteBstWouldAddEndPuncttrue
\mciteSetBstMidEndSepPunct{\mcitedefaultmidpunct}
{\mcitedefaultendpunct}{\mcitedefaultseppunct}\relax
\EndOfBibitem
\bibitem{LHCb-DP-2013-002}
LHCb collaboration, R.~Aaij {\em et~al.},
  \ifthenelse{\boolean{articletitles}}{\emph{{Measurement of the track
  reconstruction efficiency at LHCb}},
  }{}\href{https://doi.org/10.1088/1748-0221/10/02/P02007}{JINST \textbf{10}
  (2015) P02007}, \href{http://arxiv.org/abs/1408.1251}{{\normalfont\ttfamily
  arXiv:1408.1251}}\relax
\mciteBstWouldAddEndPuncttrue
\mciteSetBstMidEndSepPunct{\mcitedefaultmidpunct}
{\mcitedefaultendpunct}{\mcitedefaultseppunct}\relax
\EndOfBibitem
\bibitem{Skwarnicki:1986xj}
T.~Skwarnicki, {\em {A study of the radiative cascade transitions between the
  Upsilon-prime and Upsilon resonances}}, PhD thesis, Institute of Nuclear
  Physics, Krakow, 1986,
  {\href{http://inspirehep.net/record/230779/}{DESY-F31-86-02}}\relax
\mciteBstWouldAddEndPuncttrue
\mciteSetBstMidEndSepPunct{\mcitedefaultmidpunct}
{\mcitedefaultendpunct}{\mcitedefaultseppunct}\relax
\EndOfBibitem
\bibitem{Breton:681262}
V.~Breton, N.~Brun, and P.~Perret,
  \ifthenelse{\boolean{articletitles}}{\emph{{A clustering algorithm for the
  LHCb electromagnetic calorimeter using a cellular automaton}}, }{}
  \href{http://cdsweb.cern.ch/search?p=LHCb-2001-123&f=reportnumber&action_search=Search&c=LHCb+Reports}
  {LHCb-2001-123}\relax
\mciteBstWouldAddEndPuncttrue
\mciteSetBstMidEndSepPunct{\mcitedefaultmidpunct}
{\mcitedefaultendpunct}{\mcitedefaultseppunct}\relax
\EndOfBibitem
\bibitem{LHCb-DP-2013-001}
F.~Archilli {\em et~al.},
  \ifthenelse{\boolean{articletitles}}{\emph{{Performance of the muon
  identification at LHCb}},
  }{}\href{https://doi.org/10.1088/1748-0221/8/10/P10020}{JINST \textbf{8}
  (2013) P10020}, \href{http://arxiv.org/abs/1306.0249}{{\normalfont\ttfamily
  arXiv:1306.0249}}\relax
\mciteBstWouldAddEndPuncttrue
\mciteSetBstMidEndSepPunct{\mcitedefaultmidpunct}
{\mcitedefaultendpunct}{\mcitedefaultseppunct}\relax
\EndOfBibitem
\bibitem{LHCb-PUB-2011-003}
V.~V. Gligorov, \ifthenelse{\boolean{articletitles}}{\emph{{A single track HLT1
  trigger}}, }{}
  \href{http://cdsweb.cern.ch/search?p=LHCb-PUB-2011-003&f=reportnumber&action_search=Search&c=LHCb+Notes}
  {LHCb-PUB-2011-003}\relax
\mciteBstWouldAddEndPuncttrue
\mciteSetBstMidEndSepPunct{\mcitedefaultmidpunct}
{\mcitedefaultendpunct}{\mcitedefaultseppunct}\relax
\EndOfBibitem
\bibitem{gulin11a}
A.~Gulin, I.~Kuralenok, and D.~Pavlov,
  \ifthenelse{\boolean{articletitles}}{\emph{{Winning the transfer learning
  track of Yahoo's learning to rank challenge with YetiRank}}, }{} in {\em
  Proceedings of the Learning to Rank Challenge} (O.~Chapelle, Y.~Chang, and
  T.-Y. Liu, eds.), vol.~14 of {\em Proceedings of Machine Learning Research},
  (Haifa, Israel), pp.~63--76, PMLR, 25 Jun, 2011\relax
\mciteBstWouldAddEndPuncttrue
\mciteSetBstMidEndSepPunct{\mcitedefaultmidpunct}
{\mcitedefaultendpunct}{\mcitedefaultseppunct}\relax
\EndOfBibitem
\bibitem{Dettori:2297352}
F.~Dettori, D.~Martinez~Santos, and J.~Prisciandaro,
  \ifthenelse{\boolean{articletitles}}{\emph{{Low-$p_T$ dimuon triggers at LHCb
  in Run 2}}, }{}
  \href{http://cdsweb.cern.ch/search?p=LHCb-PUB-2017-023&f=reportnumber&action_search=Search&c=LHCb+Notes}
  {LHCb-PUB-2017-023}\relax
\mciteBstWouldAddEndPuncttrue
\mciteSetBstMidEndSepPunct{\mcitedefaultmidpunct}
{\mcitedefaultendpunct}{\mcitedefaultseppunct}\relax
\EndOfBibitem
\bibitem{Kenzie:2110638}
M.~W. Kenzie and V.~Gligorov,
  \ifthenelse{\boolean{articletitles}}{\emph{{Lifetime unbiased beauty and
  charm triggers at LHCb}}, }{}
  \href{http://cdsweb.cern.ch/search?p=LHCb-PUB-2015-026.
  CERN-LHCb-PUB-2015-026&f=reportnumber&action_search=Search&c=LHCb+Notes}
  {LHCb-PUB-2015-026. CERN-LHCb-PUB-2015-026}\relax
\mciteBstWouldAddEndPuncttrue
\mciteSetBstMidEndSepPunct{\mcitedefaultmidpunct}
{\mcitedefaultendpunct}{\mcitedefaultseppunct}\relax
\EndOfBibitem
\bibitem{LHCb-DP-2016-003}
K.~Carvalho~Akiba {\em et~al.}, \ifthenelse{\boolean{articletitles}}{\emph{{The
  HeRSCheL detector: high-rapidity shower counters for LHCb}},
  }{}\href{https://doi.org/10.1088/1748-0221/13/04/P04017}{JINST \textbf{13}
  (2018) P04017}, \href{http://arxiv.org/abs/1801.04281}{{\normalfont\ttfamily
  arXiv:1801.04281}}\relax
\mciteBstWouldAddEndPuncttrue
\mciteSetBstMidEndSepPunct{\mcitedefaultmidpunct}
{\mcitedefaultendpunct}{\mcitedefaultseppunct}\relax
\EndOfBibitem
\bibitem{LHCb-PUB-2011-016}
V.~V. Gligorov, C.~Thomas, and M.~Williams,
  \ifthenelse{\boolean{articletitles}}{\emph{{The HLT inclusive B triggers}},
  }{} \href{http://cdsweb.cern.ch/search?p=LHCb-PUB-2011-016.
  CERN-LHCb-PUB-2011-016.
  LHCb-INT-2011-030&f=reportnumber&action_search=Search&c=LHCb+Notes}
  {LHCb-PUB-2011-016. CERN-LHCb-PUB-2011-016. LHCb-INT-2011-030},
  LHCb-INT-2011-030\relax
\mciteBstWouldAddEndPuncttrue
\mciteSetBstMidEndSepPunct{\mcitedefaultmidpunct}
{\mcitedefaultendpunct}{\mcitedefaultseppunct}\relax
\EndOfBibitem
\bibitem{BBDT}
V.~V. Gligorov and M.~Williams,
  \ifthenelse{\boolean{articletitles}}{\emph{{Efficient, reliable and fast
  high-level triggering using a bonsai boosted decision tree}},
  }{}\href{https://doi.org/10.1088/1748-0221/8/02/P02013}{JINST \textbf{8}
  (2013) P02013}, \href{http://arxiv.org/abs/1210.6861}{{\normalfont\ttfamily
  arXiv:1210.6861}}\relax
\mciteBstWouldAddEndPuncttrue
\mciteSetBstMidEndSepPunct{\mcitedefaultmidpunct}
{\mcitedefaultendpunct}{\mcitedefaultseppunct}\relax
\EndOfBibitem
\bibitem{LHCb-PROC-2015-018}
T.~Likhomanenko {\em et~al.}, \ifthenelse{\boolean{articletitles}}{\emph{{LHCb
  topological trigger reoptimization}},
  }{}\href{https://doi.org/10.1088/1742-6596/664/8/082025}{J.\ Phys.\ Conf.\
  Ser.\  \textbf{664} (2015) 082025}\relax
\mciteBstWouldAddEndPuncttrue
\mciteSetBstMidEndSepPunct{\mcitedefaultmidpunct}
{\mcitedefaultendpunct}{\mcitedefaultseppunct}\relax
\EndOfBibitem
\end{mcitethebibliography}

\newpage
\end{document}